%% file: sample-acmsmall.tex
\documentclass[acmsmall,screen]{acmart}
\input{setup}

\AtBeginDocument{%
  \providecommand\BibTeX{{%
    \normalfont B\kern-0.5em{\scshape i\kern-0.25em b}\kern-0.8em\TeX}}}

\setcopyright{acmcopyright}
\copyrightyear{2025}
\acmYear{2025}
\acmDOI{XXXXXXX.XXXXXXX}

\acmJournal{JACM}
\acmVolume{37}
\acmNumber{4}
\acmArticle{111}
\acmMonth{8}

\begin{document}

\title{Assessing and Advancing Benchmarks for Evaluating Large Language Models in Software Engineering Tasks}

\author{Xing Hu}
\email{xinghu@zju.edu.cn}
\orcid{0000-0003-0093-3292}
\affiliation{%
  \institution{School of Software Technology, Zhejiang University}
  \city{Ningbo}
  \state{Zhejiang}
  \country{China}
}

\author{Feifei Niu}
\authornote{Corresponding author.}
\email{feifeiniu96@gmail.com}
\orcid{0000-0002-4123-4554}
\affiliation{%
  \institution{School of Electrical Engineering and Computer Science, University of Ottawa}
  \city{Ottawa}
  \country{Canada}
}

\author{Junkai Chen}
\orcid{0009-0000-9945-7729}
\affiliation{%
  \institution{School of Computing and Information Systems, Singapore Management University}
  \country{Singapore}
}
\email{junkaichen2000@gmail.com}

\author{Xin Zhou}
\orcid{https://orcid.org/0000-0002-4558-0622}
\affiliation{%
  \institution{School of Computing and Information Systems, Singapore Management University}
  \country{Singapore}
}
\email{xinzhou.2020@phdcs.smu.edu.sg}

\author{Junwei Zhang}
\orcid{0000-0003-4323-8951}
\affiliation{%
  \institution{College of Computer Science and Technology, Zhejiang University}
  \city{Hangzhou}
  \country{China}
}
\email{jw.zhang@zju.edu.cn}

\author{Junda He}
\orcid{0000-0003-3370-8585}
\email{jundahe.2022@phdcs.smu.edu.sg}
\affiliation{%
  \institution{School of Computing and Information Systems, Singapore Management University}
  \country{Singapore}
}

\author{Xin Xia}
\email{xin.xia@acm.org}
\orcid{0000-0002-6302-3256}
\affiliation{%
  \institution{College of Computer Science and Technology, Zhejiang University}
  \city{Hangzhou}
  \state{Zhejiang}
  \country{China}
}

\author{David Lo}
\email{davidlo@smu.edu.sg}
\orcid{0000-0002-4367-7201}
\affiliation{%
  \institution{School of Computing and Information Systems, Singapore Management University}
  \country{Singapore}
}

\begin{abstract}
Large language models (LLMs) are gaining increasing popularity in software engineering (SE) due to their unprecedented performance across various applications. These models are increasingly being utilized for a range of SE tasks, including requirements engineering and design, code analysis and generation, software maintenance, and quality assurance. As LLMs become more integral to SE, evaluating their effectiveness is crucial for understanding their potential in this field. In recent years, substantial efforts have been made to assess LLM performance in various SE tasks, resulting in the creation of several benchmarks tailored to this purpose. This paper offers a thorough review of \papers~benchmarks, addressing three main aspects: \emph{what benchmarks are available}, \emph{how benchmarks are constructed}, and \emph{the future outlook for these benchmarks}. We begin by examining SE tasks such as requirements engineering and design, coding assistant, software testing, AIOps, software maintenance, and quality management. We then analyze the benchmarks and their development processes, highlighting the limitations of existing benchmarks. Additionally, we discuss the successes and failures of LLMs in different software tasks and explore future opportunities and challenges for SE-related benchmarks. We aim to provide a comprehensive overview of benchmark research in SE and offer insights to support the creation of more effective evaluation tools.

\end{abstract}

\keywords{Large Language Models, Benchmark, Software Engineering, Evaluation}

\maketitle

\input{introduction}

\input{background}

\input{methodology}
\input{R_Requirement}

\input{R_development}

\input{R_Testing}
\input{R_AIOps}
\input{R_Code_Analysis} 
\input{R_quality}
\input{Summary}
\input{Challenges_Opportunities}
\input{threats}
\input{conclusion}

\bibliographystyle{ACM-Reference-Format}
\input{output.bbl}

\input{appendix}
\end{document}

%% file: setup.tex
\usepackage{multirow}
\usepackage[figuresleft]{rotating}
\usepackage{soul}
\usepackage{color, xcolor}
\usepackage{array}
\usepackage{enumitem}
\usepackage{comment}
\usepackage{multicol}
\usepackage{pifont}
\usepackage{makecell}
\usepackage{colortbl}
\usepackage[normalem]{ulem}
\usepackage{longtable}  
\useunder{\uline}{\ul}{}

\newcommand{\todo}[1]{\textcolor{red}{{\bf [TODO: #1]}}}
\newcommand{\hx}[1]{\textcolor{black}{#1}}
\newcommand{\zjw}[1]{\textcolor{black}{#1}}
\newcommand{\zx}[1]{\textcolor{black}{#1}}
\newcommand{\junda}[1]{\textcolor{black}{#1}}
\newcommand{\jk}[1]{\textcolor{black}{#1}}
\newcommand{\cjk}[1]{\textcolor{black}{#1}}

\newcommand{\nff}[1]{\textcolor{black}{#1}}

\newcommand{\phead}[1]{\noindent\textbf{#1}}

\newcommand{\papers}{291}

\newcommand{\highlight}{\cellcolor[HTML]{EFEFEF}}

%% file: introduction.tex
\section{Introduction}
Large language models (LLMs), such as ChatGPT~\cite{chatgpt} and GPT-4o~\cite{gpt-4o}, have recently received great attention from both academia and industry community, because of their remarkable performance across a wide range of tasks, from natural language processing (NLP)~\cite{gao2024llm,yang2024harnessing, karanikolas2023large,hu2024rag,tang2023domain, zan2023large} to reasoning~\cite{plaat2024reasoning, lee2024reasoning, huang2022towards} and even creative generation~\cite{franceschelli2024creativity, chakrabarty2023creativity, gu2024llms}. Their ability to understand and generate text has been leveraged in various domains, including software engineering (SE)~\cite{fan2023large, DBLP:journals/tse/WangHCLWW24, li2024large, yu2024where, DBLP:conf/iclr/ChenZNZLLC23, belzner2023large, tang2023domain,ren2023misuse, yu2024fight, zhang2023repocoder}. LLMs have showcased remarkable adaptability, significantly enhancing the performance of software engineering (SE) tasks, such as code generation~\cite{cassano2023multipl, DBLP:journals/corr/abs-2108-07732, codegeex, li2022alphacode, le2022coderl}, automated program repair~\cite{cao2023study, InferredBugs, DBLP:journals/corr/abs-2405-01466}, and test case generation~\cite{DBLP:journals/tse/PanichellaKT18, DBLP:journals/corr/abs-2302-10352}. These advancements highlight the potential of LLMs to transform the way we approach SE tasks.
Meanwhile, to evaluate the capability of LLMs in SE, numerous benchmarks have been developed, such as HumanEval~\cite{chen2021codex} for the code generation task, SWE-bench~\cite{jimenez2024swe} for real GitHub issues, and CodeXGLUE~\cite{lu2021codexglue} for code intelligence tasks.


Benchmarks are of paramount prominence to the successful application of LLMs in SE tasks. 
They provide a standardized framework for evaluating LLMs across different models and SE tasks, ensuring consistency and enabling direct comparisons under uniform conditions~\cite{mcintosh2024inadequacies, lee2024checkeval}. Furthermore, benchmarks are essential for understanding the performance of LLMs, as they highlight the strengths and weaknesses of current models, guiding future research and development efforts. For example, the evaluation of SWE-bench~\cite{jimenez2024swe} demonstrates that current state-of-the-art LLMs still have considerable room for improvement in addressing real GitHub issues, as they can only resolve the simplest problems. This underscores the direction for the future development of LLMs: to be more practical, intelligent, and autonomous.
In essence, benchmarks are indispensable for the rigorous, transparent, and collaborative evaluation of LLMs~\cite{DBLP:journals/tist/ChangWWWYZCYWWYZCYYX24, guo2023evaluating}. They establish a solid foundation for tracking progress, fostering innovation, and ensuring that advancements in AI are translated into meaningful improvements in real-world SE applications.

As LLMs grow in complexity and capability, a variety of benchmarks have been tailored in different ways, with different evaluation metrics, aiming for evaluating different LLMs on specific SE tasks. For example, some benchmarks are primarily constructed through manual effort~\cite{chen2021codex, codegeex}, some are tailored by automated collection methods~\cite{lu2021codexglue, DBLP:journals/corr/abs-1909-09436}, and others leverage a combination of both approaches~\cite{jimenez2024swe, swe-bench-verified, wang2023mconala, hendrycksapps2021}. However, currently, there is no systematic literature review (SLR) that provides a comprehensive overview of these benchmarks and their construction methods. Moreover, existing benchmarks may no longer suffice for assessing their performance and potential risks in SE tasks. To this end, we conduct an SLR on SE-related benchmarks. Firstly, we identify the current SE tasks being addressed by LLMs, which mainly include requirements engineering and design, coding assistant, software testing, AIOps, software maintenance, and quality management. Next, we identify and analyze the benchmarks~\footnote{\nff{A benchmark is defined here as a dataset for evaluating LLMs; some are dataset-only without explicit metrics, but are included given their wide adoption in the community.}} used to evaluate these solutions across three dimensions: 1) \emph{what
benchmarks are available}, 2) \emph{how benchmarks are constructed}, and 3) \emph{the future outlook for these benchmarks}. Specifically, ``what benchmarks are available'' encapsulates existing SE-related benchmarks that have been used for evaluating LLMs, along with the proposed evaluation metrics. ``How benchmark are constructed'' explores the methodologies behind creating these benchmarks, providing guidance for future development. ``The future outlook for these benchmarks'' addresses the resolved issues and the remaining challenges within the domain of SE-related benchmarks.

In the literature, there have been a substantial number of detailed and systematic reviews on LLMs' application in SE or benchmarking for general LLMs ~\cite{DBLP:journals/tosem/HouZLYWLLLGW24, zhang2023survey, DBLP:journals/tse/WangHCLWW24, fan2023large, DBLP:journals/corr/abs-2405-01466, cao2025toward}. However, this paper serves as the first comprehensive survey on existing SE-related benchmarks for LLMs. This study reviews current benchmarks, raise awareness within the community about their importance, and, crucially, illuminate future research directions for developing new benchmarks. The main contributions of this paper are as follows:

\begin{itemize}
    \item We are the first to present a comprehensive SLR on \papers~benchmarks released before June 2025, which focus on the evaluation of LLM-based solutions to address SE challenges. We conducted a detailed analysis of the benchmarks.
    
    \item We have classified the \papers~benchmarks based on their applicability to specific SE tasks and further summarized their usage, construction methods, and trends within each task.
    
    \item We have identified key challenges associated with LLM benchmarks in the SE domain and proposed several potential research directions for Benchmark4SE.

    \item \nff{We have released an open-source GitHub repository~\cite{LLM4SEBenchmarks} that curates and maintains high-quality benchmarks for evaluating LLMs in software engineering tasks, aiming to foster fair comparison and broad community impact.}
    
\end{itemize}

The remainder of this article is organized as follows:
In Section~\ref{sec: background}, we provide background knowledge of LLMs and related surveys on LLMs and benchmarks. 
In Section~\ref{sec:methodology}, we introduce
the methodology to conduct the survey. 
In Section 4\textasciitilde9, we introduce the
details of up-to-date benchmarks under each SE task. We comprehensively discuss the future challenges and opportunities of LLM benchmarks in the field of SE in Section~\ref{sec:challengesandopportunities}.
We discussed the threats to validity
in Sections~\ref{sec:threats}. 
Finally, we conclude the whole study and summarise future work in Section~\ref{sec:conclusion}.

%% file: background.tex
\section{Background and Related Work}
\label{sec: background}

\subsection{Large Language Models for Software Engineering}

LLMs are huge language models pre-trained on large amounts of datasets~\cite{yang2024harnessing}, and have shown great performance in various NLP tasks. 
Recent years have seen an emerging interest in training LLMs for software engineering by learning from large codebases (e.g., GitHub).
These SE-related LLMs are exploited to solve software engineering tasks, such as code generation, code search, and test generation.
According to Pan et al.~\cite{pan2023unifying} and Yang et al.~\cite{yang2024harnessing}, LLMs are primarily categorized into three types based on their architectures: encoder-only, decoder-only, and encoder-decoder models. Each of these architectures is suited to different software engineering tasks. For example, encoder-only models (e.g., BERT and its variants) excel at understanding and representing code, making them highly effective for tasks such as code search~\cite{wang2023natural} and bug detection~\cite{ni2022best}. In contrast, decoder-only models (e.g., GPT) are particularly well-suited for generation tasks, such as automatic code generation~\cite{chen2021codex}. Meanwhile, encoder-decoder models (e.g., T5~\cite{t5}) perform exceptionally well in tasks that involve transforming or summarizing code~\cite{wan2018improving}.

\subsection{Related Work}

 In addition, there are more survey studies focusing on specific software engineering tasks, such as code generation~\cite{chen2024survey, wan2024deep}, requirements-based test generation~\cite{yang2025requirements}, bug localization~\cite{niu2025deep}, software defects datasets~\cite{zhu2025bugs}.
 With the rise of LLM-based agents, several survey papers have emerged that investigate the use of agents in software engineering~\cite{jin2024llms, liu2024large, he2024llm, wang2024agents}.

The study of SE-related LLMs has gained significant attention due to their transformative impact on software engineering tasks such as code generation, bug detection, and program synthesis. 
Existing surveys focus on the general application of LLMs within SE. For example, Zhang et al.~\cite{zhang2023survey} comprehensively summarized LLM-based SE research from two key perspectives: LLM architectures and their applications. Their review spans 30 LLMs, 15 pre-training tasks, 16 downstream tasks, and 43 specific code-related tasks. 
Similarly, Wang et al.~\cite{DBLP:journals/tse/WangHCLWW24} reviewed 102 studies on the use of LLMs in software testing, analyzing them through the lenses of both testing and LLM methodologies. 
Fan et al.~\cite{fan2023large} explored the achievements and challenges of LLMs across various SE domains, including coding, design, requirements, repair, refactoring, performance optimization, documentation, and analytics. 
Hou et al.~\cite{DBLP:journals/tosem/HouZLYWLLLGW24} conducted a systematic review of LLMs for SE (LLM4SE), analyzing 395 research papers published from January 2017 to January 2024. In addition, an increasing number of survey studies have focused on specific software engineering tasks, such as code generation~\cite{chen2024survey, wan2024deep}, requirements-based test generation~\cite{yang2025requirements}, bug localization~\cite{niu2025deep}, and software defect datasets~\cite{zhu2025bugs}. Zhang et al.~\cite{DBLP:journals/corr/abs-2405-01466} focused specifically on automated program repair (APR) studies powered by advanced LLMs. With the rise of LLM-based agents, several recent surveys have investigated their application in software engineering~\cite{jin2024llms, liu2024large, he2024llm, wang2024agents}.
However, these studies generally target the broader SE/testing workflow rather than providing an in-depth analysis of benchmarks for evaluating code-related LLMs. Some surveys have addressed this gap by examining evaluation methods and benchmarks. For instance, Chang et al.~\cite{DBLP:journals/tist/ChangWWWYZCYWWYZCYYX24} reviewed evaluation methods for LLMs, organizing their discussion around three key dimensions: what, where, and how to evaluate. Hirsch et al.~\cite{DBLP:journals/jss/HirschH22} systematically reviewed 73 benchmarks for debugging, highlighting their diversity in programming languages, bug types, and available metadata, as well as the associated challenges for researchers. Wang et al.~\cite{wang2025software} provided an overview of 181 benchmarks related to code LLMs and agents. Cao et al.~\cite{cao2025should} profiled 274 code-related benchmarks to investigate how to design benchmarks that ensure quality, reliability, and reproducibility.
This paper revisits these foundational benchmarks, offering a detailed analysis of their strengths, limitations, and evolution. The goal is to deepen understanding of how current benchmarks shape the assessment of LLMs and to identify opportunities for enhancing future evaluations in this rapidly evolving field.







%% file: methodology.tex
\section{Methodology}\label{sec:methodology}
\begin{figure*}[!t]
\centering
\includegraphics[width=\textwidth]{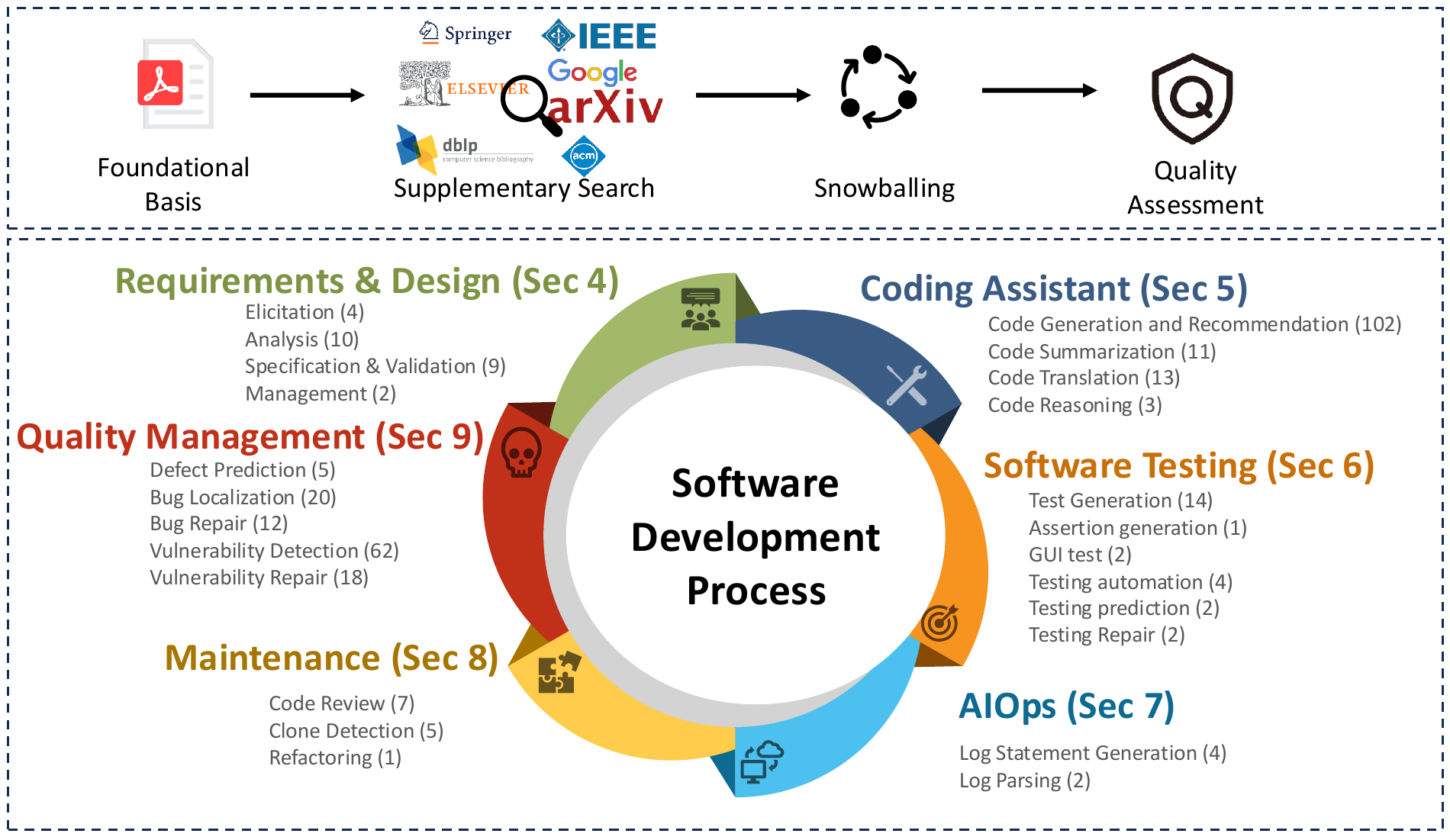}
\caption{Overview of this survey.}
\Description[]{}
\label{fig:overview}
\end{figure*}

To ensure an unbiased and repeatable procedure, this SLR follows the guidelines of Kitchenham \cite{kitchenham2004procedures} and Zhang et al. \cite{zhang2011identifying}. It was conducted in three stages: planning, execution, and analysis. During the planning stage, we identified the research questions and developed the survey protocols to align with our research aims and objectives. In the execution stage, we implemented the study according to the established protocols, performed study selection, and employed snowballing to ensure a comprehensive collection of literature on benchmarks for LLMs in SE. Finally, in the analysis stage, six of our co-authors analyzed the selected literature to address the research questions. The research process is illustrated in Figure~\ref{fig:overview}.

\subsection{Research Questions}
The primary objective of this survey is to gain insights into benchmarks for evaluating software-related LLMs. To provide a comprehensive view of this topic, the survey addresses three research questions. These questions enable us to systematically categorize and understand current research, identify limitations, and suggest future research directions. The three research questions are as follows:
    
\begin{enumerate}[label=\textbf{RQ\arabic*.}]
    \item \textbf{What benchmarks have been developed to date for evaluating LLMs in SE tasks?}

    This RQ identifies and categorizes the existing benchmarks that have been developed for evaluating LLMs in SE tasks. It seeks to provide a comprehensive overview of these benchmarks, including the types of SE tasks they address (e.g., code generation, bug fixing, testing), the programming languages covered, and the key metrics used for evaluation.



    \item \textbf{How were existing benchmarks created?}

    This RQ investigates the methodologies and data sources used to create existing benchmarks for LLMs in SE. It aims to explore the processes of data collection, task formulation, and evaluation criteria definition, highlighting common practices and potential gaps in the creation of these benchmarks.

    \item \textbf{What are the challenges of benchmarks for LLMs in SE?}

    This RQ aims to point out limitations in the existing benchmarks, with the goal of identifying key challenges. By understanding these challenges, this RQ will provide insights into potential improvements and opportunities for future benchmark development in evaluating LLMs in SE tasks.
\end{enumerate}

\subsection{Study Selection}
\subsubsection{Study Search}
To identify relevant studies for our systematic review, we leverage the existing SLRs~\cite{DBLP:journals/tosem/HouZLYWLLLGW24, fan2023large, zhang2023survey} on LLMs in SE as a foundational basis. These papers provided comprehensive overviews of LLMs and their applications in SE. Building upon these work, we specifically focus on the benchmarks used to evaluate LLMs across various SE tasks. Specifically, we thoroughly examined the papers reviewed in their SLRs, extracting the benchmarks used for evaluating LLMs in software engineering, which we use as our primary studies.

Additionally, to ensure the comprehensiveness for our review, we conducted a supplementary search in relevant academic databases (as shown in Table~\ref{tab:digitallibrary}), \nff{following the search strategy and search strings defined by Hou et al.~\cite{DBLP:journals/tosem/HouZLYWLLLGW24}}, to capture any recent benchmarks or studies that may have been published after these SLRs was conducted. We also employed snowballing techniques to identify any related benchmarks that may have been missed in the initial review, thus expanding upon the existing body of research.

This study selection process allows us to capitalize on the thoroughness of previous SLRs while ensuring that our review specifically addresses the benchmarking aspect of LLMs in SE. By combining existing knowledge with new data collection efforts, we aim to provide a comprehensive and up-to-date overview of the current benchmarks.

\begin{center}
\begin{table}[ht]%
\caption{List of Digital Libraries.\label{tab:digitallibrary}}
\centering
\begin{tabular}{ll}
\toprule
Digital   libraries & URLs \\ 
\midrule
IEEE Xplore & http://www.ieee.org/web/publications /xplore/ \\
ACM Digital Library & http://portal.acm.org \\
Science Direct      & http://www.sciencedirect.com/  \\
Springer Link       & https://link.springer.com/    \\
Wiley InterScience  & https://onlinelibrary.wiley.com/   \\
Elsevier  & http://www.elsevier.com       \\
Google Scholar      & https://scholar.google.com    \\
DBLP  & https://dblp.org \\
arXiv & https://arxiv.org \\
\bottomrule
\end{tabular}
\end{table}
\end{center}


\subsubsection{Inclusion/Exclusion Criteria.} To identify the most relevant studies for proposing benchmarks to evaluate LLMs in SE tasks, we drew inspiration from prior works~\cite{giray2023use, zakari2020multiple, croft2022data}, and established a set of inclusion (ICs) and exclusion criteria (ECs). Table~\ref{tab:criteria} outlines these criteria. We applied the ICs and ECs to the titles, abstracts, and keywords of the retrieved studies to ensure a rigorous selection process.
This process enabled us to filter out irrelevant or low-quality papers, ensuring that only the most pertinent contributions were considered. The application of these criteria was conducted iteratively, with each stage involving a review and consensus among the authors. This helped mitigate bias and improve the reliability of the selection process.
Next, we performed a full-text review of the shortlisted papers, cross-referencing their methodologies and reported benchmarks to confirm alignment with our research objectives. This multi-step filtering approach ensured comprehensive coverage of studies relevant to benchmarking LLMs in SE tasks.

\begin{table}[]
\centering
\caption{Inclusion and Exclusion Criteria}\label{tab:criteria}
\begin{tabular}{m{0.1\linewidth}<{\raggedright}m{0.8\linewidth}<{\raggedright}}
\hline
\multicolumn{2}{l}{\textbf{Inclusion Criteria}} \\ \hline
\rowcolor[HTML]{EFEFEF} 
IC1       & The paper is written in English.       \\
IC2       & The paper is published before June 2025. \\
\rowcolor[HTML]{EFEFEF} 
IC3       & The paper is published in a peer-reviewed journal, conference or a workshop, or preprint at arXiv and full-text available.  \\
IC4       & The paper contributes a new benchmark within SE tasks. \\
\rowcolor[HTML]{EFEFEF} 
IC5       & The benchmark has been used for evaluating LLMs for SE tasks.\\
IC6       & If the benchmark does not have a corresponding academic paper but has been cited or used in \nff{at least three} academic studies and holds significant influence in the field, it can be included.\\
\rowcolor[HTML]{EFEFEF} 
IC7       & The benchmark must provide a complete description of the benchmark, including datasets, evaluation metrics, and task definitions.\\
\hline
\multicolumn{2}{l}{\textbf{Exclusion Criteria}} \\ \hline
\rowcolor[HTML]{EFEFEF} 
EC1       & The benchmark is not intended for evaluating the performance of LLMs in solving SE tasks. \\
EC2       & The proposed benchmark has not yet been utilized for evaluating LLMs, \nff{except for those recently introduced since 2025}. \\
\rowcolor[HTML]{EFEFEF} 
EC3       & The paper does not introduce a new benchmark; instead, it solely evaluates LLMs relying on existing benchmarks.\\
EC4       & The paper lacks detailed instructions on how the benchmark was developed/contributed.\\ \hline
\end{tabular}
\end{table}

\subsubsection{Snowballing}
To ensure the comprehensiveness of our review, we supplemented our primary search by employing snowballing under the guideline of Wohlin~\cite{wohlin2014guidelines}. This involved:

\noindent\textbf{Backward Snowballing}: Reviewing the references of the selected papers from prior benchmark studies to identify earlier studies that discuss benchmarks for LLMs in SE.

\noindent \textbf{Forward Snowballing}: Identifying newer studies that have cited the selected papers, ensuring that recent developments in benchmarks for LLMs were not overlooked. 
For each round of snowballing, we applied the ICs and ECs to screen the articles. This iterative process was conducted until no new relevant studies were identified, signaling saturation in the search.
To maintain consistency and avoid selection bias, two authors independently conducted the snowballing process. Discrepancies in the selection of papers were resolved through discussion and, if necessary, by consulting a third author. This collaborative approach ensured thorough coverage and minimized the risk of omitting key studies.
The combination of systematic search and snowballing enabled us to capture a broad spectrum of relevant literature, ensuring that the resulting set of benchmarks for evaluating LLMs in SE tasks is both comprehensive and up-to-date.

\subsubsection{Quality Assessment} Quality assessment is a crucial process to ensure that the conclusions drawn are based on reliable and high-quality evidence~\cite{kitchenham2004procedures}.
We assessed the quality of the selected studies based on criteria such as: clarity of benchmark description, relevance to software engineering, methodological rigor, reproducibility and availability, and impact on the field. Table~\ref{tab:qualityassessment} listed the quality assessment checklist, which was applied to all primary studies by six of our authors. Discussions were conducted in cases of disagreement to reach a consensus.

Finally, a total of \papers~benchmarks were identified in the final study pool upon completing the entire study selection process.

\begin{table}[htpb]
\centering
\caption{The Quality Checklist}\label{tab:qualityassessment}
\begin{tabular}{m{0.05\linewidth}<{\raggedright}m{0.9\linewidth}<{\raggedright}}
\hline
\multicolumn{2}{l}{\textbf{Clarity of Benchmark Description}} \\ \hline
\rowcolor[HTML]{EFEFEF} 
BD1       & Does the study clearly describe the benchmark, including the tasks, datasets, and evaluation metrics? \\
BD2       & Are the components of the benchmark relevant and well-justified in the context of software engineering? \\
\hline
\multicolumn{2}{l}{\textbf{Relevance to Software Engineering}} \\ \hline
\rowcolor[HTML]{EFEFEF} 
SE1       & Is the benchmark designed for specific software engineering tasks, such as code completion, bug fixing, or software testing? \\
SE2       & How well does the benchmark align with the challenges and requirements of software engineering? \\
\hline
\multicolumn{2}{l}{\textbf{Methodological Rigor}} \\ \hline
\rowcolor[HTML]{EFEFEF} 
MR1       & Is the process of benchmark development well-documented and scientifically rigorous? \\
MR2       & Are the methods used to construct the benchmark and apply it to LLMs clearly defined and appropriate for the task? \\
\hline
\multicolumn{2}{l}{\textbf{Reproducibility and Availability}} \\ \hline
\rowcolor[HTML]{EFEFEF} 
RA1       & Are the benchmark datasets, tasks, and evaluation scripts publicly available to facilitate reproducibility? \\
RA2       & Does the study provide sufficient details for others to reproduce the experiments? \\
\hline
\multicolumn{2}{l}{\textbf{Impact on the Field}} \\ \hline
\rowcolor[HTML]{EFEFEF} 
IF1       & Has the benchmark been widely adopted or cited in subsequent work? \\
IF2       & Does the study contribute to advancing the understanding or improvement of LLMs in software engineering? \\
\hline
\end{tabular}
\end{table}

\subsection{Data Analysis}
After identifying the final study pool, we proceeded with the data analysis phase.
We systematically extracted data from the selected studies to gather the necessary information to address our research questions. A structured data extraction form was developed to ensure consistency across all studies. The form included the following key elements:
\begin{itemize}[leftmargin=*]
    \item \textbf{Information:} Author(s), publication year, title, and source.
    \item \textbf{Research Focus:} The primary objective of the benchmark, specifically the SE tasks or LLMs being evaluated.
    \item \textbf{Benchmark Description:} Details of the benchmark, including the tasks, datasets, evaluation metrics, and their relevance to software engineering.
    \item \textbf{Methodology:} The methodologies used for constructing benchmarks and evaluating LLMs, particularly in terms of preprocessing and related aspects.
    \item \textbf{Challenges and Limitations:} Any challenges or limitations reported by the authors concerning the benchmark or LLM evaluation.
    \item \textbf{Citations and Adoption:} Whether the benchmark has been cited or adopted by subsequent studies.
    \item \textbf{Reproducibility and Availability:} The availability of the benchmark dataset, tasks, and evaluation scripts for public use.
\end{itemize}
Data extraction was performed independently by six co-authors. In cases of discrepancies, discussions were held to reach a consensus, ensuring the accuracy and completeness of the extracted data.

\subsection{Results}
\begin{figure*}[!t]
\centering
\begin{minipage}{0.48\textwidth}
    \centering
    \includegraphics[width=\textwidth]{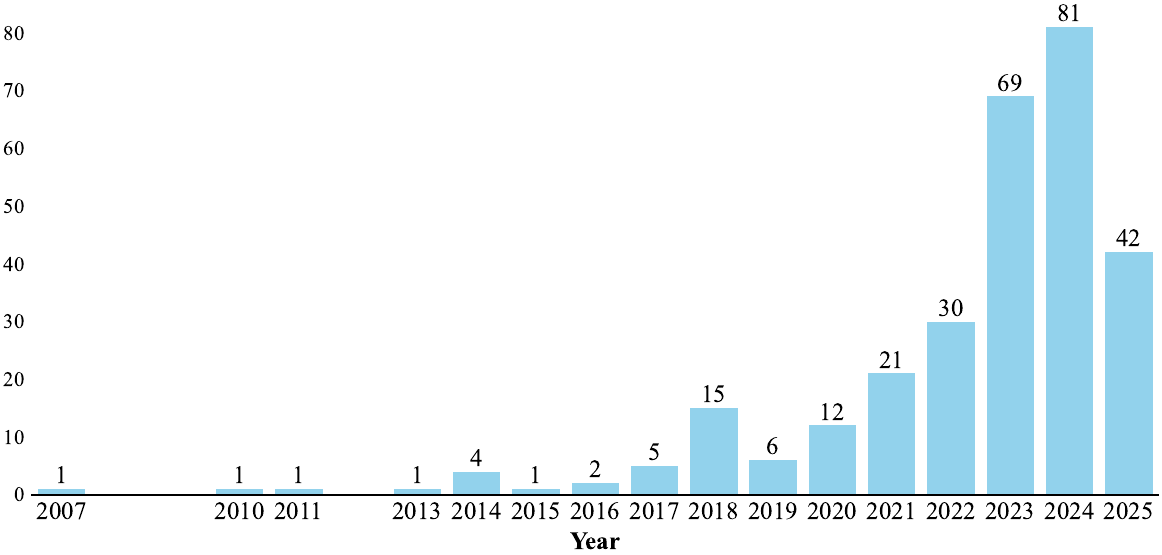}
    \caption{Number of benchmarks over the years.}
    \Description[]{}
    \label{fig:year}
\end{minipage}
\hfill
\begin{minipage}{0.46\textwidth}
    \centering
    \includegraphics[width=\textwidth]{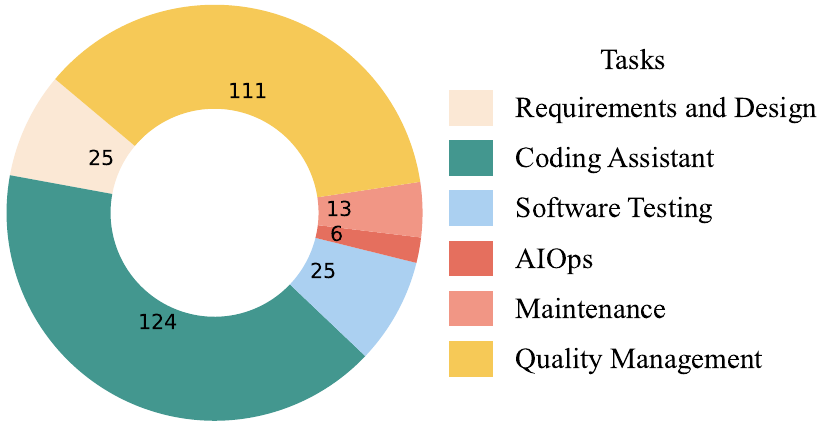}
    \caption{Benchmark Distribution by Task.}
    \Description[]{}
    \label{fig:task}
\end{minipage}
\end{figure*}

There are \papers~SE benchmarks in literature that have been applied to evaluate LLMs. Figure~\ref{fig:year} illustrates the temporal distribution of these benchmarks (\nff{A complete list of benchmarks categorized by their SE activities is provided in Appendix~\ref{appendix:A}}). The number of benchmarks has shown a rapid increase since 2022, with 30 benchmarks in 2022, 69 in 2023, 81 in 2024, and 42 in 2025, accounting for 76\% of all the benchmarks. This trend highlights a growing research interest in SE-LLM benchmarks over the past three years. It is worth noting that this survey was conducted as of June 2025; therefore, the reported benchmark data for 2025 may not reflect the full number of benchmarks released throughout the entire year. Additionally, as many benchmarks are first posted on arXiv, their arXiv publication year may differ from the year of formal acceptance. For accepted benchmarks, we adopt the acceptance year as the official benchmark year. Moreover, while some pre-LLM era datasets (e.g., Defects4J~\cite{just2014defects4j}, QuixBugs~\cite{QuixBugs}) continue to be used for evaluating LLMs, the development of new benchmarks to assess various LLM capabilities remains the dominant trend. The introduction of new benchmarks is crucial. On one hand, it mitigates the risk of data leakage during testing; on the other, it facilitates the evaluation of the diverse functionalities exhibited by LLMs.  

Figure~\ref{fig:task} illustrates the distribution of benchmarks across various software engineering tasks, \nff{while the lower part of Figure~\ref{fig:overview} presents the sub-tasks along with the corresponding benchmark counts (a detailed description of each category is listed in Appendix~\ref{appendix:B})}. The figure clearly shows that ``Coding Assistant'' holds the largest share, with 124 benchmarks, underscoring its prominence in current benchmarking efforts. ``Quality Management'' also represents substantial portions, with 111 benchmarks. Tasks such as ``Requirements and Design'' (25 benchmarks), ``Software Testing'' (25 benchmarks), and ``Maintenance'' (13 benchmarks) remain well-represented, reflecting their ongoing significance. In contrast, ``AIOps'' (6 benchmarks) exhibits comparatively lower benchmark coverage, indicating that this area may benefit from further development. The inclusion of 13 multi-faceted benchmarks~\cite{hendrycksapps2021, DBLP:journals/corr/abs-2108-07732, chen2021codex, li2022alphacode, siddiq2022securityeval, codegeex, DBLP:journals/corr/abs-2403-19287, just2014defects4j, DBLP:conf/sigsoft/WidyasariSLQPTT20, li2022automating, zhou2019devign,fan2020bigvul, liu2025vader} highlights the increasing focus on evaluating LLMs across multiple software engineering tasks, reinforcing the need for versatile benchmarks that assess a broad range of capabilities. 

\nff{While our study targets benchmarks for evaluating LLMs on SE tasks, and most curated datasets originate from or are primarily used within the SE community, we also observe a distinct subset rooted in, or widely adopted by, the ML/NLP community. Representative examples include HumanEval~\cite{openai_humaneval} and MBPP~\cite{DBLP:journals/corr/abs-2108-07732} (function-level code generation with pass@k), CodeXGLUE~\cite{lu2021codexglue} (multi-task code intelligence; NeurIPS), APPS~\cite{hendrycksapps2021} (programming problems with unit tests; NeurIPS), DS-1000~\cite{lai2023ds} (data-science coding; ICML), and xCodeEval~\cite{DBLP:conf/acl/KhanBLWPJ24} (multilingual/multitask code evaluation; ACL). By contrast, SE-focused datasets (e.g., Defects4J~\cite{just2014defects4j}, Bugs.jar~\cite{saha2018bugs}, ManyBugs~\cite{le2015manybugs}, SWE-Bench~\cite{jimenez2024swe}) emphasize executable oracles and repository context, and are most frequently cited in SE venues (ICSE, FSE, ASE, TSE).}

\begin{figure}[!t]
\centering
\includegraphics[width=\textwidth]{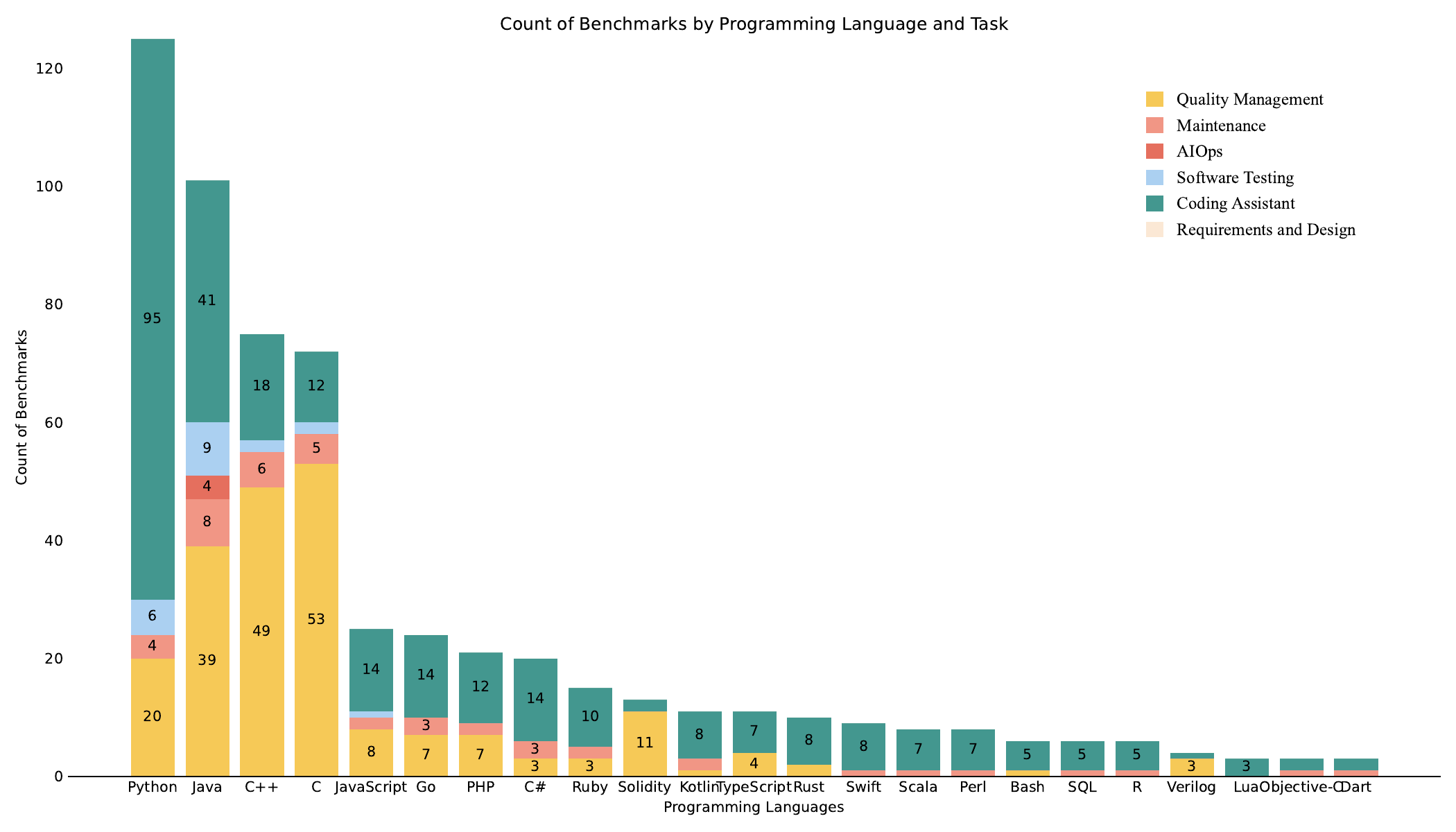}
\caption{Count of Benchmark by Programming Language and Task (Programming languages with fewer than three benchmarks are excluded).}
\Description[]{}
\label{fig:language}
\end{figure}

Figure~\ref{fig:language} presents the distribution of benchmarks across different programming languages, categorized by SE tasks (Figure~\ref{fig:language} excluded certain benchmarks that only involve natural language and do not involve programming languages. Furthermore, programming languages associated with fewer than three benchmarks were excluded from the diagram for clarity). From the figure, it is evident that Python leads by a significant margin, with the highest number of benchmarks (dominated by coding assistant, particularly code generation tasks), reflecting its widespread use and versatility in software engineering and machine learning contexts. Java, C++, and C follow, with notable contributions across multiple task categories, particularly in quality management and coding assistant. Languages such as JavaScript, Go, PHP, and C\# also feature prominently but to a lesser extent, suggesting their relevance in specific domains like web development and enterprise applications. In contrast, for newer or domain-specific languages—such as Ruby, Solidity, Kotlin, and so on—only limited benchmarks have emerged, reflecting increasing attention but underscoring the need for broader evaluation efforts. The figure further reveals that benchmarks are distributed unevenly, with a few dominant languages receiving extensive coverage, while many other languages have minimal representation. This trend suggests that while LLMs are well-supported in mainstream programming languages, further efforts may be needed to expand benchmark development for niche or emerging languages to ensure comprehensive evaluation across diverse coding environments.

%% file: R_Requirement.tex
\section{Requirements and Design}\label{sec:requirementsanddesign}

As the first phase of the software life cycle, requirements engineering (RE) involves elicitation, specification, and verification of user/customer needs~\cite{pohl1996requirements, niu2025feature}. It ensures clear communication among stakeholders, reducing the risk of miscommunication and misaligned expectations. By defining the project scope, it prevents scope creep and aids in accurate time and cost estimation, which is essential for effective project management. Well-defined requirements also serve as the foundation for testing, ensuring that the final product meets the intended functionality and helping to avoid costly rework~\cite{10.5555/1869735}. Additionally, requirements engineering enables better risk management by identifying potential issues early in the process, and it provides the adaptability needed to handle changes in a dynamic environment. Ultimately, this leads to higher quality, more reliable software, and efficient use of resources. This section presents an overview of existing benchmarks for evaluating LLMs in the context of software requirements and design.

\subsection{Benchmark Overview}
Hou et al.~\cite{DBLP:journals/tosem/HouZLYWLLLGW24} identified studies that use LLMs to address requirements engineering and design, with a total of 19 related studies. These studies mainly focus on anaphoric ambiguity treatment~\cite{ezzini2022automated, moharil2023tabasco, moharil2022identification, sridhara2023chatgpt}, requirements classification~\cite{el2023ai, helmeczi2023few, hey2020norbert, luo2022prcbert}, requirement analysis and evaluation~\cite{poudel2023leveraging, ronanki2022chatgpt}, specification generation~\cite{xie2023impact, ma2024specgen}, coreference detection~\cite{wang2020deep}, requirements elicitation~\cite{white2024chatgpt}, specification formalization~\cite{endres2023formalizing}, traceability automation~\cite{lin2021traceability}, use cases generation~\cite{zhang2024experimenting}, GUI retrieval~\cite{kolthoff2023data}, rapid prototyping~\cite{mandal2023large}, software specification synthesis~\cite{white2024chatgpt}, and system design~\cite{zhang2024experimenting}. Moreover, we also identified benchmarks on requirements generation~\cite{DBLP:journals/corr/abs-2404-01558, habib2025reqbrain, voria2025recover}, contradiction detection~\cite{DBLP:journals/ase/GartnerG24}, requirements traceability~\cite{preda2024supporting, DBLP:conf/re/KolthoffKBMP24}, verification~\cite{DBLP:journals/corr/abs-2411-11582}, and specification generation~\cite{DBLP:conf/re/KrishnaGVJ24, DBLP:journals/corr/abs-2504-20964}.
According to different requirements engineering activities, we categorize related benchmarks into requirements elicitation~\cite{wang2018aspects, DBLP:journals/corr/abs-2404-01558, habib2025reqbrain, voria2025recover}, requirements analysis~\cite{ferrari2017pure, abualhaija20203rd, ezzini2022automated, nfr2007jane, SecReq, dalpiaz2019requirements, luo2022prcbert, DBLP:journals/ase/GartnerG24, preda2024supporting, DBLP:conf/re/KolthoffKBMP24}, requirements specification and validation~\cite{svbenchmarks2024, blasi2018translating, xie2022docter, ma2024specgen, poudel2023leveraging, mandal2023large, DBLP:journals/corr/abs-2411-11582, DBLP:conf/re/KrishnaGVJ24, DBLP:journals/corr/abs-2504-20964}, and requirements management~\cite{helmeczi2023few, wang2020deep}.
\begin{table}[!htbp]
    \caption{Existing Benchmarks for Evaluating LLMs Related to Software Requirements and Design}
    \scriptsize
    \begin{tabular}{>{\centering}p{1.3cm}p{2.1cm}p{0.4cm}p{5.5cm}p{2.5cm}}
    \toprule
Task & Benchmarks & Year &  Evaluation Metrics & \# Cases\\ \hline
\multirow{7}{*}{\begin{tabular}[c]{@{}c@{}} Elicitation \end{tabular} } 
 & \highlight NFR-Review~\cite{wang2018aspects}   & \highlight 2018 & \highlight   - & \highlight 1278 user reviews \\ 
 & Rahman and Zhu~\cite{DBLP:journals/corr/abs-2404-01558} & 2024 & Readability, Understandability, Specifiability, and Technical-aspects & Seven RE documents \\
 & \highlight Habib et al.~\cite{habib2025reqbrain} & \highlight 2025 & \highlight Precision, Recall, and F & \highlight 242 high-quality requirements\\ 
 & Voria et al.~\cite{voria2025recover} & 2025 & Precision, Recall, F, Accuracy, BLUE, ROUGE, METEOR, Brevity Penalty, and Length Ratio & \\ 
 \hline
\multirow{13}{*}{Analysis} 
& \highlight PROMISE~\cite{Sayyad-Shirabad+Menzies:2005}  & \highlight 2005 & \highlight   - & \highlight 625 requirements \\
&  SecReq~\cite{SecReq}& 2010&  -&  510 requirements\\
& \highlight PURE~\cite{ferrari2017pure} & \highlight 2017 & \highlight - & \highlight 79 requirements \\
& Dalpiaz et al.~\cite{dalpiaz2019requirements}  & 2019 &  Precision, Recall, F1-score, and AUC   &   1,502 requirements \\
& \highlight ReqEval~\cite{abualhaija20203rd} &  \highlight 2020   & \highlight Precision, Recall, F2, success rate & \highlight 98 sentences \\
&  NFR-SO~\cite{luo2022prcbert}  &   2022    &  F1   & 17434 \\
& \highlight DAMIR~\cite{ezzini2022automated}  & \highlight 2022 & \highlight  Precision, Recall, F2, success rate   & \highlight 1251 sentences   \\
 
 & Gärtner and Göhlich~\cite{DBLP:journals/ase/GartnerG24} & 2024 & Accuracy, Precision, Recall and F & 5197 requirement pairs \\ 
 &\highlight  Preda et al.~\cite{preda2024supporting} & \highlight 2024 & \highlight Precision, Recall, and F & \highlight 318 high-level and 1,217 low-level requirements \\ 
 & Koltoff et al.~\cite{DBLP:conf/re/KolthoffKBMP24} & 2024 & Precision, Recall, F, and Accuracy & 60 GUIs and 231 user stories \\ 
 \hline
\multirow{13}{*}{\begin{tabular}[c]{@{}c@{}} Specification \\ \& Validation \end{tabular} } 
& \highlight Jdoctor~\cite{blasi2018translating} & \highlight 2018 & \highlight  Precision, Recall, F & \highlight 829 Javadoc comments \\
 &  DocTer~\cite{xie2022docter} & 2022&  Precision, Recall, F1  & 2,666 API documents \\ 
 &  \highlight Poudel et al.~\cite{poudel2023leveraging}   & \highlight  2023 &  \highlight    F2 and MAP &  \highlight  1,197 links \\ 
 & Mandal et al.~\cite{mandal2023large} & 2023 &  Precision, Recall, and F1 & 300 specifications  \\
 &  \highlight  SV-Benchmarks~\cite{svbenchmarks2024} &  \highlight 2024 &  \highlight   -  &   \highlight 265 class definitions  \\
 &SpecGenBench~\cite{ma2024specgen} &2024 &\#Passes, Success Probability, \#Verifier Calls, User Rating  & 120 programs \\
 & \highlight Reinpold et al.~\cite{DBLP:journals/corr/abs-2411-11582} & \highlight 2024 & \highlight Precision, Recall, and F & \highlight 27 system specifications and requirements \\
 &  Krishna et al.~\cite{DBLP:conf/re/KrishnaGVJ24} & 2024 & Unambiguous, understandable, correctness, verifiable, internal consistency, non-redundancy, completeness, and conciseness & 47 SRS  \\
 & \highlight OSVBench~\cite{DBLP:journals/corr/abs-2504-20964} & \highlight 2025 & \highlight Pass@N, Syntax Error, and Semantic Error & \highlight 245 specification generation tasks \\ 
 \hline
\multirow{2}{*}{Management} 
 &  Wang et al.~\cite{wang2020deep}  &  2020 &  Precision, Recall, and F1   & 1,949 text-entity pairs \\
& \highlight Helmeczi et al.~\cite{helmeczi2023few} & \highlight 2023 & \highlight Accuracy, F1  & \highlight 26,940 pairs \\


 \bottomrule
 \end{tabular}
 \label{tab:requirement}
\end{table}

\subsubsection{Benchmarking Scenarios}
In total, there are 25 benchmarks on software requirements and analysis (as shown in Table~\ref{tab:requirement}). These benchmarks can be categorized by task type into classification~\cite{abualhaija20203rd, ezzini2022automated, nfr2007jane, SecReq, wang2018aspects, dalpiaz2019requirements, luo2022prcbert, poudel2023leveraging, helmeczi2023few, wang2020deep, mandal2023large, DBLP:journals/ase/GartnerG24, preda2024supporting, DBLP:conf/re/KolthoffKBMP24}, recommendation~\cite{kolthoff2023data}, generation~\cite{svbenchmarks2024, ma2024specgen, blasi2018translating, xie2022docter, DBLP:journals/corr/abs-2404-01558, habib2025reqbrain, voria2025recover, DBLP:conf/re/KrishnaGVJ24, DBLP:journals/corr/abs-2504-20964}, and multi-task scenarios~\cite{ferrari2017pure}. 
In software engineering, benchmarks related to software requirements primarily focus on natural language text, with limited involvement of programming languages~\cite{liu2024your, zhang2024experimenting}. These requirements are often sourced from user reviews~\cite{wang2018aspects}, issues~\cite{ma2024specgen}, or software requirements~\cite{SecReq, helmeczi2023few, wang2020deep}. Analyzing software requirements remains one of the most challenging aspects of software engineering research, largely due to the limited availability of real-world software requirements. As confidential assets of companies, genuine software requirements are rarely accessible to academic researchers.
From a chronological perspective, most benchmarks used to evaluate LLMs in the context of requirements and design are derived from datasets predating the large model era, with the exception of HumanEval+~\cite{liu2024your}, CAASD~\cite{zhang2024experimenting}, SV-Benchmarks~\cite{svbenchmarks2024}, and SpecGenBench~\cite{ma2024specgen}, which were introduced in the recent two years. These traditional benchmarks are relatively small in scale, typically consisting of only a few hundred to a few thousand requirements or user reviews, in contrast to more recent and complex benchmarks.

\subsubsection{Evaluation Metrics}
The evaluation metrics vary depending on the task and can be categorized into classification-based, recommendation-based, and generation-based metrics.
For classification tasks, the most commonly used metrics include precision, recall, F-score, and AUC.
Recommendation tasks are evaluated using metrics such as MAP, MRR, and Precision@k.
For generation tasks, the evaluation primarily focuses on whether the generated content meets criteria, such as, pass, success, and other related metrics.
\subsection{Benchmarcks Construction}

\textbf{Elicitation.} NFR-Review~\cite{wang2018aspects} consists of 1278 user review sentences that come from 4000 randomly collected user review sentences of iBooks (iOS) and WhatsApp (Android) -- 2,000 sentences for each app. These user review sentences are manually classified into five NFR categories: reliability, usability, performance efficiency, portability, security, according to ISO/IEC 25010~\cite{international2011systems}. Rahman and Zhu~\cite{DBLP:journals/pacmpl/RahmanKCGWMDR25} collected seven mid-sized RE documents from a SE textbook~\cite{sommerville2015software} and a local software development company. These documents were used to evaluate LLMs’ performance in terms of readability, understandability, specificity, and technical aspects in user story generation. Habib et al.~\cite{habib2025reqbrain} filted 242 high-quality requirements that compliant with ISO 29148 from 10 online software requirements dataset.

\noindent
\textbf{Analysis.} 
PROMISE dataset, introduced by Shirabad and Menzies~\cite{Sayyad-Shirabad+Menzies:2005}, contains 625 requirements from 15 projects, with 255 being functional requirements and 370 being non-functional requirements (NFR). These NFRs are further partitioned into: Availability, Legal, Look and Feel, Maintainability, Operational, Performance, Scalability, Security, Usability, Fault Tolerance, and Portability. 
SecReq dataset~\cite{SecReq} contains 510 requirements, made of security-related requirements (187) and non-security related requirements (323). The requirements were collected from three industrial projects: Common Electronic Purse (ePurse), Customer Premises Network (CPN), and Global Platform Spec (GPS). The PURE dataset~\cite{ferrari2017pure} is composed of 79 publicly available natural language requirements documents collected from the Web. The dataset includes 34,268 sentences and is designed for natural language requirements processing tasks. Dalpiaz et al.~\cite{dalpiaz2019requirements} obtained eight datasets and manually identified requirements that possess functional aspect and quality aspect. 
Ezzini et al.~\cite{ezzini2022automated} curated two datasets, i.e., ReqEval~\cite{abualhaija20203rd} and DAMIR~\cite{ezzini2022automated} for evaluating the performance of the proposed SpanBERT model for anaphoric ambiguity detection. Sridhara et al.~\cite{sridhara2023chatgpt} employed the same datasets to evaluate the performance of ChatGPT for anaphora resolution. The ReqEval dataset~\cite{abualhaija20203rd} is composed of 200 sentences, out of which 102 are marked as ambiguous. The DAMIR dataset~\cite{ezzini2022automated} consists of 22 industrial requirements specifications from eight application domains: satellite communications, medicine, aerospace, security, digitization, automotive, railway, and defence. The requirements in these specifications are manually annotated as ambiguous or not. 
NFR-SO~\cite{luo2022prcbert} (non-functional requirements from StackOverflow) is crawled from the real-world technical forum StackOverflow, with 17,434 samples. Each sample is manually labeled as one of the 7 NFR categories, including availability, performance, maintainability, portability, scalability, security, and fault-tolerance.
Luo et al.~\cite{luo2022prcbert} evaluate the performance of PRCBERT on requirements classification with PROMISE NFR~\cite{Sayyad-Shirabad+Menzies:2005}, NFR-Review~\cite{wang2018aspects}, and NFR-SO~\cite{luo2022prcbert}.

\noindent\textbf{Specification \& Validation.} The Jdoctor dataset~\cite{blasi2018translating} comprises six well-maintained open-source Java systems: Common Collections, Common Math, GraphStream, Guava, JGraphT, Plume-lib. It was constructed by manually reading the Javadoc comment and writing executable specifications that correspond to each @param, @return, and @throws and @exception tag. There are 118 analyzed classes and 829 Javadoc comments in the ground truth. The DocTer dataset~\cite{xie2022docter} includes three deep learning libraries: TensorFlow 2.1.0, PyTorch 1.5.0, and MXNet 1.6.0, containing 2,666 APIs in total. DocTer categorizes specifications into four groups: \textit{dtype} for data types, \textit{structure} for data structures, \textit{shape} for the parameter's shape or the number of dimensions, and \textit{valid value} for the set of valid (enum) or valid range. Xie et al. evaluated the performance of StarCoder, GPT-3, GPT-3.5, BLOOM, CodeGen, CodeGen2, and Incoder on software specification generation~\cite{xie2023impact}.
Poudel et al.~\cite{poudel2023leveraging} proposed to evaluate the requirements satisfaction assessment architecture against the following datasets: CM1, CCHIT, Dronology~\cite{cleland2018dronology},  and the Positive Train Control (PTC) system. Each of the datasets consists of requirements and design elements. Mandal et al.~\cite{mandal2023large} collected specifications from software manuals using a keyword-based filtering approach along with human assistance. They also utilized a Python-based parser to extract code comments and employed a crawler to retrieve posts related to software specifications from Stack Overflow. The BERT-based model SPECSYN, which synthesizes software specifications from natural language sources, was assessed using this benchmark~\cite{mandal2023large}. 
The SV-Benchmarks~\cite{svbenchmarks2024} is constructed for the International Compitition on Software Verification SV-COMP, to evaluate the effectiveness and efficiency of state-of-the-art verification technology. The repository consists of verification tasks written in C and Java. Ma et al.~\cite{ma2024specgen} evaluated the performance of SpecGen with 265 Java class definition of the SV-Benchmarks. SpecGenBench~\cite{ma2024specgen} contains 120 Java programs, with manually written verifiable JML specifications. Among the 120 programs, 20 are from the GitHub repository Java-JML, and the rest of the programs are collected from LeetCode. It has been used for evaluating the performance of SpecGen~\cite{ma2024specgen}, where the evaluation metrics are number of passes, success probability, number of verifier calls, and user rating. Reinpold et al.~\cite{DBLP:journals/corr/abs-2411-11582} constructed their benchmark by modeling system specifications in SysML using CATIA Magic Systems of Systems Architect, encoding each requirement as an OCL constraint, and validating them to generate ground truth for comparison with LLM outputs.

\noindent \textbf{Management.}
Wang et al.~\cite{wang2020deep} developed a benchmark for coreference detection in requirements, which includes 1,949 text-entity pairs from 21 projects within the China Merchants Bank repository. This benchmark was constructed first by pre-processing to remove noisy tokens such as URL, HTML tags, SQL statements, as well as template words. Then they built 1,949 context-entity pairs of entities. The ground truth was recognized by requirement engineers, and audited by project management department. This benchmark has been adopted for evaluating DeepCoref, which is a BERT based model for coreference detection~\cite{wang2020deep}. Helmeczi et al.~\cite{helmeczi2023few} introduced a benchmark for evaluating four kinds of requirements relations detection: conflicts in software requirements specification (SRS) documents, duplicate bug reports, duplicate Stack Overflow questions, and bug dependencies. The SRS conflicts detection task uses an IBM proprietary dataset consisting of pairs of requirement specifications that are classified as conflicting, duplicate, or neutral. The datasets for duplicate bug reports and bug dependencies were collected from Bugzilla, with pairs of bug reports labeled as either duplicates or not, and entailments or not, respectively. For the Stack Overflow question duplication detection task, question pairs were collected from Stack Overflow and labeled as duplicate or neutral. This benchmark has been used for evaluating the performance of BERT, RoBERTa, and DeBERTa in requirements relations detection~\cite{helmeczi2023few}. 





\nff{\subsection{Discussion}}

\nff{Benchmarks for RE remain underdeveloped compared to other SE tasks. Nearly half of the existing benchmarks (11 out of 25) were introduced before 2023, meaning that many of them are already outdated. These datasets may have even been included in LLM training corpora, which raises concerns of contamination and limits their validity for evaluation. One reason for this stagnation is the difficulty of obtaining requirement data. Most open-source projects do not include requirements, while industrial requirements are typically treated as confidential assets, restricting public access. Moreover, current requirements and analysis benchmarks are still predominantly focused on natural language requirements, with far fewer datasets covering other artifacts such as requirement models or specifications. This narrow coverage limits both the diversity and representativeness of benchmarks in the field.}

\nff{Given the limited availability of datasets, evaluation in requirements and design has so far relied heavily on a few classic benchmarks such as \textbf{PURE}~\cite{ferrari2017pure} and \textbf{PROMISE}~\cite{Sayyad-Shirabad+Menzies:2005}, which were introduced in 2017 and 2005, respectively. These datasets are widely used for several reasons. First, they were among the earliest publicly released datasets in RE, giving them a historical advantage and making them de facto standards. Second, they are relatively accessible and easy to apply, with clear task formulations (e.g., classification of functional and non-functional requirements, requirements analysis). Third, their smaller size and simpler labeling schemes make them convenient for prototyping new approaches, lowering the entry barrier for researchers. Finally, the accumulation of citations and prior work on these datasets reinforces their visibility and adoption, even though they may no longer reflect current industrial practices.}

\nff{The future development of requirements and design benchmarks will be strongly shaped by advances in generative LLMs. Instead of focusing only on classification or analysis tasks, future benchmarks are expected to place greater emphasis on generative tasks grounded in requirements, such as producing UML models, design specifications, implementation code, and test cases from natural language descriptions. They may also evolve into end-to-end evaluation pipelines, using requirements as the entry point to assess the quality, consistency, and completeness of generated artifacts throughout the software lifecycle. Crucially, maintaining benchmark freshness, scalability, and industrial representativeness will be essential to prevent contamination and ensure alignment with real-world RE practices.}

\subsection{Limitations and Opportunities}

\textbf{Lack of Sufficient Benchmarks.} Since software requirements are confidential assets of a company, it is difficult for typical researchers to access actual software requirements. Many current studies related to RE still focus primarily on using software requirements constructed a long time ago, such as those from PROMISE~\cite{Sayyad-Shirabad+Menzies:2005} and PURE~\cite{ferrari2017pure}. However, these ubiquitous RE benchmarks, on one hand, cannot comprehensively evaluate the performance of LLMs, and on the other hand, may not represent real industrial data, potentially leading to significant discrepancies with actual data. As a result, existing LLMs may not be applicable to real software requirements. Therefore, a viable future research direction is to construct larger, more diverse, and higher-quality benchmarks for evaluating LLM-based requirements engineering methods.

\noindent
\textbf{Challenges in Creating a Unified Benchmark.} The diverse patterns of requirements across different companies and the varied domain-specific terminologies in software fields pose significant challenges in developing a standardized benchmark for evaluating LLM performance. For instance, stakeholder requirements, system requirements, and software requirements may each follow distinct templates and patterns. As a result, different types of requirements may necessitate different LLMs and require tailored benchmarks.


%% file: R_development.tex
\section{Coding Assistant} \label{sec:softwaredevelopment}
Benchmarks in coding assistants are essential for advancing LLMs in software engineering. They offer standardized datasets and evaluation metrics, enabling researchers to assess LLMs' performance on tasks like code generation, code summarization, and code translation. These benchmarks facilitate model comparisons, highlight strengths and weaknesses, and promote innovation. Moreover, they ensure LLMs are robust, scalable, and relevant in real-world scenarios. 
By simulating developers' tasks, benchmarks improve automation, boost productivity, and ensure the accuracy of LLMs in software engineering.
In summary, our literature review identified 124 benchmarks used to evaluate LLM performance in coding assistant tasks, including code generation (102) and recommendation (11),  code summarization (11), code translation (13), and code reasoning (3). Detailed information on these benchmarks is provided in Table~\ref{tab: code_generation} and Table~\ref{tab: benchmarks_software_development}.

\input{table_code_generation_full}

\subsection{Code Generation and Recommendation}

\input{R_Code_Generation}

\begin{table}
\caption{Existing Benchmarks for Evaluating LLMs Related to Code Summarization and Translation} 
\label{tab: benchmarks_software_development}    
\tiny
\begin{tabular}{p{0.1\linewidth}
p{0.14\linewidth}
p{0.04\linewidth}
p{0.14\linewidth}
p{0.24\linewidth}
p{0.23\linewidth}}
\toprule
Task  & Benchmarks & Year & Language &  Evaluation Metrics & \#Cases \\ \hline           
\multirow{11}{*}{\begin{tabular}[c]{@{}c@{}}Code \\ Summarization \end{tabular}} & \highlight PCSD~\cite{DBLP:conf/ijcnlp/BaroneS17} &\highlight 2017 &\highlight Python &\highlight BLEU, BLEU-4, ROUGE-L, METEOR, CIDEr &\highlight 92,545 pairs  \\
 & JCSD~\cite{DBLP:conf/ijcai/HuLXLLJ18} & 2018 & Java & Precision, Recall, F-Score,  BLEU-4, METEOR,  ROUGE-L, CIDEr, BLEU-DC  &  87,136 pairs  \\
 & \highlight Deepcom~\cite{DBLP:conf/iwpc/HuLXLJ18} & \highlight  2018 &\highlight Java & \highlight   BLEU-4, METEOR, ROUGE-L & \highlight  69,708 pairs  \\
  &   Funcom~\cite{DBLP:conf/icse/LeClairJM19} &   2019 & Java &   BLEU, ROUGE-L, METEOR &   2.1M code-comment pairs \\
& \highlight  CodeXGLUE~\cite{lu2021codexglue} & \highlight  2021 & \highlight Java, Python & \highlight    BLEU, BLEU-4, ROUGE-L,  METEOR, USE, MRR  & \highlight  see Table~\ref{tab: code_generation}  \\
& Funcom-java-long~\cite{DBLP:conf/sigsoft/SuBJGM23} & 2023 & Java &  BLEU & 8192 Java methods \\
 & \highlight  CroCoSum~\cite{DBLP:conf/coling/ZhangE24} & \highlight  2023  &  \highlight English and Chinese &  \highlight   ROUGE, bertscore & \highlight  18,857 pairs   \\
& CAPYBARA~\cite{DBLP:journals/corr/abs-2301-01701} & 2023 & C  &  EM, BLEU-4, ROUGE-L, METEOR  & 7,826 pairs   \\
  & \highlight  BinSum~\cite{DBLP:journals/corr/abs-2312-09601} & \highlight  2023  & \highlight C & \highlight    BLEU, METEOR, ROUGE-L,  semantic similarity  & \highlight   557,664 binary functions  \\
& P-CodeSum~\cite{DBLP:journals/jss/YunLGS24} & 2024 & Multiple PLs & BLEU-4, ROUGE-L & 1,500 pairs \\
 & \highlight  FILE-CS~\cite{DBLP:conf/wcre/WangHGZZ24} & \highlight  2024  & \highlight Python & \highlight   BLEU, ROUGE-L, METEOR & \highlight  98,236 pairs   \\ \hline
\multirow{13}{*}{\begin{tabular}[c]{@{}c@{}}Code \\ Translation \end{tabular} } & CodeSearchNet~\cite{DBLP:journals/corr/abs-1909-09436} & 2020 & Multiple-PLs &  BLEU, CodeBLEU, METEOR,  Extract Match  & 6,452,446 pairs \\
&  \highlight  CodeXGLUE~\cite{lu2021codexglue} &  \highlight 2021 & \highlight Java, Python & \highlight   BLEU-4, BLEU, ACC, CodeBLEU & \highlight  11,800 pairs  \\
 & CodeNet~\cite{DBLP:conf/nips/Puri0JZDZD0CDTB21} & 2021 & C++, Python &   Compilation errors, runtime errors,  functional errors,  non-terminating execution  &  4,053 problems, 13,916,828 code samples \\
 & \highlight  CoST~\cite{DBLP:conf/aaai/Zhu0R22} & \highlight  2022 & \highlight Multiple-PLs & \highlight   BLEU, CodeBLEU & \highlight  132,046 pairs \\
 & XLCoST~\cite{DBLP:journals/corr/abs-2206-08474} & 2022 & Multiple-PLs  &  CodeBLEU, BLEU, MRR & 1,002,296 pairs \\
 & \highlight  Nova~\cite{DBLP:journals/corr/abs-2311-13721} & \highlight  2023 & \highlight Binary & \highlight BLEU, Exact Match, Instruction  Longest Common Subsequence,  Instruction BLEU  & \highlight  60,600 pairs  \\
 & SUT~\cite{DBLP:conf/emnlp/QiHWYLGCS23} & 2023 & Multiple PLs &   Syntax Unit Test Accuracy,  Syntax Element Test Score  &  60k parallel pairs,  200k monolingual pairs \\
 & \highlight  xCodeEval~\cite{DBLP:conf/acl/KhanBLWPJ24} & \highlight  2023 &\highlight Multiple PLs & \highlight   Pass@K &  \highlight  25M  coding examples \\
 & CodeTransOcean~\cite{DBLP:conf/emnlp/YanTLCW23} & 2023 & Multiple PLs &   BLEU, CodeBLEU,  Exact String Match  &  45 programming  languages\\
 & \highlight  G-TransEval~\cite{DBLP:conf/kbse/JiaoYLQGS23} &  \highlight 2023 & \highlight Multiple PLs &  \highlight  BLEU, CodeBLEU,  Computational Accuracy  & \highlight  400 pairs \\
 & AVATAR~\cite{DBLP:conf/acl/AhmadTCC23} & 2023 &Java, Python &   BLEU, Syntax Match, Dataflow Match,  CodeBLEU, Execution Accuracy,  Computational Accuracy, Pass@1,  Robust Pass@1, Robust Drop@1  &  9,515 problem  statements, 62520  java-python pairs \\
 & \highlight  AVATAR-TC~\cite{jana2023cotran} & \highlight  2024 & \highlight Java, Python  &  \highlight  BLEU, CodeBLEU, Exact String Match,  Compilation Accuracy,  Functional Equivalence Accuracy,  Average First Error Position  &  \highlight  57,368 Java-Python  pairs, across 9,198  problem statements   \\ 
  & \hx{RustRepoTrans}~\cite{ou2024repository}& 2024 &  C, Java, and Python to Rust & Pass@k & 375 tasks\\
 \hline
\multirow{2}{*}{\begin{tabular}[c]{@{}c@{}}Code \\ Reasoning\end{tabular}}
\

& \jk{CRUXEval}~\cite{gu2024cruxeval} \highlight  &\highlight 2024 & \highlight Python &\highlight Pass@k &\highlight 800 \\

& \hx{REval}~\cite{REva} & 2025 &  Python & Accuracy and Incremental Consistency Score   & 3,152  \\

& \highlight \hx{DyCodeEval}~\cite{chen2025dynamic}  & \highlight 2025 & \highlight \highlight Python & \highlight Pass@K and DivPass@K & \highlight 591\\

 \hline
\end{tabular}
\end{table}

\subsection{Code Summarization and Translation}

\subsubsection{Benchmarking Details}
Table~\ref{tab: benchmarks_software_development} summarizes existing benchmarks related code summarization, code translation, and code reasoning for evaluating LLMs in software development tasks. 

\noindent
\textbf{Context.} Benchmarks are constructed under different usage settings. Some focus on natural language documentation scenarios (e.g., code summarization benchmarks like PCSD~\cite{DBLP:conf/ijcnlp/BaroneS17}, Deepcom~\cite{DBLP:conf/iwpc/HuLXLJ18}, and CroCoSum~\cite{DBLP:conf/coling/ZhangE24}), where the model is expected to generate human-readable summaries of source code. Others are situated in software migration or multilingual development contexts (e.g., AVATAR-TC~\cite{jana2023cotran}, RustRepoTrans~\cite{ou2024repository}, and CodeTransOcean~\cite{DBLP:conf/emnlp/YanTLCW23}), where cross-language translation is essential. Newer benchmarks like \jk{CRUXEval and REval~\cite{gu2024cruxeval,REva,chen2025dynamic}} simulate realistic development workflows involving reasoning over code snippets and predicting intermediate steps or outcomes.

\noindent
\textbf{Domain.} The benchmarks span a wide variety of programming languages and domains. Some focus on general-purpose programming languages such as Java, Python, and C/C++, while others extend into binary-level code (e.g., Nova~\cite{DBLP:journals/corr/abs-2311-13721}), and low-resource or emerging domains such as Rust (RustRepoTrans~\cite{ou2024repository}) or multilingual settings involving Chinese and English (CroCoSum~\cite{DBLP:conf/coling/ZhangE24}). This diversity ensures broader evaluation coverage beyond mainstream languages.

\noindent
\textbf{Task.} Benchmarks are categorized into three primary tasks:
Code Summarization – generating natural language summaries from code (e.g., Funcom~\cite{DBLP:conf/icse/LeClairJM19}, JCSD~\cite{DBLP:conf/ijcai/HuLXLLJ18}, CodeXGLUE~\cite{lu2021codexglue}).
Code Translation – translating code between programming languages or representations (e.g., CoST~\cite{DBLP:conf/aaai/Zhu0R22}, AVATAR~\cite{DBLP:conf/acl/AhmadTCC23}, G-TransEval~\cite{DBLP:conf/kbse/JiaoYLQGS23}).
Code Reasoning – reasoning over code logic to answer questions or perform evaluations, including step-wise predictions and test-based evaluation (e.g., REval~\cite{REva}, DyCodeEval~\cite{chen2025dynamic}).

Each task reflects different cognitive and functional demands, from lexical generation to semantic equivalence and logical consistency.

\subsubsection{Evaluation Metrics}
The evaluation metrics can be classified into \textit{NLP-based}, \textit{code-specific}, \textit{execution-based}, and \textit{diversity and consistency} metrics.

\phead{\ding{42} NLP-based Metrics.}
NLP-based metrics evaluate the effectiveness of models by calculating the text similarity between generated summaries and human-written summaries. 
These metrics (such as BLEU-x, ROUGE-L, and METEOR) are first proposed to evaluate machine translation and natural language summarization tasks. 
 Generally, these metrics measure different models by comparing overlapping text between the reference and the generated text. The model can achieve higher scores if there are more overlapping words~\cite{hu2022correlating}.

 \phead{\ding{42} Code-Specific Metrics.}
To better reflect the correctness and functional equivalence of generated code, several benchmarks adopt code-specific metrics. These include CodeBLEU (which incorporates syntactic, data flow, and semantic structure), Exact Match, Compilation Accuracy, and Functional Equivalence Accuracy. These metrics are particularly common in code translation and reasoning tasks, where lexical similarity alone is insufficient. For example, even if the translated code uses different variable names or control structures, it may still be functionally correct — which CodeBLEU and functional testing can better capture.

\phead{\ding{42} Execution-Based Metrics.}
Some recent benchmarks (e.g., AVATAR~\cite{DBLP:conf/acl/AhmadTCC23}, CodeNet~\cite{DBLP:conf/nips/Puri0JZDZD0CDTB21}, DyCodeEval~\cite{chen2025dynamic}) emphasize execution-based metrics such as \textit{Pass@k}, Execution Accuracy, and Robust \textit{Pass@1}. These metrics evaluate whether the generated code passes a given set of test cases or performs correctly in actual execution. This helps assess the model’s practical usability and robustness in real-world development settings.

\phead{\ding{42} Diversity and Consistency Metrics.}
In code reasoning task, certain benchmarks (e.g., REval~\cite{REva}, DyCodeEval~\cite{chen2025dynamic}) introduce Incremental Consistency, and \textit{DivPass@K} to measure the diversity of correct outputs and the logical coherence of multi-step reasoning. These metrics go beyond correctness to evaluate how models handle ambiguity, maintain internal consistency, and avoid redundant or repetitive outputs.

\subsubsection{Benchmark Construction Methods.}

Benchmark construction typically involves two key components: the source of data and the labeling process. These factors greatly influence the benchmark’s quality, coverage, and the types of abilities it can effectively evaluate.

\noindent
\textbf{Source.}
Benchmarks are constructed from a variety of code repositories and programming platforms, depending on the targeted task:
\textbf{Open-Source Repositories}: Many benchmarks collect real-world code from platforms like GitHub (e.g., Funcom~\cite{DBLP:conf/icse/LeClairJM19}, Deepcom~\cite{DBLP:conf/iwpc/HuLXLJ18}, JCSD~\cite{DBLP:conf/ijcai/HuLXLLJ18}, CodeXGLUE~\cite{lu2021codexglue}). These sources provide large volumes of naturally occurring code-comment or code-translation pairs in diverse programming languages.
\textbf{Online Programming Platforms}: Datasets like CodeNet~\cite{DBLP:conf/nips/Puri0JZDZD0CDTB21}, AVATAR~\cite{DBLP:conf/acl/AhmadTCC23}, and xCodeEval~\cite{DBLP:conf/acl/KhanBLWPJ24} draw from competitive programming platforms or educational codebases (e.g., LeetCode, Codeforces), which offer problem descriptions, code submissions, and test cases in multiple languages.
\textbf{Binary Repositories}: Some recent benchmarks (e.g., BinSum~\cite{DBLP:journals/corr/abs-2312-09601}, Nova~\cite{DBLP:journals/corr/abs-2311-13721}) focus on low-level code representations and are constructed from disassembled binaries, enabling evaluation beyond source-level models.
Manual or Synthetic Construction: Certain benchmarks (e.g., REval~\cite{REva}, DyCodeEval~\cite{chen2025dynamic}) are carefully designed or synthesized to target reasoning abilities, where the source code and context are curated for controlled evaluation.

\noindent
\textbf{Label.}
The labeling strategies also vary depending on task type:
\textbf{Natural Labels from Documentation}: In summarization tasks, labels are often derived from code comments or docstrings written by developers. These labels may be noisy but reflect realistic usage.
Heuristic or Rule-Based Matching: Some translation datasets generate labels via static analysis, syntax tree alignment, or automated pairing of similar functions across languages.
\textbf{Execution-Driven Labels}: In translation and reasoning benchmarks, labels are increasingly defined based on program behavior (e.g., passing test cases, functional equivalence). For example, AVATAR~\cite{DBLP:conf/acl/AhmadTCC23} and DyCodeEval~\cite{chen2025dynamic} use unit tests or runtime correctness as ground truth.
\textbf{Manual Annotation or Validation}: Certain benchmarks involve human annotation to ensure correctness, especially when evaluating fine-grained reasoning steps or semantic correctness (e.g., REval~\cite{REva}).
Automatic Evaluation Programs: Some datasets provide scripts or compilers to validate the functional correctness of generated code automatically, allowing scalable labeling.

\subsubsection{Discussion}
\cjk{
In the field of code summarization, \textbf{CodeXGLUE}~\cite{lu2021codexglue} (Code-Text split) has become one of the most widely used benchmarks. 
It sources data from CodeSearchNet~\cite{DBLP:journals/corr/abs-1909-09436}, offering a large-scale (about 1M code–text pairs) and multilingual (several PLs) dataset, together with detailed baseline implementations such as CodeGPT~\cite{lu2021codexglue}. 
These features enable subsequent research to easily conduct data engineering without worrying about dataset size, explore cross-language perspectives, and transparently build upon existing baselines. 
Besides, its detailed documentation and well-maintained code base are very friendly for newcomer researchers. 
Regarding code translation, \textbf{AVATAR}~\cite{DBLP:conf/acl/AhmadTCC23} is widely adopted due to its scale and the use of diverse evaluation metrics, including both similarity-based and execution-based ones, compared with previous research. 
In terms of usage trend, we further observe that the granularity of code summarization and translation tasks is becoming increasingly diverse. Recent approaches move beyond method-level tasks to address class-~\cite{xue2025classevaltevaluatinglargelanguage, 8473199}, file-~\cite{wang2024sparsecoderidentifierawaresparsetransformer}, or even project-level~\cite{Ibrahimzada_2025,DBLP:journals/corr/abs-2502-16704} code, reflecting both the growing requirements of developers and the capabilities of LLMs. 
It is expected that large-scale, higher-granularity tasks will become easier for LLMs to handle and, in turn, provide developers with greater convenience.
}

\subsubsection{Limitations and Opportunities}
Code summarization benchmarks, though valuable, face several limitations: (1) \textbf{Limited Context}: Many benchmarks concentrate on small code snippets, often ignoring broader project or file-level contexts, which are crucial for generating meaningful summaries. As a result, models may produce incomplete or inaccurate summaries that overlook the bigger picture of the code’s functionality. (2) \textbf{Lack of Domain Diversity}: Benchmark datasets often focus on popular programming languages like Java and Python, neglecting other languages and domains (e.g., scientific computing or embedded systems), limiting the generalizability of models trained on these datasets. (3) \textbf{Quality of Summaries}: Many benchmarks rely on developer-written comments, which can vary in quality. Some comments may be vague, overly detailed, or inconsistent, affecting the model’s ability to learn from high-quality examples. (4) \textbf{Evaluation Metrics}: Current metrics like BLEU and ROUGE focus on \textit{n-gram} overlaps, which do not fully capture the semantic understanding needed for high-quality code summarization. These metrics may not accurately reflect a model’s effectiveness in generating useful, accurate summaries for developers.
\cjk{
Recently, semantic similarity–based metrics have been introduced for evaluating code summarization tasks, such as SIDE~\cite{mastropaolo2024evaluating}, BERTScore~\cite{zhangbertscore}, and SimLLM~\cite{jin2024simllm}.
These approaches typically train or adapt neural networks to assess answer quality and have shown promising improvements over traditional metrics (e.g., BLEU).
This suggests a promising new direction for benchmarking code summarization.
}

Code translation benchmarks, though useful for evaluating transcompiler models, have several limitations: (1) \textbf{Lack of Granularity}: Many benchmarks assess overall translation quality but do not provide detailed feedback on specific syntactic or semantic errors, making it hard to diagnose where models struggle, such as with complex syntax or language-specific constructs. (2) \textbf{Reliance on Static Metrics}: Evaluation metrics like BLEU and CodeBLEU focus on surface-level similarities between translated and reference code, neglecting functional correctness or the behavior of the code, which may pass syntax checks but fail during execution. (3) \textbf{Limited Language Coverage}: Most benchmarks focus on a small set of widely used programming languages (e.g., Python, Java, and C++), limiting model generalization to niche or less common languages, which may have distinct syntactic and semantic features. (4) \textbf{Lack of Execution-Based Evaluation}: Many benchmarks do not check if the translated code runs correctly. This omission means models may generate code that looks correct but fails to execute or produces incorrect outputs, which is crucial for practical code translation. Overcoming these limitations requires benchmarks that focus on functional correctness, cover a wider range of languages, and offer detailed feedback on syntactic and semantic translation quality.

%% file: table_code_generation_full.tex
\renewcommand*{\arraystretch}{1.0}
\setlength{\tabcolsep}{2pt}
{\tiny\begin{longtable}
{p{0.15\linewidth}
p{0.04\linewidth}
p{0.14\linewidth}
p{0.15\linewidth}
p{0.23\linewidth}
p{0.23\linewidth}} 
\caption{Existing Benchmarks for Evaluating LLMs Related to Code Generation and Recommendation} \label{tab: code_generation}\\    
\hline    
Benchmarks & Year & Granularity & Language & Evaluation Metrics & \#Cases \\ 
\hline    
\endfirsthead    
\multicolumn{5}{c}%
{{\bfseries \tablename\ \thetable{} -- continued from previous page}} \\    
\hline    
Benchmarks & Year & Granularity & Language & Evaluation Metrics & \#Cases \\ 
\hline    
\endhead    
\hline \multicolumn{5}{r}{{Continued on next page}} \\    
\endfoot    
\hline    
\endlastfoot    
 \rowcolor[HTML]{EFEFEF} Lin et al.~\cite{chen2023effectiveness} & 2013 & Function-level   & Go, C++   & BLEU, CodeBLEU  & Six libraries \\
  Leetcode~\cite{leetcode} & {2015} & Contest-level & Multiple PLs & {Passing Test Cases, Runtime, Memory Usage} & 180 problems (100 tests for each)    \\
   \rowcolor[HTML]{EFEFEF} ExampleCheck~\cite{zhang2018code} & 2018  & API-level & Java & misuse rate  & 100 Java APIs \\
 CONCODE~\cite{iyer2018mapping} & 2018  & Function-level & Java & BLEU   & 100,000/2,000/2,000 functions   \\
  \rowcolor[HTML]{EFEFEF} CoNaLa~\cite{yin2018learning} & 2018  & Solution-level & Python,   Java & precision, recall, TPR & 33,946 pairs for   Python and 37,882 for Java \\
NL2Bash~\cite{lin2018nl2bash} & 2018  & Solution-level  & Bash & manual, BLEU & 12,609 text-command pairs  \\
  \rowcolor[HTML]{EFEFEF} Spider~\cite{yu2018spider} & 2018  & SQL-level & SQL  & Component Matching, Exact Matching, and Execution Accuracy & 10,181 problems   \\

 CodeSearchNet~\cite{DBLP:journals/corr/abs-1909-09436} & 2020  & Function-level & Multiple PLs & {NDCG, MRR} & 6,452,446 Functions  \\
 \rowcolor[HTML]{EFEFEF} APPS~\cite{hendrycksapps2021} & 2021 & Contest-level & Python & Test Case Average, Strict Accuracy & {10,000 coding problems in total, with 131,777 test cases and 232,421 ground-truth solutions }  \\
 MBPP~\cite{DBLP:journals/corr/abs-2108-07732} & 2021 & Function-level   & Python & \% solved  & 974\\
\rowcolor[HTML]{EFEFEF} CodeXGLUE~\cite{lu2021codexglue} & 2021  & Function-level & Java,   Python & EM, ES &  100,000 examples for training and 4,000 examples for validation and testing.   \\
 HumanEval~\cite{chen2021codex} & 2021 & Solution-level & Python & Pass@k & 164 human-written  \\
\rowcolor[HTML]{EFEFEF} miniF2F~\cite{zheng2021minif2f} & 2021  & Solution-level  & Lean,   Metamath, Isabelle & Pass@k  & test: 224, valid: 244      \\
Lyra~\cite{liang2021lyra} & 2021  & Turducken-Style    & Python   with embedded SQL & BLEU, code executable, ast matching in base language, ast exact match & 2,000 source code snippets \\
   \rowcolor[HTML]{EFEFEF}FC2Code~\cite{liu2022code} & 2022  & Contest-level & Python & {BLEU}  & 320 flowcharts  \\
  CodeContests~\cite{li2022alphacode} & 2022  & Contest-level  & {C++, Python, and Java} & n@k, Pass@k   & 13,328 Train, 117 Valid and, 165 Test problems \\
 \rowcolor[HTML]{EFEFEF}AixBench~\cite{hao2022aixbench} & 2022  & Function-level & Java & Correctness, Code Quality, Maintainability, Pass@1, AvgPassRatio   & 175 test dataset, 161 nl description dataset   \\
ReCode~\cite{wang2022recode} & 2022  & Function-level & Python & robust Pass\_s@k, robust drop\_s@k, robust relative\_s@k & {5,690 problems} \\
\rowcolor[HTML]{EFEFEF}SecurityEval~\cite{siddiq2022securityeval} & 2022  & Function-level & Python & percentage & 130 prompts    \\
MathEquations~\cite{jones2022capturing} & 2022  & Function-level& - & Functional accuracy  & 12 combinations   \\
 \rowcolor[HTML]{EFEFEF}MBXP~\cite{athiwaratkun2022multi} & 2022  & Function-level & Multiple PLs & Pass@k & like MBPP    \\
NumpyEval~\cite{zan2022cert} & 2022  & Library-level & Python & Pass@k & 101 problems  \\
\rowcolor[HTML]{EFEFEF}PandasEval~\cite{zan2022cert} & 2022  & Library-level& Python & Pass@k & 101 problems  \\
TorchData-Eval~\cite{zan2022language} & 2022  & Library-level & Python & {Pass@k}  & 50 problems    \\
\rowcolor[HTML]{EFEFEF} MonkeyEval~\cite{zan2022language} & 2022  & Library-level & Python & {Pass@k}  & 101 problems   \\
BeatNumEval~\cite{zan2022language} & 2022  & Library-level & Python & {Pass@k}  & 101 problems     \\

 \rowcolor[HTML]{EFEFEF}MTPB~\cite{nijkamp2022codegen} & 2022  & Solution-level & Python & {Pass Rate, PPL}  & 115 problems  \\
 Multi-HumanEval~\cite{athiwaratkun2022multi} & 2022  & Solution-level & Multiple PLs & Pass@k & like HumanEval  \\
\rowcolor[HTML]{EFEFEF}DSP~\cite{chandel2022training} & 2022  & Statement-level & Python & {Pass@k}  & 1,119 problems     \\
ExeDS~\cite{huang2022execution} & 2022  & Statement-level  & Python & {BLEU, CodeBLEU, EM, OutputEM} & 534 problems    \\
 \rowcolor[HTML]{EFEFEF} XLCoST~\cite{DBLP:journals/corr/abs-2206-08474} & 2022  & Function-level, Statement-level & Multiple PLs & {BLEU, CodeBLEU, MRR} & 446K/22K/41K for Statement-level;  50K/3K/5K for function-level  \\
Qing et al.~\cite{huang2023adaptive} & 2023  & API-level  & Kotlin & Success Rate & 200 APIs  \\
\rowcolor[HTML]{EFEFEF}  ClassEval~\cite{du2023classeval} & 2023  & Class-level & Python & Pass@k, DEP(F), DEP(M) & 100 task, 33.1 test, 45.7 loc, 123.7 tokens\\
TACO~\cite{li2023taco} & 2023  & Contest-level    & Python & {Pass@k}  & 25,433 train, 1,000 test problems  \\
 \rowcolor[HTML]{EFEFEF} xCodeEval~\cite{DBLP:conf/acl/KhanBLWPJ24} & 2023  & Contest-level & Multiple PLs  & F1, Pass@k, Accuracy & 5,539,899 problems     \\
CodeApex~\cite{fu2023codeapex} & 2023  & Function-level & {C++}   & {AC@1, AC@all, AC Rate}&476 problems  \\
 \rowcolor[HTML]{EFEFEF} CloverBench~\cite{sun2023clover} & 2023  & Function-level & Dafny  &  Accept@k & 60 programs   \\
Mastropaolo et al.~\cite{mastropaolo2023robustness} & 2023  & Function-level  & Java & CodeBLEU, Levenshtein Distance & 892 methods from 33 repos  \\
 \rowcolor[HTML]{EFEFEF} CoderEval~\cite{yu2024codereval} & 2023  & Function-level & Python & Pass@k, Acc@k   & {230 Python functions  and 230 Java methods} \\
 EvalPlus~\cite{liu2024your} & 2023  & Function-level & Python & Pass@k & {1,138 tasks}    \\
  \rowcolor[HTML]{EFEFEF} Shapkin et al.~\cite{shapkin2023entity} & 2023  & Function-level & Python & CodeBLEU, acc  & 14,000 samples  \\
 CrossCode-Bench~\cite{niu2023crosscodebench} & 2023  & Function-level & Java,   Python & {EM, BLEU, ROUGE-L} & 19,511,282 instance \\
\rowcolor[HTML]{EFEFEF}MultiPL-E~\cite{cassano2023multipl} & 2023  & Function-level & Multiple PLs  & Pass@k & HumanEval and MBPP in 19 languages   \\
 StudentEval~\cite{babe2023studenteval} & 2023  & Function-level  & Multiple PLs & {Pass@1}  & 1,749 prompts    \\
\rowcolor[HTML]{EFEFEF}TorchData-ComplexEval~\cite{zan2023private} & 2023  & Library-level  & Python & Pass@k & 50 problems  \\
DS-1000~\cite{lai2023ds} & 2023  & Line-level & Python & Pass@1 & 1,000 problems from 451 StackOverflow problems.\\
\rowcolor[HTML]{EFEFEF}ML-Bench~\cite{tang2023ml} & 2023  & Repo-level & Bash, Python & {Pass@k}  & 18 repos  \\
LowCoder~\cite{rao2024ai} & 2023  & Snippet-level & Python & Accuracy   &  79,372 NL-Code pairs (64,779 train samples, 7,242 validation samples and 7,351 test samples)\\
  \rowcolor[HTML]{EFEFEF}Ren et al.~\cite{ren2023misuse} & 2023  & Solution-level  & Java & Time Consumption, answer Correctness & 3,709 Java coding   tasks \\
CodeAlpaca  (Py)~\cite{chaudhary2023code} & 2023  & Solution-level & Python & - & The  2,192/314/628 samples as the training/validation/test sets \\
 \rowcolor[HTML]{EFEFEF}CoLadder~\cite{yen2023coladder} & 2023  & Solution-level & {Python}   & Self-Perceived Task Completion and Completion Time, time, correctness,   satisfaction, confidence, usability, perceived cognitive load & 14 programming tasks   \\
VeriGen~\cite{thakur2024verigen} & 2023  & Solution-level  & Verilog  & Pass@k & set 1 with 17   problems and set 2 with 181 problems \\
 \rowcolor[HTML]{EFEFEF}SOEval~\cite{siddiq2023lightweight} & 2023  & Solution-level & Java, Python & NDCG@K & 1,151 prompts for Java (594) and Python (557) \\
  Decept-Prompt~\cite{wu2023deceptprompt} & 2023  & Solution-level & C,   Python & {ASR,WFR} & 25   CWE types with 40 different test case  \\
 \rowcolor[HTML]{EFEFEF} HumanEval-X~\cite{codegeex} & 2023  & Solution-level & Multiple PLs  & Pass@k, Pass@k\_pie & 820 problems  \\
ARCADE~\cite{yin2023natural} & 2023  & Statement-level  & Python   & {Pass@k}  & 1,087 problems  \\
\rowcolor[HTML]{EFEFEF}MCoNaLa~\cite{wang2023mconala} & 2023  & Statement-level & Python   & {BLEU} & 341/210/345 pairs in Spanish/Japanese/Russian  \\
CrossCode-Eval~\cite{ding2024crosscodeeval} & 2023  & Token-level & Multiple PLs & Code Match, Identifier Match  & {1,002 repos}\\
\rowcolor[HTML]{EFEFEF} Pisces~\cite{yang2023syntax} & 2023  & Turducken-Style & Java with embedded SQL   & {BLEU, Weight-BLEU, CrystalBLEU, Syntax-Match, SyntaxExact-Match, Code-BLEU, and Code-Executable}  & 2,000 code snippets   \\

 MBPP/HumanEval/ APPS-ET~\cite{dong2023codescore} & 2023  & Contest-level, Function-level & Python & CrystalBLEU, BERTScore, COMET, CodeBERTScore, AvgPassRatio, Kendall-tau, Spearman R, Pearson R, Mean Absolute Error & 323,457 examples \\
\rowcolor[HTML]{EFEFEF}LiveCode-Bench~\cite{jain2024livecodebench} & 2024  & Contest-level & Python   & {Pass@k}  & {1432 samples}   \\
Mercury~\cite{du2024mercury} & 2024  & Contest-level  & Python   & {Beyond Pass} & 1,889 tasks\\
\rowcolor[HTML]{EFEFEF}EffiBench~\cite{huang2024effibench} & 2024  & Contest-level & Python & {ET, NET, MU, NMU, TMU, NTMU, Pass@k}  & 171 easy, 589   medium, 240 hard \\
MBPP-san-DFY~\cite{misu2024towards} & 2024  & Function-level  & Dafny  & verify@k & 178 problems \\
\rowcolor[HTML]{EFEFEF} CoderUJB~\cite{DBLP:journals/corr/abs-2403-19287} & 2024  & Function-level & {Java} & {Pass@k, Count@n, Coverage@n, accuracy}   & 2,239 programming questions derived from 17 real open-source Java projects    \\
PythonSaga~\cite{yadav2024boldly} & 2024  & Function-level & Python & {Pass@k} & 185 problems  \\
\rowcolor[HTML]{EFEFEF} DevEval~\cite{li2024deveval} & 2024  & Function-level & Python & Pass@k, Recall@k  & 117 repos, 1874 functions  \\


Exec-CSN~\cite{xie2024codebenchgen} & 2024  & Function-level & Python & {Pass@k}  & 1,931 examples  \\
\rowcolor[HTML]{EFEFEF} Wang et al.~\cite{wang2024teaching} & 2024  & Repo-level & Python & BLEU-4, CodeBLEU, edit sim, dependency coverage, static validity rate & 12,744 Python function  \\
EvoCodeBench~\cite{li2024evocodebench} & 2024  & Repo-level & Python   & {Pass@k, Recall@k}  & 275 samples \\
\rowcolor[HTML]{EFEFEF}RustEval~\cite{pan2024enhancing} & 2024  & Repo-level & Rust & {Pass@k}  & 13 repos, 90   functions \\
Devbench~\cite{li2024devbench} & 2024  & Repo-level  & Python, C,C++,Java, JS  & General Principles Faithfulness, Pass@k, Oracle Test Coverage & 22 repos   \\
\rowcolor[HTML]{EFEFEF} BigCode-Bench~\cite{zhuo2024bigcodebench} & 2024  & Solution-level & Python & {Pass@k}  & 1,140 tasks  \\
OOPEval~\cite{wang2024oop} & 2024  & Solution-level & Python & Pass@k, Pass@o  & 431 samples of OOP   \\
\rowcolor[HTML]{EFEFEF}ODEX~\cite{wang2022execution} & 2024  & Solution-level & Python & {Pass@k}  & 945 samples  \\
NaturalCode Bench~\cite{zhang2024naturalcodebench} & 2024  & Solution-level & Python,   Java  & {Pass@k}  & 402 problems  \\
\rowcolor[HTML]{EFEFEF} PAREval~\cite{nichols2024can} & 2024  & Solution-level & C++, CUDA, HIP &  speedup\_max-@k, speedup\_n@k, efficiency\_n@k, Pass@k   &420 prompts covering 12 computational problem types and seven execution models.  \\ 
CAASD~\cite{zhang2024experimenting} & 2024  & Solution-level & Multiple PLs   & pass rate  & 72 software   development task \\
\rowcolor[HTML]{EFEFEF}CodeScope~\cite{DBLP:conf/acl/YanLWL0W0ZZSD24} & 2024  & Solution-level & Multiple PLs & {Pass@k}  & 803 samples  \\
CodeAgent-Bench~\cite{zhang2024codeagent} & 2024  & Function-level, Class-level & Python & Pass@k & 101 samples   \\
\rowcolor[HTML]{EFEFEF} \hx{JavaBench}~\cite{cao2024javabench} & 2024 & Repo-level & Java &Completion@k, Compilation@k, and Pass@k & 389 methods in 106 Java classes\\
Chart2Code-160k~\cite{DBLP:conf/icse/Rodriguez-Cardenas24} & 2024  & Function-level & Python & Execution/pass rate, text match & 160k chart-code pairs  \\
\rowcolor[HTML]{EFEFEF} PoorCodeSumEval~\cite{DBLP:conf/kbse/HuCZM0024} & 2024 & Function-level & Java, Python, Go & BLEU and BERTScore & 101,520 functions \\
ComplexCodeEval~\cite{DBLP:conf/kbse/FengLGCWG024} & 2024  & Function-level & Python and Java & BLEU, Syntax Match, Data Flow Match, and Average Score & 3,897 Java samples and 7,184 Python samples  \\
\rowcolor[HTML]{EFEFEF} StackEval~\cite{DBLP:conf/nips/ShahGA24} & 2024  & Solution-Level & Multiple PLs & Acceptance Score & 925 questions \\
Code-Vision~\cite{DBLP:journals/corr/abs-2502-11829} & 2025 & Function-level & Python & Pass@k & 438 problems \\
\rowcolor[HTML]{EFEFEF} CodeIF-Bench~\cite{DBLP:journals/corr/abs-2503-22688} & 2025  & Instruction-level & Python & Pass@k & 122 programming tasks with 879 instructions  \\
CodeIF~\cite{DBLP:journals/corr/abs-2502-19166} & 2025 & Instruction-level & Java, Python, Go, C++ & Completely Satisfaction/Consistent Continuity Satisfaction Rate & 1,200 tasks \\
\rowcolor[HTML]{EFEFEF} LibEvolutionEval~\cite{DBLP:conf/naacl/KuharAWJQRRMD25} & 2025  & Function-Level & Python & F1-score, MRR & 34.7 examples and 65 documentations  \\
COFFE~\cite{DBLP:journals/corr/abs-2502-02827} & 2025 & Function and File-level & Python & Efficienct@k & 756 problems \\
\rowcolor[HTML]{EFEFEF} Deep-Bench~\cite{DBLP:journals/corr/abs-2502-18726} & 2025  & Function-level & Python & Pass@k & 520 instances of AI and DL data points  \\
DynaCode~\cite{DBLP:journals/corr/abs-2503-10452} & 2025 & Function-level & Python & Pass@K & 189,263,141 problems \\
\rowcolor[HTML]{EFEFEF} FEA-Bench~\cite{DBLP:journals/corr/abs-2503-06680} & 2025  & Repository-Level & Python & Resolved ratios, Precision, Recall & 83 repositories and 1401 tasks  \\
MaintainCoder~\cite{DBLP:journals/corr/abs-2503-24260} & 2025 & Function-level & Python & Pass@k, Codedif f, ASTsim & 500 Python programming data \\
\rowcolor[HTML]{EFEFEF} mHumanEval~\cite{DBLP:conf/naacl/RaihanAZ25} & 2025  & Function-Level & Python & BERTScore & 204 class  \\
REPOEXEC~\cite{DBLP:conf/naacl/HaiNB25} & 2025 & Repository-level & Python & Functional correctness, Dependency utilization & 355 problems \\
\rowcolor[HTML]{EFEFEF} Plot2Code~\cite{DBLP:conf/naacl/WuLGGLWSL25} & 2025  & Function-Level & Python and R & code pass rate, text-match ratio, and GPT-4V rating judgement & 368 plots  \\
ProjectEval~\cite{DBLP:journals/corr/abs-2503-07010} & 2025 & Project-level & Python & Pass@K & 20 tasks \\
\rowcolor[HTML]{EFEFEF} SolEval~\cite{DBLP:journals/corr/abs-2502-18793} & 2025  & Repository-Level & Solidity & Pass@K, Compile@k, Gas Consumption, Vulnerability Rate & 1,125 samples from 9 repositories \\
ConvCodeWorld~\cite{DBLP:conf/iclr/HanHSH25} & 2025 & Function-level & Python & Pass@K, MRR, Recall & 1,140 problems \\
\rowcolor[HTML]{EFEFEF} Web-Bench~\cite{xu2025web} & 2025  & Repository-Level & HTML, CSS, and JavaScript & Pass@K & 50 projects, each consisting of 20 tasks, 3.6 cases per task and 72.4 cases per project \\
\jk{REPOCOD}~\cite{repocod} & 2025  & Repository-Level & Python & Pass@K & 11 projects, 980 functions in total \\
\end{longtable}
\nff{* Definitions of granularity levels are in Appendix~\ref{appendx:C}}\\
}

%% file: R_Code_Generation.tex


Automated code generation and recommendation systems supported by code-related LLMs have become an integral part of modern software development~\cite{DBLP:journals/tosem/HouZLYWLLLGW24}. 
To evaluate the capability boundary of such automatic programming tools, both academia and industry are continuously updating benchmarks for different evaluation aspects with various (i) programming scenarios (Section ~\ref{subsec: scenario}), (ii) evaluation metrics (Section ~\ref{subsec: metric}). We will introduce them from these two perspectives in the following sections, followed by the construction methods of the benchmark (Section ~\ref{subsec: construction}). Then, we will discuss the progress and the challenges of the present benchmarks (Section~\ref{subsec: limitations}).

\subsubsection{Benchmarking Scenarios}
\label{subsec: scenario}

We discuss the programming scenarios these benchmarks assess from three aspects: context (i.e., the scope of input), domain (i.e., the range of application), and task (i.e., the desired expectation of model's behavior) of a benchmark.

\phead{Context.}
For token and line level code completion, the model is typically required to complete several tokens, given a small segment of code as the context.
To evaluate the bimodal ability (e.g., NL and PL) of code LLMs, benchmarks designed for NL2Code task were introduced. Provided with a description in natural language, a code snippet like a complete function is expected to be generated. Popular code generation benchmarks like MBPP~\cite{DBLP:journals/corr/abs-2108-07732} need the model to generate the function corresponding to the natural language intent. 
Some research attempted to solve competition-level programming problems with code-related LLMs with problem description and input/output examples. APPS~\cite{hendrycksapps2021} and CodeContests~\cite{li2022alphacode} belong to this category. 
Apart from the aforementioned in-file and standalone context, some benchmarks made the attempt to utilize cross-file context or even the whole repository with aid of tools like retriever. CrossCodeEval~\cite{ding2024crosscodeeval} explored the problem of cross-file code completion. RepoBench~\cite{liu2023repobench} aims to benchmarking repository-level code completion abilities of LLMs. 
Currently, there are cutting-edge initiatives that leverage not only natural language and code within code repositories but also relevant information from open-source platforms such as issues and pull requests. These efforts have taken a significant step forward in automating programming. SWE-Bench~\cite{jimenez2024swe} serves as the first benchmark to make LLMs solve real-world problems.

\phead{Domain.}
Popular benchmarks like HumanEval~\cite{chen2021codex} and CodeXGLUE~\cite{lu2021codexglue} generally focus on top programming languages such as Java and Python. Therefore, some researchers identified this issue and developed benchmarks for low-resource programming languages like Julia and Ruby~\cite{athiwaratkun2022multi}. 
Moreover, multilingual benchmarks are also becoming popular in multiple tasks not limited to code generation and translation. For instance, MultiPL-E, Multilingual HumanEval, and HumanEval-X~\cite{cassano2023multipl, athiwaratkun2022multi, codegeex} put efforts into extending existing benchmarks to various languages. 
Some programming languages specialized for specific domains is receiving more attention too. In the realm of data science, DS-1000~\cite{lai2023ds} is used to benchmark model's ability to generate programs for data science. MathEquation~\cite{jones2022capturing} is proposed to evaluate the mathematical capabilities of LLMs. How to generate SQL from natural language is also popular, and some benchmarks are proposed to target it. 
For the popular application of some libraries like Numpy~\cite{harris2020array}, benchmarking code generation abilities on specific libraries is attractive and benchmarks like NumpyEval, TorchDataEval, and PandasEval~\cite{zan2022cert, zan2022language} are being proposed. 
A set of benchmarks aims to assess code-related LLMs beyond accuracy such as privacy~\cite{zan2022language}, security~\cite{siddiq2022securityeval}, efficiency~\cite{du2024mercury, huang2024effibench}, and robustness~\cite{mastropaolo2023robustness, wang2022recode}.

\phead{Task.}
Generally, classic code completion tasks need the code model to predict the next token of an incomplete program or complete the rest of the current line of code~\cite{lu2021codexglue}. 
With the growing capabilities of LLMs, more challenging tasks like generating the whole function and completing the whole class are being proposed to challenge them. 
ClassEval~\cite{du2023classeval} make the first step to evaluate LLMs on class-level code generation. 
At the current time, the problem of repository level programming tasks including generation, editing, and refactoring is becoming one of the most popular topics in software engineering. SWE-bench~\cite{jimenez2024swe} is aimed at repository-level programming. DevBench~\cite{li2024devbench} tries to extend the scope of benchmarking to more general software development process.

\subsubsection{Evaluation Metrics}
\label{subsec: metric}

To assess the quality of generated code, a wide range of metrics are proposed and utilized. Based on the foundation of evaluation, we summarize them into \textit{match-based} (e.g., Exact Match), \textit{execution-based} (e.g., Pass@K~\cite{chen2021codex}), and \textit{LLM-based} ones (e.g., CodeBERTScore~\cite{zhou2023codebertscore}).

 \phead{\ding{42} Match-Based Metrics.}
Match-based metrics examine the degree of alignment by matching tokens of code representations between the prediction and the ground truth. As the idea is simple and straightforward, many benchmarks and research leverage them for the code evaluation. 
For \textbf{token-level completion}, metrics like Accuracy and MRR~\cite{chen2024code} are used for evaluation. These metrics take a simple token as an individual element, and estimate the quality of generated code from the perspective of suggestion systems. 
With the advance of model's generation ability, the length of code recommendation is increasing. Therefore, \textbf{string metrics} are introduced for the evaluation. 
For example, an Exact Match is used to judge if the hypothesis is exactly the same as the reference. 
Edit distance~\cite{levenshtein1966binary} defines three operations, including insertion, deletion, and replacement, thereby counting the minimum number of operations one string needs to be another string. A variant of Edit Distance is Edit Similarity~\cite{svyatkovskiy2020intellicode}, which is standardized by the length.  
Besides, some metrics originally designed for \textbf{machine translation} are also applied for evaluation of code generation. 
BLEU score~\cite{papineni2002bleu} is computed based on the n-gram similarity between hypothesis and reference. 
ROUGE-L~\cite{lin2004rouge} considers the longest common subsequence of two strings to estimate the similarity. 
METEOR~\cite{banerjee2005meteor} motivated by BLEU score and takes precision and recall on the whole corpus into consideration. 
To provide a precise measure for code generation benchmarks, some metrics \textbf{designed specifically for code} are proposed like CodeBLEU~\cite{ren2020codebleu}. These metrics typically utilize the distinct features of programming language such as grammar and structure to achieve better evaluation beyond generic metrics for natural language. 
CodeBLEU is a variant of BLEU score. It leverages AST and data flow to evaluate grammatical and logical correctness of code on top of texual similarity. 
CrystalBLEU~\cite{eghbali2022crystalbleu} addresses the limitation of vanilla BLEU score in code, which gives equal weight to both trivial and essential n-gram overlaps of programming languages. 
RUBY~\cite{tran2019does} is devised to assess code generation with program dependency graphs of the reference and the candidate.

\phead{\ding{42} Execution-Based Metrics.}
The executable nature of programming languages differs from natural language. Thus, some research utilizes this feature to capture the similarity between the two programs. 
For example, $Compile@k$ is a metric that measures the success rate of compilation in $k$ generated programs. 
In like manner, $Pass@k$~\cite{chen2021codex} shows how many generated programs could pass the test suite in all $k$ candiate code. 
As the test suite could exactly capture the functional correctness of a given program, Pass@k is becoming very popular, and we think it is the current \textit{de facto} standard for the correctness of code generation in most benchmarks~\cite{chen2021codex, liu2024your}.

\phead{\ding{42} LLM-Based Metrics.}
LLMs present powerful potential in various domains, and some research explored assessing the generated code by utilizing LLMs themselves as a proxy for evaluation~\cite{zhou2023codebertscore, dong2023codescore}. 
Zhou et al.~\cite{zhou2023codebertscore} proposed a novel measure CodeBERTScore. This metric compares pairwise cosine similarity between generated and reference code by their neural representations generated by a pre-trained code-related LLM CodeBERT~\cite{feng2020codebert}. 
Dong et al.~\cite{dong2023codescore} presented CodeScore, accompanied by a unified framework aimed at learning code execution. CodeScore estimates the functional correctness of the hypothesis and supports three different types of input. These works used proposed metrics to evaluate the generated code on benchmarks like HumanEval and CoNaLa.

\phead{\ding{42} Other Metrics.}
Some metrics are used from the perspective of human and usage. 
For example, the acceptance rate is proposed to measure the usefulness of code suggestion provided by a coding assistant~\cite{ziegler2022productivity}.
The Time and Space Complexity is used to estimate how much resource could be occupied by the generated program in some coding benchmarks~\cite{liu2024no}.

\subsubsection{Benchmark Construction Methods}
\label{subsec: construction}
Generally, the code generation and recommendation benchmarks are built in the following steps: choosing a specific task and domain, collecting source code, labeling the codes, and designing the evaluation metrics. For these steps, we will discuss the methods of benchmark construction with respect to their source and label.

\phead{Source.}
Open-source platforms such as GitHub~\cite{github} store billion lines of code, which makes them ideal source for code generation benchmarks. Typically, researchers crawl repositories from GitHub based on specific filtering rules, and take them as the basis for dataset curation. A lot of popular benchmarks like CodeSearchNet~\cite{DBLP:journals/corr/abs-1909-09436}, BigCodeBench~\cite{zhuo2024bigcodebench}, and CoderEval~\cite{yu2024codereval} use GitHub as their data source. 
Apart from Open-Source Platforms, Q\&A communities (e.g., StackOverflow~\cite{stackoverflow}) are also welcomed as the source of benchmark data~\cite{lai2023ds, liu2024llms, yin2018learning}. There are real and complex programming problems with real users' conversations and corresponding code snippets. These data is valuable for their close connection between code and NL. 
Some benchmarks are constructed by existing datasets and projects, and they are further modified and processed by subsequent researchers. For example, a mixed benchmark CodeXGLUE~\cite{lu2021codexglue} consists of different data sources and postprocesses them to fit its requirements.

\phead{Label.}
The labeling strategies adopted in code generation benchmarks vary significantly depending on dataset design goals, task granularity, and the intended evaluation focus. Many benchmarks provide \textbf{human-written ground truth} solutions, either crafted by experts or verified crowdworkers. These labels serve as reliable references for functional correctness and style quality, as seen in datasets like HumanEval, MBPP, and CoNaLa.
A large portion of benchmarks are \textbf{mined from real-world code repositories or programming platforms}, such as GitHub or LeetCode. In these cases, accepted submissions or popular implementations are treated as ground-truth labels, sometimes with multiple correct solutions per task (e.g., APPS, CodeContests, StudentEval, CrossCode-Bench).
To support automated evaluation, many benchmarks use \textbf{execution-based} labels, where a generated solution is deemed correct if it passes predefined unit tests or functional evaluations. This approach is widely adopted due to its scalability and objective measurement (e.g., EvalPlus, DevEval, CoderEval, MBPP-san-DFY).
Beyond execution, some benchmarks utilize \textbf{static or structural matching}, evaluating correctness through syntactic similarity, abstract syntax trees (AST), or data flow representations (e.g., CodeBLEU-based datasets, ComplexCodeEval, Code-Vision). These enable evaluation even in the absence of executable test environments.
In summarization-style or documentation-linked tasks, \textbf{automatically derived labels} are obtained by aligning code snippets with associated natural language descriptions, such as docstrings or comments (e.g., CodeXGLUE, PoorCodeSumEval, Chart2Code-160k).
More recent benchmarks expand evaluation beyond correctness by incorporating \textbf{behavioral and efficiency-oriented labels}, such as runtime, memory usage, and dependency coverage, providing more holistic assessments of model utility (e.g., EffiBench, Devbench, Mercury, SolEval).
Finally, some datasets focus on robustness and safety by using \textbf{heuristic or rule-based labeling} to detect risky or incorrect usage patterns. These labels are derived from static analysis or vulnerability patterns (e.g., SecurityEval, Decept-Prompt, ExampleCheck), targeting misuse detection and adversarial robustness.

\subsubsection{Discussion}
~\cjk{
Among these code generation and recommendation benchmarks, \textbf{HumanEval}~\cite{chen2021codex} and \textbf{MBPP}~\cite{DBLP:journals/corr/abs-2108-07732} are the most widely adopted and influential. We would like to discuss three key factors we identify that contributed to their success: 
(i) Appropriate level of difficulty. 
Both benchmarks were proposed around 2021–2023, a period when large pretrained models were rapidly advancing from token- or line-level completion to function-level code generation. HumanEval and MBPP were well-aligned with this transition: they focus on function-level tasks that are neither unrealistically difficult nor trivially easy, making them suitable for evaluating models at that stage of development. 
(ii) Well-designed and user-friendly evaluation. 
These benchmarks provide clear evaluation metrics and accessible evaluation suites. For instance, the HumanEval repository~\cite{openai_humaneval} has attracted nearly 3K GitHub stars, reflecting both its popularity and ease of use. Its implementation is simple to read, extend, and adapt to other scenarios—features that make it especially convenient for researchers. Moreover, both benchmarks adopt reliable execution-based metrics (i.e., Pass@K), which are generally more trustworthy than similarity-based metrics (e.g., CodeBLEU). 
This also enables a single comparable score across models.
(iii) Strong institutional backing. 
HumanEval and MBPP were introduced by leading AI companies (OpenAI and Google, respectively). Benchmarks released by such organizations tend to receive significant attention and quickly become community standards, as subsequent researchers often follow their evaluation practices. The involvement of these resource-rich institutions also lends credibility and ensures dataset quality. For example, HumanEval is human-annotated, which makes it more trustworthy than benchmarks curated by individuals with limited resources.}

\nff{Recently, \textbf{BigCodeBench}~\cite{zhuo2024bigcodebench} has emerged as a highly influential benchmark for evaluating LLMs in code generation. It incorporates diverse function calls, complex instructions, and a rigorous evaluation mechanism, thereby providing a more realistic and challenging assessment of model capabilities. In contrast to earlier benchmarks such as HumanEval~\cite{chen2021codex} and MBPP~\cite{DBLP:journals/corr/abs-2108-07732}, which primarily consist of simple and self-contained tasks, BigCodeBench more accurately reflects the complexity of real-world programming scenarios. Furthermore, its ease of use, open accessibility, and strong community support have contributed to its wide adoption, extensive citations, and growing popularity, as evidenced by its numerous GitHub stars.}

\cjk{
However, with the rapid progress of foundation models and novel training techniques, agentic coding is becoming mainstream, and function-level benchmarks are becoming saturated and no longer sufficient to distinguish performance differences among state-of-the-art models. As a result, agentic benchmarks, with a representative of SWE-bench~\cite{yang2024swe}, are emerging as the new standard for software engineering tasks. 
We believe that these benchmarks, which better reflect real-world developer experiences, will play a central role in future evaluations and gain widespread adoption.}

\subsubsection{Limitations and Opportunities} \label{subsec: limitations}
Despite the rapid development of code generation and recommendation benchmarks, current datasets still face several limitations that constrain comprehensive and robust evaluation. At the same time, these limitations also reveal opportunities for future benchmark design.

\noindent
\textbf{Over-reliance on Pass@k and Execution-Based Metrics.}
Many benchmarks adopt \textit{Pass@k} as the dominant evaluation metric, which solely measures whether generated code passes a predefined set of test cases. 
\cjk{
For example, MBPP~\cite{DBLP:journals/corr/abs-2108-07732} provides only three functional test cases for each problem.}
This often overlooks partial correctness, code quality, and semantic nuance, and is sensitive to the number and diversity of test cases. 
~\cjk{
The enhancement of the existing test suite is one promising direction, such as EvalPlus~\cite{liu2024your} and UTBoost~\cite{yu2025utboost}, which provide a more rigorous assessment of LLM's solutions. 
Also, evaluating coding assistants beyond functional correctness (e.g., maintainability, readability, efficiency) is attracting more attention~\cite{zheng2024correctnessbenchmarkingmultidimensionalcode,wang2025maintaincodermaintainablecodegeneration}.
}

\noindent
\textbf{Limited Diversity of Tasks and Contexts.}
Existing benchmarks predominantly focus on short, isolated programming tasks.
\cjk{
In the scope of surveyed benchmarks, a large proportion of them focus on function-level, self-contained programming tasks. 
For example, HumanEval~\cite{chen2021codex} requires LLMs to generate a standalone function based on a short piece of function signature, without other dependencies (i.e., third-party packages) or context (i.e., cross-file reference). 
}
There is a lack of coverage for real-world software engineering scenarios, such as multi-file projects, GUI or API development, or maintenance-oriented tasks. 

\noindent
\textbf{Insufficient Coverage of Programming Languages and Paradigms.}
Most benchmarks target high-resource, imperative languages like Python, Java, and C++. Functional, logic-based, and domain-specific languages (e.g., Rust, Prolog, Verilog) remain underrepresented, limiting the generalizability of evaluation results. 
\cjk{
Benchmarks targeted at low-resource programming languages (e.g., OCaml) or domain-specific scenarios (e.g., scientific and high-performance computing) are required to perform a holistic and unbiased evaluation for LLMs. Recently, some efforts~\cite{thakur2023benchmarking, mora2024synthetic} have begun to focus on them. 
}

\noindent
\textbf{Shallow Reasoning and Low-Level Semantics.}
Benchmarks often fail to capture deeper logical reasoning, algorithmic insight, or mathematical precision. Few benchmarks evaluate symbolic reasoning, problem decomposition, or invariant preservation. 
\cjk{
Some research has commenced investigating this from the perspective of program comprehension~\cite{chakraborty2023ranking,wei2025equibenchbenchmarkinglargelanguage, allamanis2025disprovingprogramequivalencellms}, however, it remains unclear how these semantic reasoning abilities can be utilized and assessed in a coding assistant.
}

\noindent
\textbf{Label Noise and Ground Truth Ambiguity.}
Some benchmarks rely on mined or auto-aligned ground truth solutions, which can be noisy or incomplete. For example, there may be multiple valid implementations for a task, but benchmarks assume a single correct reference or oracle. 
\cjk{
A recent research~\cite{evalarena} systematically investigates this issue on several popular benchmarks such as LiveCodeBench~\cite{jain2024livecodebench} and DS1000~\cite{lai2023ds}. They find that widely adopted benchmarks can also contain much noise, including wrong reference answers and underspecified requirements. 
To address these challenges, some benchmarks provide a human-validated subset beyond the original datasets to mitigate this issue, e.g., the SWE-bench team collaborated with OpenAI and released a verified version~\cite{openai2024swebench}, which could be used in other benchmarks.
}

\noindent
\textbf{Lack of Human-Centric Evaluation Dimensions.}
Code generation quality also involves readability, maintainability, efficiency, and developer satisfaction, which are difficult to capture through automated metrics alone. 
\cjk{
Although a few works pay some attention to the human aspect in coding assistants~\cite{miah2024usercentricevaluationcode,mozannar2025the}, how to fairly and systematically assess this perspective remains largely unaddressed compared with other points of view.
}

%% file: R_Testing.tex
\begin{table}[]
\centering
\caption{Existing Benchmarks for Evaluating LLMs Related to Software Testing}
    \resizebox{\linewidth}{!}{
\begin{tabular}{ccccll}
\toprule
Task & Benchmarks & Year &  Language & Evaluation Metrics & \# Cases  \\ \midrule
\rowcolor[HTML]{EFEFEF}
\cellcolor[HTML]{FFFFFF} & Evosuite SF110~\cite{DBLP:journals/tosem/FraserA14} & 2011 &  Java & \begin{tabular}[c]{@{}l@{}}Line coverage, branch coverage, \\ and test correctness\end{tabular} & 8,784 public classes \\
 & Defects4J~\cite{just2014defects4j} & 2014 &  Java & \begin{tabular}[c]{@{}l@{}}The number of test case is executable, \\ CodeBLEU, line coverage, \\ branch coverage, and \\ the number of detected bugs\end{tabular} & 395 patches \\
 \rowcolor[HTML]{EFEFEF}
\cellcolor[HTML]{FFFFFF} & DynaMOSA~\cite{DBLP:journals/tse/PanichellaKT18} & 2018 & Java & \begin{tabular}[c]{@{}l@{}}Line coverage, branch coverage, \\ number of detected bugs\end{tabular} & 346 Java classes \\
 & BugsInPy~\cite{DBLP:conf/sigsoft/WidyasariSLQPTT20} & 2020 &  Python & \begin{tabular}[c]{@{}l@{}}Line coverage, branch coverage,\\ and number of detected bugs\end{tabular} & 493 real bugs \\
\rowcolor[HTML]{EFEFEF}
\cellcolor[HTML]{FFFFFF} Test Generation & HumanEval~\cite{chen2021codex} & 2021 &  Python & \begin{tabular}[c]{@{}l@{}}Mutation score, Pass@K, \\ the number of killed mutants, \\ line coverage, and branch coverage\end{tabular} & 164 pairs \\
& MBPP~\cite{DBLP:journals/corr/abs-2108-07732} & 2021 &  Python & Pass@K & 964 instances \\
\rowcolor[HTML]{EFEFEF}
\cellcolor[HTML]{FFFFFF} & APPS~\cite{hendrycksapps2021} & 2021 & Python & Pass@K & \begin{tabular}[c]{@{}l@{}}232,421 programs, \\ 131,777 test cases\end{tabular} \\
 & CodeContests~\cite{li2022alphacode} & 2022 &  Python & Pass@K & - \\
\rowcolor[HTML]{EFEFEF}
\cellcolor[HTML]{FFFFFF} & HumanEval-X~\cite{codegeex} & 2023 & Multiple PLs & Pass@K & 820 pairs \\
 & CoderUJB~\cite{DBLP:journals/corr/abs-2403-19287} & 2024 & Java & \begin{tabular}[c]{@{}l@{}}Syntax correctness rate, \\ compile passing rate, \\ line coverage,\end{tabular} & 2,239 questions \\
\rowcolor[HTML]{EFEFEF}
\cellcolor[HTML]{FFFFFF} & \hx{SWT-Bench}~\cite{swt-bench} &2024 & Python &Success
rate and change coverage  & 2,294 issues \\
&TestBench~\cite{DBLP:journals/corr/abs-2409-17561}& 2024 &Java & \begin{tabular}[c]{@{}l@{}}Syntax/compilation/execution correctness rate, \\ coverage/defect detection rate,\end{tabular} & 108 Java programs \\ 
\rowcolor[HTML]{EFEFEF} 
\cellcolor[HTML]{FFFFFF} &\hx{TestEval}~\cite{wang-etal-2025-testeval}& 2025 &Python & overall/line/branch/path coverage& 210 problems \\

\cellcolor[HTML]{FFFFFF} &ProjectTest~\cite{DBLP:journals/corr/abs-2502-06556}& 2025 &Python, Java, and JavaScript & Compilation/correctness/coverage rate& 59 projects \\

 \midrule
\rowcolor[HTML]{EFEFEF}
\cellcolor[HTML]{FFFFFF} Assertion Generation & ATLAS~\cite{DBLP:conf/icse/WatsonTMBP20} & 2020 & Java & \begin{tabular}[c]{@{}l@{}}Exact match, edit distance, and \\ longest common subsequence\end{tabular} & \begin{tabular}[c]{@{}l@{}}188,154 test \\ assertions\end{tabular} \\ \midrule
\multirow{4}{*}{GUI Test} & Themis~\cite{DBLP:conf/sigsoft/0001WS21} & 2021 & Android Apps & \begin{tabular}[c]{@{}l@{}}The number of detected bugs, and \\ activity coverage\end{tabular} & \begin{tabular}[c]{@{}l@{}}20 Apps with \\ 34 bugs\end{tabular} \\
 & \highlight QTypist~\cite{DBLP:conf/icse/LiuCWCHHW23} &\highlight  2021 &\highlight  Android Apps &\highlight  \begin{tabular}[c]{@{}l@{}}Passing rate, coverage metrics,\\ activity number, and page number\end{tabular} &\highlight  106 Apps\\ \midrule
\multirow{6}{*}{Testing Automation} & LAVA-M~\cite{DBLP:conf/sp/Dolan-GavittHKL16} & 2016 &  C/C++ & Coverage, Unique bug & - \\
& \highlight Unibench~\cite{DBLP:conf/uss/LiJCLLCLWBCL021} & \highlight 2021 & \highlight C/C++ & \highlight \begin{tabular}[c]{@{}l@{}}Quality of bugs, stability of finding bugs,\\ speed of finding bugs, and overhead\end{tabular} & \highlight \begin{tabular}[c]{@{}l@{}}20 programs and 12+ \\ vulnerability types\end{tabular} \\
 & FuzzBench~\cite{DBLP:conf/sigsoft/MetzmanSSSA21} & 2021 &  Multiple PLs & Coverage, Unique bug & - \\
& \highlight  FuzzGPT~\cite{DBLP:journals/corr/abs-2304-02014} &\highlight  2024 &\highlight  Python &\highlight  \begin{tabular}[c]{@{}l@{}}Code coverage, API coverage,\\ number of unique crashes\end{tabular} &\highlight  \begin{tabular}[c]{@{}l@{}}1750, 633 snippets for \\ PyTorch,TensorFlow\end{tabular} \\
\midrule
\multirow{2}{*}{Testing Prediction} & IDoFT~\cite{DBLP:conf/icst/LamOSM019} & 2019 & Java & Precision, Recall, F1-Score & 3,742 Flaky test cases\\
& \highlight FlakeFlagger~\cite{DBLP:conf/icse/AlshammariMH021} & \highlight 2021 &\highlight  Java &\highlight  Precision, Recall, F1-Score & \highlight 21,661 test cases \\ 
\midrule
\multirow{2}{*}{Testing Repair} & TARBENCH~\cite{DBLP:journals/tse/YaraghiHKB25} & 2025 & Java & CodeBLEU, BLEU, exact match, repair accuracy & 45,373 broken test repairs\\
& \highlight Syn-Bench~\cite{DBLP:journals/pacmpl/RahmanKCGWMDR25} & \highlight 2025 &\highlight  Python &\highlight Syntactic/semantic correctness, code coverage  & \highlight 721 test cases \\ 
\bottomrule
\end{tabular}}
\label{tab:benchmarks_test_generation}
\end{table}

\section{SOFTWARE TESTING}
Software testing is a fundamental activity throughout the software development life cycle, serving as a critical mechanism for early fault detection, functional correctness verification, and the enhancement of overall software reliability and quality~\cite{myers2011art, everett2007software}.
In our analysis, we identified 25 benchmarks that have been used to evaluate LLM-based software testing models across various tasks, including test generation (14), assertion generation (1), GUI testing (2), test automation (4), test prediction (2), and testing repair (2).
A summary of these benchmarks is presented in Table~\ref{tab:benchmarks_test_generation}.


\subsection{Benchmarks Overview}
\subsubsection{Benchmarking Scenarios}
Benchmarks for evaluating LLMs in software testing span a diverse set of task scenarios, each targeting a specific aspect of the testing process. Among these, test generation stands out as the most prominent. \zjw{Test generation refers to the automated creation of executable test cases, which typically include input values, test drivers, and the overall structure required to execute the target program under test.}
Leveraging the advanced natural language understanding and code synthesis capabilities of LLMs, test generation can be significantly accelerated while enhancing the quality of the produced tests~\cite{DBLP:journals/corr/abs-2406-00515}.
In recent years, the research community has introduced a variety of high-quality benchmarks to rigorously evaluate the effectiveness of LLMs in this task. Notable examples include EvoSuite SF110~\cite{DBLP:journals/tosem/FraserA14}, Defects4J~\cite{just2014defects4j}, DynaMOSA~\cite{DBLP:journals/tse/PanichellaKT18}, and HumanEval~\cite{chen2021codex}, which are widely used to assess a model’s ability to generate executable and effective test cases with high code coverage or strong bug detection capabilities. Another important category is assertion generation, represented by datasets like ATLAS~\cite{DBLP:conf/icse/WatsonTMBP20}, which focus on generating correct and meaningful test assertions. \zjw{Assertion generation focuses on automatically producing the oracles within those test cases, i.e., the conditions that determine whether the observed program outputs or states are correct. Unlike test generation, which mainly focuses on creating inputs and execution paths, assertion generation is characterized by its emphasis on producing test oracles that check correctness.}
GUI testing benchmarks such as Themis~\cite{DBLP:conf/sigsoft/0001WS21} and QTypist~\cite{DBLP:conf/icse/LiuCWCHHW23} evaluate the ability to explore and verify user interfaces, often in Android applications. Testing automation and fuzzing are covered by benchmarks like LAVA-M~\cite{DBLP:conf/sp/Dolan-GavittHKL16}, FuzzBench~\cite{DBLP:conf/sigsoft/MetzmanSSSA21}, Unibench~\cite{DBLP:conf/uss/LiJCLLCLWBCL021}, and FuzzGPT~\cite{DBLP:journals/corr/abs-2304-02014}, emphasizing code coverage, bug discovery, and execution efficiency in automated or adversarial testing environments. Finally, test flakiness prediction is addressed by benchmarks such as IDoFT~\cite{DBLP:conf/icst/LamOSM019} and FlakeFlagger~\cite{DBLP:conf/icse/AlshammariMH021}, which involve predicting flaky test cases using precision, recall, and F1-score. These benchmarks collectively support comprehensive evaluation of LLMs across various software testing dimensions.

\subsubsection{Evaluation Metrics}
Software testing benchmarks employ a variety of evaluation metrics to assess the performance of test generation models and tools. The most common metrics include:

\phead{\ding{42} Coverage-based metrics}, such as line coverage and branch coverage, are widely used across test generation (e.g., Evosuite, Defects4J~\cite{just2014defects4j}, DynaMOSA~\cite{DBLP:journals/tse/PanichellaKT18}, FuzzGPT) and fuzzing benchmarks (e.g., FuzzBench~\cite{DBLP:conf/sigsoft/MetzmanSSSA21}), reflecting how thoroughly the generated tests explore the codebase.

\phead{\ding{42} Test effectiveness metrics}, such as the number of detected bugs, number of killed mutants, or unique crashes, measure the fault-revealing power of the generated test cases. These are crucial in benchmarks like Defects4J~\cite{just2014defects4j}, BugsInPy~\cite{DBLP:conf/sigsoft/WidyasariSLQPTT20}, FuzzGPT, and LAVA-M~\cite{DBLP:conf/sp/Dolan-GavittHKL16}.
Pass@K is a dominant metric in code generation-oriented benchmarks (e.g., HumanEval~\cite{chen2021codex}, MBPP~\cite{DBLP:journals/corr/abs-2108-07732}, APPS, CodeContests~\cite{li2022alphacode}), evaluating whether at least one of the top-K generated test cases passes the associated test suite.

\phead{\ding{42} Mutation score}, another effectiveness-oriented metric, evaluates how well test cases can detect injected faults (mutants).

\phead{\ding{42} Syntax and compilation correctness}, used in benchmarks like CoderUJB~\cite{DBLP:journals/corr/abs-2403-19287}, assess whether the generated code is syntactically valid and compiles successfully.

\phead{\ding{42} Assertion quality metrics}, including exact match, edit distance, and longest common subsequence, are used in assertion-focused benchmarks such as ATLAS~\cite{DBLP:conf/icse/WatsonTMBP20} to evaluate the correctness of generated test oracles.

\phead{\ding{42} GUI-specific metrics}, like activity coverage, page number, and passing rate, appear in mobile testing benchmarks such as Themis~\cite{DBLP:conf/sigsoft/0001WS21} and QTypist~\cite{DBLP:conf/icse/LiuCWCHHW23} to assess interaction completeness and behavior correctness.

\phead{\ding{42} Prediction metrics} such as precision, recall, and F1-score are adopted in flaky test prediction benchmarks (IDoFT~\cite{DBLP:conf/icst/LamOSM019}, FlakeFlagger~\cite{DBLP:conf/icse/AlshammariMH021}) to evaluate classification accuracy.

These diverse metrics capture multiple dimensions of software testing quality—including structural coverage, functional correctness, fault detection, and behavioral validity—enabling comprehensive evaluation of LLM-based testing approaches.

\subsection{Benchmarks Construction} 
\subsubsection{Test Generation Benchmark}
MBPP~\cite{DBLP:journals/corr/abs-2108-07732} is derived from programming problems written by crowdsourcing participants, including a problem statement, a Python function that solves the problem, and three test cases that check the semantic correctness of the function.
APPS~\cite{hendrycksapps2021} is collated from open-source coding websites such as Codewars, AtCoder, Kattis, and Codeforces. The researchers optimized parts of the questions such as formulas and lists. Questions that relied on image graphics and duplicate data were removed. Further manual inspection was used to ensure the benchmark's accuracy.
CodeContests~\cite{li2022alphacode} consists of problems, solutions, and test cases scraped from the Codeforces platform. It combines new data scraped from Codeforces with existing datasets such as Description2Code and CodeNet. This mix ensures diversity and comprehensiveness of competitive programming problems. This benchmark undergoes thorough cleaning to remove duplicates and ensure data quality. Problems with insufficient test coverage in both validation and test splits are filtered out. Only problems with at least five hidden or generated test cases leading to at least two different outputs are retained. In addition, this benchmark generates additional test cases by mutating existing test inputs, including bit flipping, random increment or decrement of integers, and mutation of string characters. Correct solutions are validated by running them on these mutated inputs to ensure consistency and correctness. To avoid data leakage and ensure dataset integrity, a strict temporal split is adopted. All training data was publicly released on or before July 14, 2021, while validation problems appeared between July 15, 2021 and September 20, 2021. The test set contains problems posted after September 21, 2021. This split ensures no information from training affects the evaluation problems.
OpenAI expanded HumanEval~\cite{chen2021codex} to include Python problems only, manually rewriting the prompts, canonical solutions, and test cases for each problem in C++, Java, JavaScript, and Go. Adjustments were made to ensure that the code adheres to idiomatic programming style and language-specific behaviors. For example, naming conventions such as CamelCase or snake\_case were applied according to the language specification. In addition, the location of docstrings and modifications to symbols and keywords were tailored to fit the target language's syntax. Test cases were refined to account for language-specific behaviors and ensure functional correctness.
The initial version of Defects4J~\cite{just2014defects4j} includes 357 real bugs from five large open-source Java projects. These projects were chosen for their size and diversity, ensuring a wide range of real-world bugs. Bugs were identified by analyzing version control and bug tracking systems. A commit was considered as fixing a bug if it was labeled as such by the developer or referenced in the bug tracking system. Only source code-related bugs were included, excluding issues related to documentation, tests, or build configurations. For each identified bug, a buggy and fixed source code version was obtained. The buggy version contains the bug, while the fixed version includes the bug fix. 
DynaMOSA~\cite{DBLP:journals/tse/PanichellaKT18} includes a random selection of classes from four test benchmarks: the SF110~\cite{DBLP:journals/tosem/FraserA14} corpus, the SBST tool contest 2015, the SBST tool contest 2016, and a previous benchmark used in their conference paper. 
BugsInPy~\cite{DBLP:conf/sigsoft/WidyasariSLQPTT20} includes 493 real bugs from 17 open-source Python projects. These projects were chosen based on their popularity (each with more than 10,000 stars on GitHub) and their representation of diverse domains such as machine learning, developer tools, scientific computing, and web frameworks. Bugs were identified by investigating commits that modified or added test files. A heuristic approach was used to identify test files, specifically looking for files with ``test'' in their names and those that imported testing libraries such as unittest or pytest. Each commit was manually examined to determine if it fixed a bug, using commit messages, source code changes, and linked information such as GitHub issues. To ensure reproducibility, each identified bug had to have at least one test case that exposed the bug. This means the test case should fail on the buggy version and pass on the fixed version. Two independent authors checked the commits to ensure they contained only the changes necessary to fix the bug, without unrelated changes such as refactoring or new feature additions. The database includes metadata for each bug, such as the Git revision hash, the patch that fixes the bug, and the test cases that expose the bug. A unique identifier was assigned to each bug, mapping to the corresponding Git revision hash in the original repository. 
CoderUJB~\cite{DBLP:journals/corr/abs-2403-19287} is built on 17 real open-source Java projects, chosen due to Java's prevalence in real-world software production. These projects were carefully selected to cover a wide range of practical programming tasks and ensure the generated solutions are executable within a complete program context. From these projects, 2,239 programming questions were extracted. The extraction process involved analyzing the abstract syntax trees and test coverage relationships of the project source code. The resulting benchmark includes both the code under test and the associated test cases, providing a complete program context for each question.

\subsubsection{Assertion Generation Benchmark}
ATLAS~\cite{DBLP:conf/icse/WatsonTMBP20} was constructed through a detailed and systematic process to generate meaningful assertion statements for unit test cases using a Neural Machine Translation (NMT) model. The first step involved mining GitHub for Java projects utilizing the JUnit testing framework. The search targeted projects with at least one pom file indicating dependency on JUnit version 4. From the identified projects, all test methods annotated with @Test and declared methods were mined. This resulted in a large dataset comprising over 9,000 projects and approximately 2.5 million examples of developer-written assertion statements within test methods. Each assertion statement was paired with its corresponding test method and the focal method it tests. 

\subsubsection{GUI Test Benchmark}
Themis~\cite{DBLP:conf/sigsoft/0001WS21} benchmark was constructed to provide a comprehensive and empirical evaluation of automated GUI testing tools for Android against real-world bugs. The initial step involved selecting open-source Android apps from GitHub and F-Droid. The selection aimed to include as many candidate apps as possible, resulting in 1,829 unique Android apps. Critical bugs were identified based on issue labels assigned by developers, using a set of predefined keywords to filter issues. This process resulted in 409 valid apps with critical issues. Collecting raw data on critical issues involved manual inspection of each issue from the filtered apps. The criteria for selecting issues included the presence of crash bugs, explicit reproducing steps, and issues submitted after January 1, 2017. The validation process required reproducing the bugs, which involved reviewing bug reports, locating buggy code versions, building the app binaries, and manually reproducing the bugs. This extensive effort resulted in the collection of 52 valid critical issues from 20 apps. The validated issues were archived, including an executable APK file, a bug-reproducing video, the exception stack trace, and other necessary information such as login scripts. The entire evaluation process involved running these tools on the collected bugs, resulting in over 10,920 CPU hours of testing. This extensive evaluation provided insights into the effectiveness and limitations of each tool.

\subsubsection{Testing Automation Benchmark}
Unibench~\cite{DBLP:conf/uss/LiJCLLCLWBCL021} consists of 20 real-world programs selected to represent various functionalities, sizes, and vulnerability types. Each fuzzer is manually built and tested. Flaws are identified and reported to developers, with fixes integrated into the platform. Dockerfiles are created to ensure consistent execution environments. 
FuzzGPT~\cite{DBLP:journals/corr/abs-2304-02014} benchmark adopts an HTML crawler to collect issues and pull requests from GitHub repositories of target deep learning libraries. Relevant code snippets that triggered bugs are extracted from these reports and pull requests. A self-training approach using few-shot learning is employed to label the buggy APIs in the extracted code snippets. This involves manual annotation of a small number of examples, which are then used to prompt the LLM to automatically label the rest. 
FuzzBench~\cite{DBLP:conf/sigsoft/MetzmanSSSA21} uses a diverse set of real-world programs from OSS-Fuzz, a community fuzzing service. The selected benchmarks are representative of a wide range of applications and coding styles, ensuring broad applicability of the evaluation results. The benchmark programs process various input formats, including JSON, JPEG, XML, and more. This variety helps ensure that tools optimized for these benchmarks will perform well in general scenarios. FuzzBench~\cite{DBLP:conf/sigsoft/MetzmanSSSA21} ensures reproducibility by pinning fuzzers and benchmarks to specific versions and providing detailed version information in reports. This allows experiments to be replicated accurately. 

\subsubsection{Testing Prediction Benchmark}
The IDoFT~\cite{DBLP:conf/icst/LamOSM019}  benchmark starts by collecting a list of project URLs and specific commits from various sources. This includes 44 projects from previous related work and the 150 most popular Java projects on GitHub up to October 2018. Additionally, 500 popular Java projects updated in November 2018 were included. In total, 683 projects were considered, comprising 5171 modules and over 1.88 million tests. For each project, a Docker image is created to ensure an isolated and reproducible environment. FlakeFlagger~\cite{DBLP:conf/icse/AlshammariMH021} involved rerunning the test suites of 24 projects 10,000 times each. The dataset was constructed by capturing various behavioral features of each test. These features include test execution time, test lines of code, number of assertions, coverage of source code, coverage of changed lines, usage of third-party libraries, and more. FlakeFlagger~\cite{DBLP:conf/icse/AlshammariMH021} represents a significant step forward in automating the detection of flaky tests, providing a more efficient and reliable approach compared to traditional rerun-based benchmark.

\zjw{\subsection{Discussion}}
\zjw{\textbf{Defects4J}~\cite{just2014defects4j} is a widely used benchmark for evaluating unit test generation, as it provides a collection of real, reproducible bugs from well-maintained open-source Java projects, along with automated test suites and reproducible build environments, enabling rigorous, consistent, and comparable evaluation of testing and debugging techniques. However, its dataset is small, outdated, and limited to a few well-maintained Java projects, failing to capture the diversity of modern software. Similarly, \textbf{BugsInPy}~\cite{DBLP:conf/sigsoft/WidyasariSLQPTT20} extends the ideas of Defects4J to the Python ecosystem, but faces data imbalance. Another popular benchmark, \textbf{HumanEval}~\cite{chen2021codex}, widely adopted for LLM-based testing, risks outdated samples and label noise, since many of its hand-crafted tasks represent simple programming puzzles rather than realistic, evolving test scenarios. These issues suggest that existing benchmarks may not generalize well to more complex, real-world settings.}

\zjw{In terms of usage trends, benchmarks such as Defects4J~\cite{just2014defects4j} and HumanEval~\cite{chen2021codex} continue to dominate largely due to their early introduction, ease of use, and broad community adoption. Researchers often prefer these datasets because they provide standardized evaluation process and are supported by extensive prior studies, which facilitates comparability across new approaches. By contrast, newer datasets like BugsInPy~\cite{DBLP:conf/sigsoft/WidyasariSLQPTT20}, despite addressing real-world tests, are still gaining traction, partly because they require more sophisticated setup and evaluation pipelines. This trend highlights a tension between convenience and realism: older benchmarks are easier to use but less representative, whereas newer ones are more realistic but face slower adoption.}

\subsection{Limitations and Opportunities}
\zjw{\textbf{Single-Language Limitation.} Defects4J~\cite{just2014defects4j}, SF110~\cite{DBLP:journals/tosem/FraserA14}, DynaMOSA~\cite{DBLP:journals/tse/PanichellaKT18}, ATLAS~\cite{DBLP:conf/icse/WatsonTMBP20}, and CoderUJB~\cite{DBLP:journals/corr/abs-2403-19287} target only Java programs. In fact, over half of the widely adopted testing benchmarks are Java-specific, which reduces their utility for evaluating tools that support multiple programming languages. Similarly, HumanEval~\cite{chen2021codex}, MBPP~\cite{DBLP:journals/corr/abs-2108-07732}, APPS~\cite{hendrycksapps2021}, BugsInPy~\cite{DBLP:conf/sigsoft/WidyasariSLQPTT20}, and CodeContests~\cite{li2022alphacode} are limited to Python. These benchmarks focus on Python and Java, making them unsuitable for systematically evaluating multi-language software testing. Furthermore, these problems are often simple single-function generation tasks, which do not reflect real-world development scenarios (e.g., writing code across multiple files in a software project). \textbf{Opportunities:} Future benchmarks should broaden language coverage by systematically collecting problems from multiple ecosystems (e.g., C++, JavaScript, Rust). Constructing multilingual or cross-language benchmarks would not only better reflect industrial settings but also provide evidence on how well LLMs generalize across programming languages.}

\noindent
\zjw{\textbf{Project Selection Bias.} Several benchmarks rely on a narrow range of projects. For instance, Defects4J~\cite{just2014defects4j}, BugsInPy~\cite{DBLP:conf/sigsoft/WidyasariSLQPTT20} and CoderUJB~\cite{DBLP:journals/corr/abs-2403-19287} are selected from popular open-source repositories. This selection risks overlooking more complex or less-maintained systems, thereby biasing evaluations toward cleaner and less error-prone projects. \textbf{Opportunities:} Incorporating projects from diverse domains, including legacy systems, industrial codebases, and under-maintained repositories, would lead to more realistic and representative benchmarks. This diversification would enable LLM evaluations to capture edge cases and failure modes that are critical in practice but often absent from curated datasets.}

\noindent
\zjw{\textbf{Data Leakage Limitation.} Existing benchmarks for evaluating test generation models based on LLMs face severe data leakage risks. Such leakage inflates performance metrics and weakens conclusions. Besides, many benchmarks do not apply systematic deduplication~\cite{DBLP:journals/pacmse/ZhangHGXLL25}, which increases the likelihood of overlapping or duplicated entries between training and testing sets. \textbf{Opportunities:} To mitigate data leakage, future work should design benchmarks with rigorous dataset curation pipelines, including automated deduplication, cross-referencing with known pretraining corpora, and version-controlled dataset releases. Transparent reporting of overlap rates would further strengthen the reliability of benchmark-driven evaluations.}

%% file: R_AIOps.tex
\section{AIOps}\label{sec:aiops}

AIOps (Artificial Intelligence for IT Operations) employs machine learning and data analytics to streamline and improve IT operations. It automates crucial tasks such as log parsing, log generation, and log statement generation, boosting efficiency and minimizing downtime. By processing and analyzing large volumes of data from diverse sources, AIOps enables proactive issue identification and resolution, often before users are affected.

\subsection{Benchmarks Overview}

\subsubsection{Benchmarking Scenarios}
Benchmarks for evaluating LLMs in AIOps span two primary task scenarios, each addressing a critical aspect of system observability and log intelligence. The first and most studied scenario is log statement generation, where benchmarks such as LANCE~\cite{mastropaolo2022using}, \nff{LEONID~\cite{mastropaolo2024log}}, LogBench~\cite{li2023exploring}, SCLoger~\cite{li2024go}, and AL-Bench~\cite{tan2025bench} are designed to assess a model's ability to generate accurate, context-aware logging statements within source code. These benchmarks typically consist of Java methods annotated with developer-written log statements, and are evaluated using metrics such as accuracy, precision, recall, F1, BLEU, and ROUGE. The second scenario is log parsing, which focuses on extracting structured templates or key fields from raw log data. Benchmarks like Loghub~\cite{zhu2023loghub} and Loghub-2.0~\cite{jiang2024large} provide large-scale, real-world log datasets from a variety of systems including distributed systems, operating systems, and supercomputers. These benchmarks are used to evaluate both traditional log parsing tools and emerging LLM-based approaches, with accuracy and F1-score as common metrics. Together, these benchmarks provide a foundation for systematically evaluating LLM capabilities in addressing core AIOps challenges such as automated observability, failure diagnosis, and log understanding.

\subsubsection{Evaluation Metrics}
AIOps benchmarks employ a variety of evaluation metrics to assess LLM performance across log-related tasks.

\phead{\ding{42} Accuracy-based metrics}, such as correct prediction ratio and accuracy, are commonly used in log statement generation (e.g., LANCE, LogBench, SCLoger) and log parsing (e.g., Loghub, Loghub-2.0) to evaluate output correctness.

\phead{\ding{42} Similarity metrics}, including BLEU and ROUGE, measure the lexical and semantic similarity between generated and reference log statements.

\phead{\ding{42} Classification metrics}, such as precision, recall, and F1-score, are applied to both log generation and parsing tasks to assess structural and semantic prediction quality.

These metrics collectively support comprehensive evaluation of LLMs on generation fidelity, semantic alignment, and structured log understanding in AIOps scenarios.

\begin{table}[t]
    \centering
    \caption{Existing Benchmarks for Evaluating LLMs Related to AIOps}
    \resizebox{0.7\linewidth}{!}{
    \begin{tabular}{cccccc}
    \toprule
        Task & Benchmarks & Year &  Language & Evaluation Metrics & \# Cases   \\ 
       \hline
       \rowcolor[HTML]{EFEFEF}
    \cellcolor[HTML]{FFFFFF} \multirow{8}{*}{\makecell{Log Statement \\ Generation}} & LANCE~\cite{mastropaolo2022using} & 2022 &  Java & correct prediction ratio &  76,421 \\ 
    & LEONID~\cite{mastropaolo2024log} & 2024 &  Java & \begin{tabular}[c]{@{}l@{}}correct prediction ratio,\\ BLEU, ROUGE, METEOR,\\ LEVENSHTEIN Distance\end{tabular} &  1.7 Million \\
     & LogBench~\cite{li2023exploring} & 2024 &  Java & \begin{tabular}[c]{@{}l@{}}Accuracy, Precision, Recall,\\ F1, BLEU, and ROUGE\end{tabular}  & 6,849 \\ 
    & \highlight  SCLoger~\cite{li2024go} & \highlight 2024 & \highlight Java & \highlight \begin{tabular}[c]{@{}l@{}}Accuracy, Precision, Recall,\\ F1, BLEU, and ROUGE\end{tabular} & \highlight 31,170 \\ 
    &\hx{AL-Bench}~\cite{tan2025bench} &2025 & Java &\begin{tabular}[c]{@{}l@{}} Position Accuracy, Level Accuracy, \\ Average Level Distance, Message Accuracy, \\ Dynamic Expression Accuracy, Static Text Similarity\end{tabular}&  39,600 \\
    \hline

    \multirow{2}{*}{Log Parsing} & \highlight Loghub~\cite{zhu2023loghub} & \highlight 2023 &  \highlight Multiple PLs & \highlight Accuracy  & \highlight -   \\ 
     &  Loghub-2.0~\cite{jiang2024large} & 2024 &  Multiple PLs & Accuracy, F1-score  & 3.6 Million \\ 
         \bottomrule
    \end{tabular}}
    \label{tab: aiops}
\end{table}

\subsection{Benchmarks Construction}

\subsubsection{Log statement generation.}
For the log statement generation task, the LANCE benchmark~\cite{mastropaolo2022using} collection begins with Java projects mined from GitHub. They applied selection criteria including projects with at least 500 commits, 10 contributors, 10 stars, and not being forks, resulting in 5,459 successfully cloned projects. Only projects declaring a dependency on Apache Log4j, a well-known Java logging library, were further considered. They identified 1,465 Java projects with an explicit dependency on Log4j.
The authors then used srcML to extract all Java methods from these projects. Approximately 96\% of these methods do not contain log statements, leaving around 320k methods that do feature at least one log statement. Methods with $\#$tokens $\geq$ 512 or \#tokens $<$ 10, where $\#$tokens represent the number of tokens composing a method (excluding comments), were then filtered out.
Lastly, the authors remove the original log statement from the Java method and insert a <LOG> placeholder in its place. The modified method, with the placeholder, is used as the input, while the original Java method, which includes the log statement, serves as the ground truth.
\nff{The LANCE benchmark assumes all methods require logging and restricts injection to a single log. To address the limitations, the LEONID benchmark~\cite{mastropaolo2024log} is proposed to identify methods that need logging and enable multiple log insertions when necessary. Compared to LANCE, LEONID introduces a new dataset for multi-log injection: given a method $M$ with $k$ log statements, it randomly removes $y$ log statements, where $1 \leq y \leq k$, and trains the model to reconstruct the removed logs. In addition, to support training a classifier that determines whether a method requires logging, LEONID introduces another dataset: given a method $M$ with $k$ log statements, it randomly removes $y$ log statements, where $0 \leq y \leq k$. Unlike the dataset used for multi-log injection, this dataset also includes instances where no log statements are removed ($y = 0$). Each processed method is then labeled to indicate whether it requires logging or not (True if $y \geq 1$, False if $y = 0$). 
}
The LogBench benchmark~\cite{li2023exploring} focuses on high-quality and well-maintained Java files with logging statements, mined from open-source repositories on GitHub. The authors select projects that have more than 20 stars, more than 100 commits, and at least five contributors.
They then filter projects based on whether their POM file includes popular logging utility dependencies (e.g., Log4j, SLF4J), resulting in 3,089 repositories. Java files containing at least one logging statement are extracted by matching them with regular expressions, as logging statements follow a specified syntax (e.g., \texttt{log.info()}). Subsequently, the authors randomly sample these files across various repositories, creating a dataset of 2,420 files containing 3,870 methods and 6,849 logging statements, referred to as LogBench-O.
In addition to LogBench-O, the authors create LogBench-T to avoid the potential data leakage to LLMs' pre-training data. They develop a code transformation tool that generates semantics-preserving and readability-preserving variations of the original code from LogBench-O. As the transformed code is no longer collected from GitHub, it is less likely to be encountered by LLMs during their pre-training.
The SCLoger benchmark~\cite{li2024go} selected ten open-source Java projects from various domains, including storage, cloud platforms, and computation engines, as the subject projects. They then analyzed the invocation of popular logging APIs (i.e., Log4j and SLF4J) at the Abstract Syntax Tree (AST) level to extract all log statements from the original samples. 
To complete the datasets, the extracted logging statements were tagged with line numbers (e.g., `<Line Number$\#$>') and the corresponding logging statements (e.g., `<Statement>: log.info(msg)') to indicate their position within the initial method and provide the complete logging statement. These tags served as the ground truth labels for their respective methods.
\zx{Recently, Tan et al. proposed AL-Bench~\cite{tan2025bench}, a new benchmark consisting of 21,804 code snippets and 39,600 log statements, extracted from the ten popular open-source repositories. When choosing the ten target repositories, Tan et al. require repositories to have at least 10,000 GitHub stars and 500 log-related issues, ensuring both community validation and logging relevance. Candidate repositories are then ranked by the total number of log statements to prioritize those with extensive logging activity. From the top of this ranked list, they manually curate a final set of 10 target repositories.}

\subsubsection{Log parsing}
For the log parsing task, LogHub~\cite{zhu2023loghub} was collected by aggregating logs from running several 19 software systems in the lab environment. 
The Loghub-2.0~\cite{jiang2024large} benchmark is an enhanced version of the original Loghub, featuring extensive annotation efforts facilitated by a proposed annotation framework. To construct Loghub-2.0, the authors selected 14 log datasets from Loghub, covering various types of systems, including distributed systems, supercomputers, and operating systems.
An annotation team of five skilled data annotators was assembled to carry out the process. The team first removed logs that did not contain any alphabetical characters, such as those composed solely of numerical figures or punctuation marks, due to their lack of parseable content. They also temporarily removed duplicate content to minimize manual effort in subsequent steps.
The annotators then used logging levels and component names, along with the most frequent tokens, to group logs into categories with similar log templates. Manual template annotation was employed to derive ground-truth log templates for each group. Finally, each template was converted into a regular expression, and each log message was matched against these templates, halting when a match was found.

\subsection{Discussions}

\nff{\textbf{LANCE}~\cite{mastropaolo2022using} has been widely adopted because it is one of the first large-scale, high-quality corpora specifically designed for log statement generation. It contains around 6.9 million Java methods from diverse open-source repositories, covering a wide variety of domains, coding styles, and logging conventions. The dataset bridges a critical gap between general code generation benchmarks and domain-specific tasks in software maintenance, enabling systematic evaluation of models that aim to understand and automate logging behavior.
For log parsing, the \textbf{Loghub}~\cite{zhu2023loghub} benchmark is the most widely used, because it provides 19 real-world log datasets collected from a wide range of software systems, including distributed systems, supercomputers, operating systems, mobile systems, server applications, and standalone software. Moreover, it offers free accessibility, open usability, and has been downloaded tens of thousands of times by hundreds of organizations, making it both a very practical and community-trusted resource for evaluating log parsing techniques.}

\nff{Regarding usage trends, benchmarks like LANCE~\cite{mastropaolo2022using} and Loghub~\cite{zhu2023loghub} remain dominant, primarily due to their early adoption and wide community support. Researchers frequently prefer these datasets as they are supported by extensive prior studies, enabling consistent comparisons across old and new approaches. However, over time, we expect the adoption of improved datasets (such as Loghub 2.0) to grow. As newer tools leverage these datasets and demonstrate their advantages over older versions, their usage may eventually surpass that of the original datasets, especially if recent baselines all adopt the improved data.}

\subsection{Limitations and Opportunities}

\textbf{Benchmark Data Contamination.} A significant issue is the risk of benchmark data contamination, where overlaps between training and test data, or between pre-training and test data, can lead to inflated performance metrics. Due to the lack of retraining capabilities and inaccessibility to the pre-training data of many LLMs, it is challenging to assess the extent of benchmark data contamination. This limitation applies to all benchmarks in AIOps. Although LogBench has implemented a list of code transformations to modify the source code, it is difficult to determine how effectively these measures address the benchmark data contamination issue.

\noindent
\textbf{Generalizability to Diverse Software.} Despite the vast amount of log data available, existing benchmarks in log parsing typically focus on a limited set of well-known and popular software. While the volume of logs is substantial, the number of studied projects is relatively small, which may impact the generalizability of the benchmarks to a broader range of projects.

\noindent
\textbf{Short Length Limit.}
In log statement generation, LANCE \nff{and LEONID} exclude methods longer than 512 tokens. This restriction could cause the LANCE \nff{and LEONID} benchmarks to differ from the real distribution of methods in the real software, as a number of methods may exceed 512 tokens but are not represented in those benchmarks.


%% file: R_Code_Analysis.tex
\section{Maintenance}

In software development, maintaining high-quality code and ensuring efficient maintainability rely on several key practices, including code review, code clone detection, and refactoring.  
Code review involves the systematic examination of source code by peers to ensure quality, correctness, and adherence to coding standards. It helps identify bugs early, promotes knowledge sharing, and improves overall code reliability.  
Code clone detection focuses on identifying and managing duplicated code fragments, which, if left unchecked, can increase maintenance overhead and introduce inconsistencies. Detecting clones enables better code reuse and helps prevent redundant changes.  
Refactoring refers to the process of restructuring existing code to enhance its internal structure and readability without altering its external behavior. It improves maintainability, reduces complexity, and facilitates future development. However, refactoring can disrupt code history~\cite{niu2024rat} and influence tasks such as bug localization~\cite{niu2023rat} and defect prediction~\cite{niu2025refactoring}. Therefore, it is essential to thoroughly understand the role and impact of refactoring.

\subsection{Benchmarks Overview}
\subsubsection{Benchmarking Scenarios}

In our literature review, we identified 13 benchmarks commonly used to evaluate the performance of LLMs on software maintenance tasks, as summarized in Table~\ref{tab: analysis}.

\noindent
\textbf{Task.} The benchmarks encompass three primary software maintenance tasks: code review, code clone detection, and refactoring. Code review benchmarks assess models’ ability to generate or evaluate review comments (e.g., CodeReview~\cite{tufano2022using}, CodeReviewer~\cite{li2022automating}), with recent additions focusing on explainability (e.g., Review-Explaining~\cite{widyasari2023explaining}) and comment rewriting (e.g., Code-Review-Assist~\cite{sghaier2023multi}). Clone detection benchmarks (e.g., BigCloneBench~\cite{svajlenko2015evaluating}, Curated CodeNet~\cite{dou2023towards}) evaluate a model’s capacity to identify semantically similar code segments. Refactoring benchmarks (e.g., JavaRef~\cite{liu2023refbert}) test how well models can transform code to improve structure and readability without changing functionality.

\noindent
\textbf{Context.} The benchmarks vary widely in context and evaluation format. Some involve natural language generation tasks, such as review comment generation, while others focus on code similarity detection or transformation accuracy. This variation reflects the diverse ways in which LLMs can be applied—ranging from generative to discriminative tasks.

\noindent
\textbf{Domain.} Most benchmarks target general-purpose programming languages. Java is the predominant language in code review and refactoring benchmarks (e.g., CodeReview, JavaRef), while clone detection benchmarks span a broader set of languages, including Java, C, C++, and Python (e.g., POJ-104~\cite{mou2016convolutional}, Company-C/C++~\cite{chochlov2022using}, Curated CodeNet~\cite{dou2023towards}).

\noindent
\textbf{Scale.} The size of the benchmarks ranges from small datasets (e.g., Company-C/C++ with 180 instances) to large-scale corpora (e.g., BigCloneBench and Curated CodeNet, each with over one million examples). This range allows researchers to assess model performance across various levels of data availability and generalizability.

\subsubsection{Evaluation Metrics}
To assess the performance of LLMs on software maintenance tasks, existing benchmarks employ a diverse set of evaluation metrics tailored to the nature of each task. For code review, metrics such as Exact Match, BLEU, and ROUGE are commonly used to measure the similarity between model-generated and reference review comments. Some benchmarks also evaluate semantic correctness and explanation type classification to capture deeper understanding. For clone detection, traditional information retrieval metrics like precision, recall, F1 score, MAP, and MRR are widely adopted to quantify the accuracy of clone identification. In code refactoring tasks, metrics such as accuracy, Exact Match, Edit Distance, and Character Error Rate (CER) are used to evaluate the syntactic and semantic equivalence of refactored code. These metrics collectively enable a comprehensive evaluation of model capabilities across syntactic, semantic, and structural dimensions of software maintenance.

\begin{table}[t]
    \centering
    \caption{Existing Benchmarks for Evaluating LLMs Related to Code Review, Code Clone, and Code Refactoring}
    \resizebox{0.9\linewidth}{!}{
    \begin{tabular}{llllll}
    \toprule
        Task & Benchmarks & Year &  Language & Evaluation Metrics & \# Cases  \\ 
       \hline
    \multirow{9}{*}{Code Review}  & \highlight CodeReview~\cite{tufano2022using} & \highlight 2022 &  \highlight Java & \highlight Exact Match  & \highlight 382,955   \\ 
    & CodeReviewer~\cite{li2022automating} & 2022 &   Multiple PLs  &  Exact Match and BLEU & \texttt{\textasciitilde} 642,000 \\ 
   &\highlight AUGER~\cite{li2022auger} &\highlight 2023 & \highlight Java & \highlight ROUGE, Perfect Prediction Rate & \highlight 79,344 \\
     & Review-Explaining~\cite{widyasari2023explaining} & 2023 &  Multiple PLs & \begin{tabular}[c]{@{}l@{}}explanation type correctness,\\ the semantic meaning correctness\end{tabular}   &  1,325 \\
   & Code-Review-Assist~\cite{sghaier2023multi}   & 2023 &   Multiple PLs & Precision, Recall, and F1 score  &342,742  \\ 
    & \highlight CodeReview-New~\cite{guo2024exploring} & \highlight 2024 &   \highlight Multiple PLs  & \highlight \begin{tabular}[c]{@{}l@{}}Exact Match Trim, \\ Exact Match, BLEU\end{tabular}  & \highlight 14,568  \\
    &  \zx{ManualReviewComment}~\cite{DBLP:journals/corr/abs-2502-02757} &  2025 &    Multiple PLs  &  \begin{tabular}[c]{@{}l@{}}Precision, Recall, F1\end{tabular}  & 270  \\
   \hline
\multirow{5}{*}{Clone Detection} & \highlight BigCloneBench~\cite{svajlenko2015evaluating} & \highlight 2014 & \highlight Java & \highlight Precision, Recall, F1   &\highlight 1,732,000  \\
    & POJ-104~\cite{mou2016convolutional}  & 2016 &  C, C++ & Precision, Recall, MAP  &52,000  \\ 
    & \highlight Company-C/C++~\cite{chochlov2022using}   & \highlight 2023 & \highlight C, C++ & \highlight MRR, Precision, Recall  &\highlight 180  \\ 
 & GPTCloneBench~\cite{alam2023gptclonebench} & 2023 &  Multiple PLs & Precision, Recall  &77,207 \\ 
& \highlight Curated CodeNet~\cite{dou2023towards} & \highlight 2023 &  \highlight C++, Python & \highlight Precision, Recall &\highlight 1,160,000 \\ 
   \hline
   Refactoring &  JavaRef~\cite{liu2023refbert} & 2023 &  Java & \begin{tabular}[c]{@{}l@{}}Accuracy, Exact Match,\\ Edit Distance,\\ Character Error Rate\end{tabular} &17,910 \\ 
        \bottomrule
    \end{tabular}}
    \label{tab: analysis}
\end{table}

\subsection{Benchmark Construction}

\subsubsection{Automatic Code Review Benchmarks}
The Review-Explaining~\cite{widyasari2023explaining} benchmark began by collecting 4,617 reviews with 9,169 revisions from Google open-source projects available on Gerrit, covering the period from October 5 to October 9, 2022.
They obtained the data from accessible listings of Google open-source projects, specifically selecting projects like Android, Bazel, Chromium, Dart, and Flutter. Google was chosen because it provides a comprehensive guide emphasizing the importance of detailed review explanations, which aligns with their benchmark goals.
To ensure the quality and relevance of the data, they filtered the reviews to include only useful comments using a decision tree based on criteria established by Bosu et al.~\cite{bosu2015characteristics}.  Additionally, they retained only inline code review comments, as these provide a clear connection to specific sections of code, enhancing the dataset's clarity and usefulness for analysis.

The CodeReview~\cite{tufano2022using} benchmark is mined from Java open-source projects hosted on GitHub and Gerrit. From GitHub, they selected projects with at least 50 pull requests, 10 contributors, and 10 stars, resulting in 4,901 projects. They also extracted data from 6 Gerrit installations used in prior research, totaling 6,388 projects.
They extracted triplets $<method, suggestion, revised\underline{\hphantom{v}}version>$, where the $method$ is a method submitted for review, the $suggestion$ is a reviewer's code change recommendation, and the $revised\underline{\hphantom{v}}version$ is the updated method incorporating the suggestion. They focused on accepted pull requests and excluded comments by code authors and inline comments not related to code lines. In total, they obtained 382,955 valid triplets for the dataset.
In addition, they performed several cleaning steps. They replaced any links with special tokens. They removed instances where the combined length of method+suggestion or revised version exceeded 512 tokens. They removed non-relevant comments, i.e., comments not recommending code change suggestions (e.g., ``looks good to me''). They excluded triplets with non-English suggestion comments. Finally, they removed all duplicates.

The CodeReviewer~\cite{li2022automating} benchmark gathers pull request data from high-quality, publicly available open-source repositories on GitHub across nine popular programming languages, such as C/C++, C\#, Go, and Ruby. It selects the top 10,000 projects for each language, excluding those that do not explicitly allow data redistribution. Additionally, projects with fewer than 1,500 pull requests are filtered out. Comments containing external URLs, which often link to other commits or pull requests, are also excluded. Additionally, comments not written in English, as well as those that are both short and positive, are removed.
\zx{Recently, Liu et al.\cite{DBLP:journals/corr/abs-2502-02757} manually labeled a random subset of 270 review comments from the CodeReviewer dataset\cite{li2022automating} to classify them as either valid or noisy. We refer to the resulting labeled data as the ManualReviewComment dataset~\cite{DBLP:journals/corr/abs-2502-02757}.}

The CodeReview-New~\cite{guo2024exploring} benchmark was developed to address the issues of low-quality code review data in the existing CodeReview dataset, as noted by the authors. To ensure the relevance of review comments to the code changes, the original code containing the review comments was collected. Data collection was restricted to after January 1, 2022, to prevent overlap with ChatGPT’s pre-training data, which only extends up to 2021, when the authors created their benchmark. Additionally, code reviews were gathered from 1,400 extra repositories (top 200 per language for seven languages not previously included in CodeReview), bringing the total to 2,029 repositories—829 from the existing CodeReview dataset and 1,200 from new sources. After applying filtering rules and time-based selection criteria, 467 repositories remained from the initial 2,029. The exclusion of the other 1,562 repositories was due to stricter filtering and the focus on pull requests created on or after January 1, 2022. The final dataset consists of 9,117 samples from 232 repositories already in the CodeReview dataset and 5,451 samples from 240 new repositories with different programming languages. Some languages, like SQL and Perl, have smaller datasets due to fewer pull requests or reviews.
The AUGER~\cite{li2022auger} benchmark is gathered from 11 prominent Java repositories on GitHub, each among the top 60 by star count. Each of these projects has at least 4,000 pull requests and 100 contributors.
The Code-Review-Assist~\cite{sghaier2023multi} benchmark collects pull requests from 19 popular projects via the GitHub REST API. These projects were selected based on previous research and each includes multiple pull requests with review comments related to specific parts of the source code. The dataset comprises 119,994 pull requests and 342,742 reviews. Each entry in the dataset is a triplet consisting of <original code, review, revised code>, where the review represents a comment made by the reviewer on the original code. The dataset is further augmented with issue types for use in subsequent analyses.

\subsubsection{Code Clone Benchmarks}

The Company-C/C++~\cite{chochlov2022using} benchmark is manually designed bugs provided by a partner company of a research study. They manually designed more representative datasets of clones, consisting of both C and C++ source code and known clones embedded in that code, spanning the four clone types. 
The BigCloneBench~\cite{svajlenko2015evaluating} benchmark is mined from the large inter-project repository IJaDataset 2.0 and includes manually checked clones of ten common functionalities. Initially, search heuristics were used to automatically identify code snippets in IJaDataset that might implement a target functionality. These candidate snippets were then manually tagged by judges as true or false positives of the target functionality.
The GPTCloneBench~\cite{alam2023gptclonebench} benchmark collection begins with extracting code functions from SemanticCloneBench. The authors then devised an automated script to detect potential functions that met the desired functionality. Once identified, a query was formulated using the function from the previous stage to extract a response from the GPT-3 language model. The output was processed with an automated script to generate input and corresponding output files based on the input log. To ensure high-quality clone pairs, textually similar pairs (input vs. output) with a similarity greater than 75\% were eliminated. Following this, manual validation was conducted to further refine the results by removing any additional clone pairs that did not meet the established criteria.
The POJ-104~\cite{mou2016convolutional} benchmark is sourced from a pedagogical programming open judge (OJ) system. In this system, students submit their source code as solutions to given problems, and the system judges the code's validity by executing it. The POJ-104 Benchmark includes the source code and the corresponding programming problems, represented as IDs, downloaded from the OJ system.
The Curated-CodeNet~\cite{dou2023towards} benchmark is sourced from the CodeNet dataset. Since the clone types were not pre-classified in CodeNet, the authors performed a classification following the standards set by BigCloneBench. This involved using lexical analyzers specific to Python and C++ for code tokenization, followed by calculating Jaccard indices to measure syntactical similarity scores. Based on these scores, they categorized the code clones for each language.

\subsubsection{Refactoring Benchmark}				
JavaRef~\cite{liu2023refbert} leverages commit data and open-source Java projects provided by Tsantalis et al. and Claes et al. From these sources, 114 popular Java GitHub projects with a total of 8,451 commits were selected. The authors initially extracted refactoring history, including commit hashes and refactoring types, but this did not include information on changed source code files or code diffs. To address this, they used PyDriller to obtain URLs for changed source code files and line numbers and then crawled the actual code changes from GitHub.
Additionally, five Computer Science students manually verified the collected data over three months to accurately link code diffs to refactoring types. Verified function-level refactoring data was retained, while other refactorings, duplicate code, and functions shorter than three lines were excluded.


\subsection{Discussions}

\nff{\textbf{BigCloneBench}~\cite{svajlenko2015evaluating} is a large-scale and high-quality benchmark for code clone detection, containing over eight million validated clones from 25,000 open-source Java systems. It covers the four major clone types across both intra- and inter-project scenarios, enabling systematic evaluation of tool performance under varying syntactic similarity levels. With its real-world data, comprehensive evaluation dimensions, and rigorous validation, BigCloneBench has become one of the most representative and authoritative benchmarks for assessing clone detection techniques. However, it exhibits bias due to its inclusion of clones for only ten common functionalities (e.g., Web Download and Copy a File), which limits its diversity. Moreover, it has been found that over 50\% of its samples were present in the pre-training data of an open-source LLM~\cite{zhou2025lessleak}, further limiting its suitability for evaluating LLM-based methods. For code review, the \textbf{CodeReview}~\cite{tufano2022using} benchmark is widely used because the data is mined from real-world code review data (pull requests, code diffs and comments) from GitHub and covers multiple programming languages.
But it exhibits bias: it excludes instances where the combined length of a method and its suggestion or revised version exceeds 512 tokens, resulting in a dataset that does not fully reflect the distribution of methods in real-world software.}

\nff{Regarding usage trends, benchmarks like BigCloneBench~\cite{svajlenko2015evaluating} and CodeReview~\cite{tufano2022using} remain dominant, primarily due to their early adoption, user-friendliness, and wide community support. Researchers frequently prefer these datasets as they are backed by extensive prior studies, enabling consistent comparisons across new approaches.}

\subsection{Limitations of Existing Benchmarks}

\noindent
\textbf{Benchmark Data Contamination.} A significant issue is the risk of benchmark data contamination, where overlaps between training and test data, or between pre-training and test data, can lead to inflated performance metrics. Due to the lack of retraining capabilities and inaccessibility to the pre-training data of many LLMs, it is challenging to assess the extent of benchmark data contamination. This limitation applies to all benchmarks in code analysis. 
\nff{Evidence of data contamination in maintenance benchmarks was reported in a recent comprehensive study on data leakage~\cite{zhou2025lessleak}. For example, the authors found that 55.7\% of BigCloneBench samples appear in the pre-training data of the fully open-source LLM StarCoder, highlighting a severe contamination issue. To mitigate this risk, researchers can collect newer datasets to prevent overlap.}

\noindent
\textbf{\nff{Bias in Benchmarks.}}
\nff{The BigCloneBench benchmark suffers from bias due to its inclusion of clones for only ten common functionalities (e.g., Web Download and Copy a File), resulting in limited diversity given the numerous functionalities present in the software engineering domain. The GPTCloneBench benchmark also exhibits bias in its collection process, as it relies on LLM synthesis. Consequently, the syntactic clones generated may follow a different distribution than those written by developers.
POJ-104 is biased in terms of both functionality coverage and code quality, as it is based on programming problems that are relatively simpler than real-world software systems, and the code written by students may be less mature than that of experienced developers. These biases can limit the generalizability of findings from these benchmarks, potentially resulting in overly optimistic performance estimates on real-world software.
}

\noindent\textbf{Reliance on Exact Match and Similarity Metrics.} 
Another limitation is the absence of comprehensive measures to evaluate the accuracy of predictions. For example, code review benchmarks often rely on Exact Match or similarity metrics like BLEU scores. However, correct answers can vary, and these metrics only assess how closely predictions align with a single ground truth, failing to capture the full range of possible correct solutions.



%% file: R_quality.tex
\section{Quality Management}
Quality management primarily aims to ensure software quality. In our study, we categorize it into two key areas: bug management~\ref{sec:bug} and vulnerability management~\ref{sec:vulnerability}.
\input{R_Bug}

\input{R_Vulnerability}

%% file: R_Bug.tex
\subsection{Bug Management} \label{sec:bug}

Effective bug management is a cornerstone of the software development process, involving several interconnected tasks: defect prediction, bug detection/localization, and bug repair (also known as program repair). Defect prediction is particularly valuable as it anticipates areas of the code that are more likely to contain defects, enabling developers to prioritize their efforts on these high-risk zones. Within this context, just-in-time (JIT) defect prediction is especially crucial, as it identifies defective code changes immediately after they are committed, allowing developers to address potential issues before they propagate further.
Following defect prediction, bug detection and localization are essential for identifying and precisely locating the sources of anomalies within the code, enabling targeted and efficient corrective actions. The final phase, bug repair, focuses on rectifying these identified issues, ensuring the software is restored to a stable and functional state. The seamless integration of these processes not only improves the overall quality of software products but also streamlines the maintenance workflow, significantly reducing both time and resource expenditures.

\subsubsection{Benchmark Scenarios.}
We identified 36 benchmarks for evaluating the performance of LLMs in bug management, spanning three primary tasks: defect prediction (5), bug (fault) localization (20), and program repair (12). These benchmarks vary significantly in terms of data composition, programming language coverage, and task formulation, offering a diverse landscape for model evaluation. Table~\ref{tab:bug} summarizes the key characteristics of these benchmarks.

Five benchmarks~\cite{zeng2021deep, ni2022best, saha2018bugs, zeng2021deep, madeiral2019bears} are widely used for evaluating the ability of models to predict defect-prone code. These datasets focus on identifying buggy commits based on code changes, commit messages, and historical metadata.
Bug localization tasks are supported by 20 benchmarks, including: Defects4J~\cite{just2014defects4j} includes 835 Java bugs with ground truth locations. BugsInPy~\cite{DBLP:conf/sigsoft/WidyasariSLQPTT20} offers 493 real-world Python bugs. Devign~\cite{zhou2019devign} contains 25,872 C-language commits labeled with vulnerability presence and locations. Ciborowska et al.~\cite{ciborowska2022fast} provides a large-scale dataset of 422,000 Java code hunks for ranking-based localization. Ma et al.~\cite{ma2023capturing} contributes 29,208 Java files annotated with suspicious lines.
The repair task includes 12 benchmarks: Defects4J~\cite{just2014defects4j}, QuixBugs~\cite{QuixBugs}, LMDefects~\cite{fan2023automated}, ARHE~\cite{kang2023explainable}, API-Misuse-Repair~\cite{API-Misuse-Repair}, DebugBench~\cite{tian2024debugbench}, and SWE-Bench~\cite{jimenez2024swe} offer Java and Python bugs along with their corresponding fixes. InferredBugs~\cite{InferredBugs} focuses on large-scale repair tasks in Java and C\#. Leetcode-debug~\cite{sakib2023extending} provide repair scenarios in educational settings. DebugBench~\cite{tian2024debugbench} is a recent large-scale benchmark spanning C++, Java, and Python, with 4,253 bugs. SWE-bench Multimodel~\cite{yang2024swe} provides a benchmark for JavaScript, while SWE-Lancer~\cite{miserendino2025swe} includes benchmarks for both JavaScript and TypeScript. Multi-SWE-bench~\cite{zan2025multi} covers a broader range of programming languages, including Python, Java, TypeScript, JavaScript, Go, Rust, C, and C++.
These benchmarks vary in format—some include ground-truth patches, while others rely on automated test suites for validation. They also differ in bug types (e.g., API misuse, logic error) and repair granularity (statement-level vs. function-level).

Overall, these benchmarks collectively cover a wide range of programming languages including Java, C/C++, Python, Golang, and C\#, supporting a broad evaluation of LLM capabilities in real-world bug management tasks across multiple modalities and scales.

\subsubsection{Evaluation Metrics.}
The benchmarks used for evaluating LLMs in bug management adopt a variety of evaluation metrics, reflecting the diverse nature of the tasks involved—namely, defect prediction, bug localization, and program repair.

\phead{\ding{42} Defect prediction} benchmarks typically adopt classification and ranking metrics to measure the model’s ability to identify defect-prone code changes. Common metrics include:
\begin{itemize}
    \item Accuracy, Recall, F1-score – To assess the correctness and completeness of predictions.
    \item AUC-ROC / AUC-PR – To evaluate the model's ability to distinguish defective from clean code across thresholds.
    \item Effort-aware metrics – Such as Recall@20\%Effort, Effort@20\%Recall, and P$_{opt}$, which consider the cost of code inspection relative to predictive gain.
    \item Top-N Accuracy – The proportion of true defects captured within the top-N predicted items.
\end{itemize}

\phead{\ding{42} Bug localization} tasks are typically formulated as ranking problems, and their metrics reflect how well the model prioritizes faulty code locations:
\begin{itemize}
    \item Precision@K / Recall@K / F1-score@K – Evaluate the model’s ability to rank buggy code snippets among the top-K candidates.
    \item Mean Reciprocal Rank (MRR) – Measures the rank position of the first correct prediction.
    \item Mean Average Precision (MAP) – Averages precision over all relevant positions.
    \item Top@N / ACC@K – Measures whether the buggy location appears within the top-N or top-K predictions.
    \item False Positive Rate (FPR) – Used to quantify erroneous predictions.
\end{itemize}

\phead{\ding{42} Program repair} benchmarks evaluate whether the generated patches are correct and executable, using both generation-level and test-based metrics:
\begin{itemize}
    \item Exact Match – Checks whether the generated patch exactly matches the ground truth.
    \item BLEU / CodeBLEU – Measures token-level or syntax-aware similarity between generated and reference code.
    \item Pass Rate – The percentage of generated patches that pass all test cases.
    \item \# Fixed Bugs / Fix Rate – The number or ratio of bugs that are successfully fixed.
    \item Acceptance Rate – The proportion of patches accepted in downstream systems or by human reviewers.
    \item Accuracy – In some benchmarks, used to indicate correctness in classification-style repair tasks.
\end{itemize}

These metrics collectively enable both qualitative and quantitative assessment of LLMs across the full bug management lifecycle—from predicting defects, to locating faults, to repairing code.

\subsubsection{Benchmark Construction.}
Five benchmarks have been identified for defect prediction: 
JIT-Defects4J~\cite{ni2022best} is a large-scale manually labeled dataset designed for just-in-time defect prediction. This dataset is derived from LLTC4J~\cite{herbold2022fine} and encompasses 21 Java projects. The dataset includes a varying number of commits per project, ranging from 544 to 4,026, with a bug ratio between 1.76\% and 18.75\%. Additionally, the dataset features a varying number of lines of code per project, ranging from 472 to 20,006, with a corresponding bug ratio between 3.77\% and 18.43\%.
Overall, JIT-Defects4J contains a total of 2,332 buggy commits and 24,987 clean commits, along with 12,594 buggy lines and 119,368 clean lines. It was subsequently refined to exclude commits involving purely refactoring-related changes~\cite{niu2025refactoring}.
Zeng et al.~\cite{zeng2021deep} curated a dataset to evaluate their proposed method, LApredict. The final dataset consists of six projects with a total of 310,370 real-world commits, including 81,300 defects. The projects include QT and OpenStack (primarily written in C++), Eclipse JDT, Eclipse Platform, and Gerrit (mainly programmed in Java), along with the Go project, which represents a modern programming language. 

We identified 20 benchmarks for bug localization/fault localization:
Ciborowska et al.~\cite{ciborowska2022fast} refined a dataset comprising bugs and their inducing changesets, which was previously validated by Wen et al.~\cite{wen2016locus}, to enhance bug localization efforts. The dataset includes six software projects, i.e., AspectJ, JDT, PDE, SWT, Tomcat, and ZXing. The changes are divided into different granularities, supporting commit-level, file-level, and hunk-level bug localization.
In total, this dataset contains 657 Bugs, 47,301 commits, 174,330 files, and 422,581 Hunks.
Ma et al.~\cite{ma2023capturing} created a dataset from four widely-used real-world open-source software projects for evaluating file-level bug localization methods. These projects include the Eclipse Platform, the Plug-in Development Environment (PDE), the Java Development Tools (JDT), and AspectJ. In total, there are 29,208 files, 13,056 bug reports. 
Defects4J~\cite{just2014defects4j} is a database of real-world reproducible software bugs. The latest version (v2.0.1) contains 835 bugs from 17 open-source Java projects. For each of these bugs, the dataset provides both the buggy and the fixed versions of the code.
Devign is a C benchmark~\cite{zhou2019devign} that primarily focuses on security vulnerabilities within software. Originally, the dataset included 15,512 security vulnerabilities from four major open-source projects—Linux, FFmpeg, Qemu, and Wireshark.
BugsInPy~\cite{DBLP:conf/sigsoft/WidyasariSLQPTT20} is a Python benchmark with 493 bugs from 17 different projects. For each bug, BugsInPy provides detailed artifacts and metadata, including revisions in the version control system with a unique bug ID mapped to the Git revision hash of its original GitHub repository. Additionally, the benchmark includes the original patches that fix the bugs, derived from the differences in source code files (excluding test files) between the faulty and fixed versions. BugsInPy also furnishes a list of test cases that effectively expose each bug, enhancing the utility and accessibility of this benchmark for researchers and developers aiming to improve bug detection and repair techniques in Python projects.
\zx{LINUXFLBENCH~\cite{zhou2025benchmarkingenhancingllmagents} is a recent bug localization benchmark for Linux Kernel. Its construction consists of three main phases. First, the authors collected 2,138 bug reports for the Linux kernel from Kernel.org Bugzilla, focusing on reports marked as “CLOSED” and “CODE\_FIX”. Next, they identified buggy locations by analyzing developer-submitted patches. They retained only those reports involving a single buggy file to avoid ambiguity, narrowing the set to 635 bug reports. Finally, they conducted a manual inspection to ensure data quality. Three annotators reviewed each report to exclude those without actual bugs, retain those with clear descriptions or logs, and filter out any that directly revealed the fix or buggy location. This process resulted in a final dataset of 250 high-quality bug localization tasks.}

\begin{table}[]
    \centering
    \caption{Existing Benchmarks for Evaluating LLMs Related to Defect Prediction, Bug (Fault) Localization, and Repair}
    \label{tab:bug}
    \resizebox{\textwidth}{!}{%
    \begin{tabular}{clllll}
    \toprule
    Task     & Benchmarks     & Year & Language & Evaluation Metrics      & \# Cases     \\ \midrule
    \rowcolor[HTML]{EFEFEF}
    \cellcolor{white} \multirow{6}{*}{Defect Prediction}
     & \highlight \junda{Bugs.jar}~\cite{saha2018bugs}& 2018 & \highlight Java  & \highlight Precision, Recall, F1, Accuracy, MCC&  \highlight21k functions\\
   & \junda{Bears}~\cite{madeiral2019bears} & 2019 & Java  & Precision, Recall, F1, Accuracy, MCC& 251 reproducible bugs  \\
    & \highlight Zeng et al.~\cite{zeng2021deep}   & \highlight 2021 & \highlight C++, Java, Golang & \highlight \begin{tabular}[l]{@{}l@{}}Accuracy, Recall, False Discovery Rate,\\ AUC-ROC, AUC-PR\end{tabular}      & \highlight 310, 370 changes\\
    &  JIT-defects4j~\cite{ni2022best}  &  2022 &  Java  & \begin{tabular}[l]{@{}l@{}}F1-score, AUC, Recall@20\%Effort,\\ Effort@20\%Recall, P$_{opt}$, Top-N Accuracy\end{tabular} &  27,319 commits  \\ 
    & \highlight \junda{Opu et al.}~\cite{opu2025llm}   & \highlight 2025 & \highlight Java    & \highlight Precision, Recall, F1, Accuracy, MCC & \highlight 1.7k function change   \\ \midrule
    \multirow{22}{*}{Bug Localization}
    & \junda{Ye et al.}~\cite{ye2014learning}   & 2014 & Java & Accuracy, MRR, MAP& 16,314 bugs      \\
    & \highlight Defects4J~\cite{just2014defects4j} & \highlight 2014 & \highlight Java  & \highlight ACC@K, FPR, Top@N     & \highlight 835 bugs \\ 
    & \junda{Bench4BL}~\cite{lee2018bench4bl}    & 2018 & Java & MRR, MAP, HIT@K   & 4,683 bug reports\\
    \rowcolor[HTML]{EFEFEF} \cellcolor{white}
    & Devign~\cite{zhou2019devign} & 2019 & C     & Top@N& 25,872 commits  \\
    & BugsInPy~\cite{DBLP:conf/sigsoft/WidyasariSLQPTT20} & 2020 & Python& ACC@K,Top@N   & 493 bugs     \\ 
    \rowcolor[HTML]{EFEFEF} \cellcolor{white} & \junda{Zhu et al.}~\cite{zhu2021syntax} & 2021 & Java & Accuracy& 1,083,185 commits\\
    & \junda{CodeReviewer}~\cite{li2022automating} & 2022 & 9 languages & Accuracy& 183,881 bug-fixing samples \\
    & \highlight Ciborowska et al.~\cite{ciborowska2022fast} & \highlight 2022 & \highlight Java  & \highlight \begin{tabular}[l]{@{}l@{}}Precision@K, Recall@K, \\ F1-score@K, MRR, MAP\end{tabular} & \highlight 422K hunks\\ 
    & Ma et al.~\cite{ma2023capturing}    & 2023 & Java  & MAP, MRR, Top@N  & 29,208 files \\ 
    & \highlight \junda{RTLLM}~\cite{lu2024rtllm}  & \highlight 2024 & \highlight Verilog     & \highlight Hit Rate, pass@k  & \highlight 102 cases \\
    & \junda{BeetleBox}~\cite{chakraborty2024blaze} & 2024 &\begin{tabular}[l]{@{}l@{}}C++, Java, Python, Go,\\ Python, and JavaScript\end{tabular}  & Accuracy, MRR, MAP& 26,321 bugs      \\
    \rowcolor[HTML]{EFEFEF} \cellcolor{white}& \junda{SWE-Bench}~\cite{jimenez2024swe}    & 2024 & Python      & Accuracy, MRR, MAP, TopN, Precision  & 2,294 issues     \\
    & \junda{Chandramohan et al.}~\cite{chandramohan2024supporting}& 2024 & Java, C/C++, Golang & Accuracy, MRR, MAP& 14,441 bugs      \\
    \rowcolor[HTML]{EFEFEF} \cellcolor{white}& \junda{Stracquadanio et al.}~\cite{stracquadanio2024veribug} & 2024 & Verilog     & top-1 bug coverage& 120 bugs \\
    & \junda{Manke et al.}~\cite{manke2024leveraging}     & 2024 & Python      & TP, FP  & 75 bugs \\
    \rowcolor[HTML]{EFEFEF} \cellcolor{white}& \junda{D58}~\cite{yan2024better} & 2024 & Java & Recall, MRR, CandiAvg      & 580 crash reports\\
    & \junda{Saha et al.}~\cite{saha2024toward} & 2024 & Java, Kotlin& MRR, MAP, HIT@K   & 87 bug reports   \\
    \rowcolor[HTML]{EFEFEF} \cellcolor{white} & \junda{Widyasari et al.}~\cite{widyasari2024demystifying}    & 2024 & Python & Top-K & 324 files \\
    & \zx{LINUXFLBENCH}~\cite{zhou2025benchmarkingenhancingllmagents}  & 2025 & C     & Recall@k,MRR  & 250 tasks   \\ 
    \rowcolor[HTML]{EFEFEF} \cellcolor{white}& \junda{ACPR}~\cite{dai2025less}   & 2025 & Python      & Accuracy& 52,168 samples   \\
    \midrule
    \multirow{12}{*}{Repair}
    & Defects4j~\cite{just2014defects4j} & 2014 & Java & \# fixed bugs  & 835   \\
    \rowcolor[HTML]{EFEFEF} \cellcolor{white} & QuixBugs~\cite{QuixBugs} & 2017 & Java, Python    & \# fixed bugs & 40  \\ 
    & LMDefects~\cite{fan2023automated} &  2023 &  Java & \# fixed bugs  &  113   \\ 
    \rowcolor[HTML]{EFEFEF} \cellcolor{white} & InferredBugs~\cite{InferredBugs} & 2023 & Java, C\#& Ratio of fixed bugs     & 11,595\\ 
      & ARHE~\cite{kang2023explainable} & 2023 & Python & accuracy      & 200   \\ 
    & \highlight Leetcode-debug~\cite{sakib2023extending} & \highlight 2023 & \highlight -     & \highlight acceptance rate  & \highlight 128 \\ 
    & API-Misuse-Repair~\cite{API-Misuse-Repair} & 2023 & Java  & Eaxct Match, BLEU, CodeBLEU      & 172,700      \\ 
    &\highlight DebugBench~\cite{tian2024debugbench} &\highlight 2024 &\highlight C++, Java, Python &\highlight Pass Rate     & \highlight 4,253 \\ 
    &\hx{SWE-Bench}~\cite{jimenez2024swe}& 2024 & Python &Resolution rate &2,294 issues \\
    \rowcolor[HTML]{EFEFEF} \cellcolor{white} &\zx{SWE-bench Multimodal~\cite{yang2024swe}}& 2024 & JavaScript &Resolution rate & 617 tasks \\
    &\zx{SWE-Lancer}~\cite{miserendino2025swe}& 2025 & JavaScript, TypeScript &Resolution rate &1,488 tasks \\
    \rowcolor[HTML]{EFEFEF} \cellcolor{white} &\zx{Multi-SWE-bench}~\cite{zan2025multi}& 2025 & Multiple PLs    &Resolution rate & 1,632 tasks \\
    \bottomrule
    \end{tabular}%
    }
\end{table}

We have identified 12 benchmarks for evaluating LLMs related to bug repair: API-Misuse-Repair Benchmark~\cite{API-Misuse-Repair} consists of 172,700 API misuse samples, originally sourced from Li et al.~\cite{DBLP:conf/icst/LiJB0Z21}, with low-quality data removed. 
LMDefects~\cite{fan2023automated} contains 113 Java bugs collected from LeetCode.
InferredBugs~\cite{InferredBugs} contains 11,595 Java and C\# bugs collected from GitHub projects. GPT-3 is evaluated on this benchmark using the ratio of fixed bugs as the metric. Defects4j~\cite{just2014defects4j} contains 835 Java bugs collected from real-world open-source software. 
QuixBugs~\cite{QuixBugs} contains 40 Java and Python bugs based on problems from the Quixey Challenge.
In addition, there are three benchmarks that aid in debugging before bug repair. 
These benchmarks focus on analyzing buggy code, identifying the root cause of the bugs, and ultimately repairing them.
ARHE~\cite{kang2023explainable} contains 200 instances sourced from the HumanEval benchmark. This benchmark has been used to evaluate CodeGen, Codex, and ChatGPT, with accuracy as the metric.
Leetcode-debug~\cite{sakib2023extending} consists of 128 instances collected from Leetcode. This benchmark has been used to evaluate ChatGPT, with the acceptance rate as the evaluation metric.
DebugBench~\cite{tian2024debugbench} contains 4,253 instances originally from Leetcode. This benchmark has been used to evaluate GPT-3.5 and GPT-4, with Pass Rate as the evaluation metric.
\zx{SWE-bench~\cite{jimenez2024swe} is an issue-fixing benchmark that has gained significant attention recently. SWE-bench comprises 2,294 issues collected from 12 open-source Python libraries, where LLMs are required to generate patches based on issue descriptions and the buggy repositories.
SWE-bench Verified\footnote{\url{https://openai.com/index/introducing-swe-bench-verified/}} is a curated subset of 500 issues from SWE-Bench with human-validated ground truth, which offers more reliable evaluation.
There are several benchmark variants related to SWE-bench. For instance, the SWE-Lancer~\cite{miserendino2025swe} dataset consists of 1,488 real-world freelance software engineering tasks derived from the Expensify open-source repository. Multi-SWE-bench~\cite{zan2025multi} is a recently introduced large-scale multilingual benchmark for issue resolution, comprising 1,632 human-validated GitHub issues spanning seven widely used programming languages. SWE-bench Multimodal~\cite{yang2024swe} is a benchmark to evaluate systems on their ability to fix bugs in visual, user-facing JavaScript software, consisting of 617 task instances collected from 17 JavaScript libraries.}

\subsubsection{Discussions}

\nff{
\textbf{Defects4J}~\cite{just2014defects4j} has long been the dominant benchmark in bug management research. Its widespread adoption stems from several key reasons. 
(i) First, it provides a high-quality dataset at the time it was proposed. Each bug fix is carefully separated from other code changes, such as refactoring. This process ensures the ground truth is reliable and accurate for experiments.
(ii) Its early introduction and ease of use drove broad community adoption. This becomes part of the standardized evaluation process for bug management, allowing SE researchers to consistently experiment with their methods and replicate experiments across the community with minimal overhead.
(iii) Its versatility makes it suitable for diverse tasks, including bug localization and automated program repair. 
Together, these attributes have fostered a vast body of prior work, establishing Defects4J as the essential default benchmark for researchers aiming to compare novel techniques.}

\nff{While Defects4J remains foundational the popular benchmarks, the rise of LLMs has created an urgent need for new bug benchmarks. 
Two primary concerns have emerged with these older benchmarks. First, their long-standing popularity increases the risk of data contamination, as they were likely included in the training corpora of modern LLMs~\cite{kong2025demystifying}. High performance may therefore reflect memorization rather than genuine problem-solving ability. Second, as LLMs become more capable, older benchmarks are often no longer challenging enough to differentiate state-of-the-art models, as they tend to feature isolated, single-file bugs.}

\nff{In response to these limitations, the research community is increasingly utilizing \textbf{SWE-bench}~\cite{swe-bench-verified}. To counter the risk of data contamination, its primary strength is its foundation in recent, real-world GitHub issues, making it less likely to have been part of LLM training sets. To provide a greater challenge, SWE-bench presents complex faults whose fixes span multiple files and non-contiguous code blocks (hunks). These bugs exist within large, active codebases, requiring a deep architectural understanding that offers a more realistic and difficult evaluation for today's powerful LLMs.}

\nff{Consequently, SWE-bench is exceptionally difficult. Initial studies show that even powerful models like GPT-4 can solve only a small fraction of its tasks~\cite{yang2025swe, swe-bench-verified}. This high degree of difficulty is a key attraction, offering a long runway for impactful research and providing a reliable way to measure true progress in the field. As a result, the shift towards SWE-bench is a necessary evolution. It reflects the community's need for contamination-free, highly realistic problems that align with the powerful capabilities of modern LLMs. It provides a challenging, standardized environment to push the boundaries of automated fault localization and program repair.
}

\subsubsection{Limitations and Opportunities}

Despite the advancements in bug management techniques, encompassing defect prediction, bug localization, and bug repair, several limitations still constrain their effectiveness:

\noindent \textbf{Limited Bug Type Coverage.}
Current benchmarks often lack diversity in the types of bugs they cover, focusing predominantly on well-known and easily identifiable issues. This limited scope can result in models that are less effective at handling rarer or more complex bugs encountered in real-world software development.
\nff{For example, Defects4J~\cite{just2014defects4j} primarily contains Java functional logic errors detectable by unit tests, overlooking critical issues related to performance, concurrency, or configuration. Similarly, QuixBugs~\cite{QuixBugs} is restricted to simple, ``toy'' algorithmic bugs in small programs.}

\noindent \textbf{Data Imbalance.}
Many datasets exhibit an uneven distribution of bug types and project contexts, leading to models that perform well on common issues but struggle with less frequent or more nuanced problems. 
\nff{For example, benchmarks such as Bugs.jar~\cite{saha2018bugs} and Defects4J~\cite{just2014defects4j} exhibit highly imbalanced distributions of bug types and project contexts, resulting in models that perform well on frequent defect patterns but struggle with rare or domain-specific issues.} This imbalance can reduce the overall robustness and generalizability of bug management tools.

\noindent \textbf{Narrow Language and Framework Focus.}
A significant portion of existing research and benchmarks is centered around a few programming languages, particularly Java, with less attention given to other languages and frameworks. This focus limits the applicability of the developed models across diverse software ecosystems.

\noindent \textbf{Reliance on Simplified Scenarios.}
Bug management benchmarks often rely on isolated and simplified scenarios, such as those captured by specific test cases or within highly curated datasets. This approach can omit complex, real-world conditions where bugs are more challenging to identify and fix, thus reducing the practical relevance of the models.
\nff{QuixBugs~\cite{QuixBugs}, for example, uses tiny, isolated programs with a single synthetic bug, which is far from the intricate and interconnected nature of production codebases. Even datasets based on real bugs, such as Defects4J~\cite{just2014defects4j}, curate them around specific unit-test failures, which simplifies the debugging context and may not capture the full complexity of the original problem.}

\noindent \textbf{Potential Data Contamination.}
There is a risk of overlap between training and test data, or between pre-training and test data, especially in LLMs. This contamination can lead to inflated performance metrics that do not accurately reflect the model's effectiveness in real-world applications, complicating the evaluation process. 
\nff{For instance, in 32.67\% of the SWE-Bench~\cite{jimenez2024swe} cases, the solution is available directly within the issue report or comments, allowing LLMs to copy the answer rather than generating a genuine fix. A recent study~\cite{zhou2025lessleak} found that the QuixBugs benchmark had a 100\% data leakage rate with StarCoder's pertaining data. The study also found that BugsInPy~\cite{DBLP:conf/sigsoft/WidyasariSLQPTT20} had a leakage rate of 11\% with StarCoder.}

%% file: R_Vulnerability.tex
\subsection{Vulnerability Management} \label{sec:vulnerability}

Software vulnerabilities are defined as flaws or weaknesses within a software system~\cite{krsul1998software}. When exploited or triggered by attackers, these vulnerabilities can lead to significant security risks. Such risks include unauthorized access~\cite{marques2019vulnerability}, service disruption~\cite{bell1987service}, and unintended behaviors~\cite{pistoia2007survey}, compromising the system’s integrity, availability, and confidentiality~\cite{krsul1998software}.
Numerous studies indicate that many open-source code repositories contain high-risk vulnerabilities~\cite{li2021sysevr,lipp2022empirical}. Therefore, managing these vulnerabilities is crucial for enhancing software security.

\subsubsection{Benchmark Scenarios.}
In the context of LLM, vulnerability management is divided into two main activities. The first is vulnerability detection~\cite{wang2010taintscope, zhou2024large, zhang2024prompt}, which focuses on identifying vulnerabilities before they are exploited. The second is vulnerability repair~\cite{fu2022vulrepair, zhou2024large, fu2024vision}, which involves implementing fixes or patches to mitigate and resolve these vulnerabilities.

    

To evaluate LLM-based techniques for vulnerability management, researchers have introduced a diverse set of benchmarks spanning both vulnerability detection and vulnerability repair tasks. These benchmarks vary in terms of task formulation, data size, programming language coverage, and data construction methodology, providing comprehensive scenarios for assessing LLM performance across different stages of secure software development.

\noindent \textbf{Vulnerability Detection.}
As summarized in Table~\ref{tab:vul-detection}, vulnerability detection benchmarks cover a wide range of dataset scales and programming languages. Early benchmarks such as Juliet~\cite{black2018juliet}, VulDeePecker~\cite{li2018vuldeepecker}, and Draper~\cite{russell2018draper} offer synthetic and curated real-world samples in C/C++ and Java, often targeting CWE-classified vulnerabilities.
Later benchmarks like Devign~\cite{zhou2019devign}, BigVul~\cite{fan2020bigvul}, and CVEFixes~\cite{bhandari2021cvefixes} focus on real-world vulnerable functions and commits from open-source projects.
Recent large-scale benchmarks such as D2A~\cite{zheng2021d2a}, VulEval~\cite{wen2024vuleval}, FalconVulnDB~\cite{ferrag2023securefalcon}, and ReposVul~\cite{wang2024reposvul} aggregate millions of functions, commits, or files across multiple languages (C, C++, Java, Python), enabling cross-project and inter-procedural vulnerability detection.
Additionally, domain-specific benchmarks like SmartBugs~\cite{ferreira2020smartbugs}, Web3Bugs~\cite{zhang2023web3bugs}, and DefiHacks~\cite{defi_hacks_analysis} focus on Solidity smart contracts and DeFi security vulnerabilities.
Recent efforts such as TreeVul~\cite{pan2023fine}, MegaVul~\cite{ni2024megavul}, and InterPVD~\cite{li2024effectiveness} incorporate inter-procedural and commit-level information, while SecLLMHolmes~\cite{ullah2024llms} targets LLM-based reasoning over critical vulnerabilities. 
\zx{PrimeVul~\cite{ding2024vulnerability} is a vulnerability benchmark constructed using a strategy that labels a function as vulnerable only when it is the sole function modified in a vulnerability-fixing commit. This strategy helps to reduce the risk of mislabeling non-security-related functions as vulnerable.
SECVULEVAL~\cite{ahmed2025secvuleval} focuses on real-world C/C++ vulnerabilities at the statement level, comprising 25,440 function samples associated with 5,867 unique CVEs in C/C++ projects.}

\noindent \textbf{Vulnerability Repair.}
Vulnerability repair benchmarks (Table~\ref{tab:vul-repair}) evaluate the ability of models to generate or predict secure patches for known vulnerabilities. Many detection datasets are reused for repair tasks—for example, BigVul~\cite{fan2020bigvul}, VulDeeLocator~\cite{li2021vuldeelocator}, and CVEFixes~\cite{bhandari2021cvefixes} serve as the foundation for downstream repair benchmarks like Chen et al.~\cite{chen2022neural} and VulRep~\cite{wei2023vulrep}.
Java-specific benchmarks such as Vul4J~\cite{bui2022vul4j} and VJBench~\cite{wu2023vjbench} include manually validated patches with compilation and plausibility checks.
SVEN~\cite{he2023sven} and SecurityEval~\cite{siddiq2022securityeval} focus on prompt-based repair evaluation, often in the context of LLM code completions.
Pearce et al.~\cite{pearce2023examining} offer a cross-language benchmark combining synthetic, handcrafted, and real-world vulnerabilities with labeled patch quality.
\zx{Wang et al. introduce CVE-Bench~\cite{wang-etal-2025-cve}, a benchmark comprising 509 CVEs collected from 120 open-source repositories. In addition to providing the code, the benchmark offers a test environment that simulates the real-world vulnerability repair process.}

\noindent \textbf{Overall Scope.}
Across both detection and repair tasks, these benchmarks span a wide range of programming languages (C, C++, Java, Python, Solidity, PHP, Verilog), vulnerability types (e.g., CWE-787, CWE-89), and data sources (e.g., GitHub commits, CVE/NVD entries, synthesized samples).
They support evaluation in various forms—including classification, ranking, generation, and reasoning—making them essential assets for systematic assessment of LLM capabilities in vulnerability management.

\begin{table}[]
\centering
\caption{Existing Benchmarks for Evaluating LLMs Related to Vulnerability Detection}
\label{tab:vul-detection}
\resizebox{\textwidth}{!}{%
\begin{tabular}{ccccc}
\hline
\textbf{Benchmarks}                                                & \textbf{Year} & \textbf{Language}     & \textbf{Evaluation Metrics}                                       & \textbf{\# Cases}                                                                                                                                                   \\ \hline
\junda{Choi et al.}~\cite{jeong2017end}                                   & 2017          & C/C++                 & Accuracy, F1, AUC                                                 & 10,000 functions                                                                                                                                                    \\ \hline
\rowcolor[HTML]{EFEFEF} \junda{Lin et al.}~\cite{lin2017poster}           & 2017          & C/C++                 & Top-k Recall                                                      & 6486 functions                                                                                                                                                      \\ \hline
\junda{DGBBench}~\cite{bohme2017developers}                               & 2017          & C/C++                 & Precision, Recall, F1, Accuracy                                   & 27 applications                                                                                                                                                     \\ \hline
\rowcolor[HTML]{EFEFEF} Juliet~\cite{black2018juliet}             & 2018          & C/C++, Java           & Precision, Recall, MCC                                            & 106k                                                                                                                                                                \\ \hline
VulDeePecker~\cite{li2018vuldeepecker}                            & 2018          & C/C++                 & FN, FP, TN, TP, Precision, Recall, F1, AUC, MCC                   & 61.6k code gadgets                                                                                                                                                  \\ \hline
\rowcolor[HTML]{EFEFEF} Draper~\cite{russell2018draper}           & 2018          & C/C++                 & FN, FP, TN, TP, Precision, Recall, F1, AUC, MCC                   & 1.27k functions                                                                                                                                                     \\ \hline
Devign~\cite{zhou2019devign}                                      & 2019          & C/C++                 & Accuracy, Precision, Recall, F1, FPR, AUC, Precision@K,MCC        & 15.6k functions                                                                                                                                                     \\ \hline
\rowcolor[HTML]{EFEFEF} Ponta et al.~\cite{ponta2019manually}     & 2019          & Java                  & AUC, F1                                                           & 1.3k commits                                                                                                                                                        \\ \hline
BigVul~\cite{fan2020bigvul}                                       & 2020          & C/C++                 & Accuracy, Precision, Recall, F1, FPR, AUC, Precision@K, MCC       & 189k functions                                                                                                                                                      \\ \hline
\rowcolor[HTML]{EFEFEF} ReVeal~\cite{chakraborty2021reveal}       & 2020          & C/C++                 & Accuracy, Precision, Recall, F1, FPR, AUC, Precision@K            & 23k                                                                                                                                                                 \\ \hline
SmartBugs~\cite{ferreira2020smartbugs}                            & 2020          & Solidity              & Precision, Recall, F1, Top-N Accuracy, MAR, MFR                   & 48k contracts                                                                                                                                                       \\ \hline
\rowcolor[HTML]{EFEFEF} \junda{Great}~\cite{hellendoorn2019global}        & 2020          & Python                & Precision, Recall, Accuracy                                       & 322 functions                                                                                                                                                       \\ \hline
\junda{Magma}~\cite{hazimeh2020magma}                                     & 2020          & C/C++                 & ROC-AUC                                                           & 138 real-world CVEs from 9 projects                                                                                                                                 \\ \hline
\rowcolor[HTML]{EFEFEF} \junda{SolidiFI}~\cite{ghaleb2020effective}       & 2020          & Solidity              & FN, FP                                                            & 50 smart contracts                                                                                                                                                  \\ \hline
SySeVR~\cite{li2021sysevr}                                        & 2021          & C/C++                 & FPR, FNR, Precision, Recall, F1                                   & 15.6K                                                                                                                                                               \\ \hline
\rowcolor[HTML]{EFEFEF} D2A~\cite{zheng2021d2a}                   & 2021          & C/C++                 & Precision, Recall, MCC                                            & 1.29M functions                                                                                                                                                     \\ \hline
PatchDB~\cite{wang2021patchdb}                                    & 2021          & C/C++                 & Precision, Recall, F1                                             & 12k patches                                                                                                                                                         \\ \hline
\rowcolor[HTML]{EFEFEF} CVEFixes~\cite{bhandari2021cvefixes}      & 2021          & C/C++, Java           & Accuracy, Precision, Recall, F1, FPR                              & 98k functions                                                                                                                                                       \\ \hline
CrossVul~\cite{nikitopoulos2021crossvul}                          & 2021          & 40+                   & Accuracy, Precision, Recall, F1, FPR                              & 27k files                                                                                                                                                           \\ \hline
\rowcolor[HTML]{EFEFEF} VCmatch~\cite{wang2022vcmatch}            & 2022          & C/C++, Java, PHP      & AUC, F1                                                           & 1,750k commits                                                                                                                                                      \\ \hline
\junda{VUDENC}~\cite{wartschinski2022vudenc}                              & 2022          & Python                & Precision, Recall, F1, Accuracy                                   & 812 Repository                                                                                                                                                      \\ \hline
\rowcolor[HTML]{EFEFEF} SARD~\cite{black2017sard}                 & 2023          & Multiple-PLs          & Accuracy, Precision,Recall, F1                                    & 101k                                                                                                                                                                \\ \hline
DiverseVul~\cite{chen2023diversevul}                              & 2023          & C/C++                 & Accuracy, Precision, Recall, F1, FPR                              & 349k functions                                                                                                                                                      \\ \hline
\rowcolor[HTML]{EFEFEF} Web3Bugs~\cite{zhang2023web3bugs}         & 2023          & Solidity              & TP,TN, FP, FN                                                     & 72 Projects                                                                                                                                                         \\ \hline
DeFi Hacks~\cite{defi_hacks_analysis}                             & 2023          & Solidity              & TP,TN, FP, FN                                                     & 13 Projects                                                                                                                                                         \\ \hline
\rowcolor[HTML]{EFEFEF} VulBench~\cite{gao2023vulbench}           & 2023          & C/C++                 & Precision, Recall, F1                                             & 455 functions                                                                                                                                                       \\ \hline
OWASP~\cite{owasp}                                                & 2023          & Java                  & Accuracy                                                          & 21k                                                                                                                                                                 \\ \hline
\rowcolor[HTML]{EFEFEF} TreeVul~\cite{pan2023fine}                & 2023          & C/C++, Java, PHP      & F1, Macro-F1, MCC                                                 & 6,541 commits                                                                                                                                                       \\ \hline
FormAI~\cite{tihanyi2023formai}                                   & 2023          & C                     & Precision, Recall, F1, Accuracy                                   & 112k programs                                                                                                                                                       \\ \hline
\rowcolor[HTML]{EFEFEF} \junda{Hu et al.}~\cite{hu2023large}              & 2023          & Solidity              & Hit \#                                                            & 13 smart contracts                                                                                                                                                  \\ \hline
FalconVulnDB~\cite{ferrag2023securefalcon}                        & 2024          & C/C++                 & Precision, Recall, F1, Accuracy                                   & 1,750k samples                                                                                                                                                      \\ \hline
\rowcolor[HTML]{EFEFEF} FormAI-v2~\cite{tihanyi2024neutral}       & 2024          & C                     & Average Property Violations Per File/Line                         & 331k programs                                                                                                                                                       \\ \hline
MoreFixes~\cite{akhoundali2024morefixes}                          & 2024          & Multiple PLs          & Accuracy, Precision, Recall, F1                                   & 26k CVEs                                                                                                                                                            \\ \hline
\rowcolor[HTML]{EFEFEF} VulEval~\cite{wen2024vuleval}             & 2024          & C/C++                 & Precision, Recall, F1, MCC, Precision@k, Recall@k                 & 232k functions                                                                                                                                                      \\ \hline
InterPVD~\cite{li2024effectiveness}                               & 2024          & C/C++                 & FPR, FNR, Accuracy, Precision, F1                                 & 769 vulnerabilities                                                                                                                                                 \\ \hline
\rowcolor[HTML]{EFEFEF} ReposVul~\cite{wang2024reposvul}          & 2024          & C/C++, Java, Python   & Accuracy                                                          & 262k functions                                                                                                                                                      \\ \hline
MegaVul~\cite{ni2024megavul}                                      & 2024          & C/C++                 & Accuracy, Precision, Recall, F1                                   & 17k functions                                                                                                                                                       \\ \hline
\rowcolor[HTML]{EFEFEF} SecLLMHolmes~\cite{ullah2024llms}         & 2024          & C, Python             & Response Rate, Accuracy, Correct Reasoning Rate                   & 228 scenarios                                                                                                                                                       \\ \hline
\junda{VulDetectBench}~\cite{liu2024vuldetectbench}                       & 2024          & C/C++                 & F1, Accuracy                                                      & 1000 entries                                                                                                                                                        \\ \hline
\rowcolor[HTML]{EFEFEF} \junda{SC-LOC}~\cite{zhang2024empirical}          & 2024          & Solidity              & Precision, Recall, Accuracy, F1-Score                             & 1,369 entries                                                                                                                                                       \\ \hline
\junda{Ma et al.}~\cite{ma2024combining}                                  & 2024          & Solidity              & Precision, Recall, Accuracy, F1-Score                             & 1,734 positive and 1,810 negative functions                                                                                                                         \\ \hline
\rowcolor[HTML]{EFEFEF} \junda{FELLMVP}~\cite{luo2024fellmvp}             & 2024          & Solidity              & Precision, Recall, Accuracy, F1-Score                             & 15,637 real-world smart contracts                                                                                                                                   \\ \hline
\junda{Yıldırım et al.}~\cite{yildirim2024evaluating}                     & 2024          & Python                & Accuracy                                                          & 40 API code samples.                                                                                                                                                \\ \hline
\rowcolor[HTML]{EFEFEF} \junda{Vulcorpus}~\cite{kouliaridis2024assessing} & 2024          & Java                  & Accuracy, Improvement Suggestion                                  & 103 code samples                                                                                                                                                    \\ \hline
\junda{Fang et al.}~\cite{fang2024llm}                                    & 2024          & Go, Java, PHP         & Not vulnerability detection                                       & 15 real-world one-day vulnerabilities                                                                                                                               \\ \hline
\rowcolor[HTML]{EFEFEF} \junda{SLFHunter}~\cite{ye2024detecting}          & 2024          & C/C++                 & TP,TN, FP, FN, F1-score                                           & 100 Linux-based embedded firmware samples                                                                                                                           \\ \hline
\junda{Guo et al.}~\cite{guo2024outside}                                  & 2024          & C/C++                 & Precision, Recall, F1-Score                                       & 13,532 total code functions                                                                                                                                         \\ \hline
\rowcolor[HTML]{EFEFEF} \junda{VulnPatchPairs}~\cite{risse2024uncovering} & 2024          & C/C++                 & Precision, Recall, F1, Accuracy, FPR, FNR                         & 26,200 C functions                                                                                                                                                  \\ \hline
\junda{Real-Vul}~\cite{chakraborty2024revisiting}                         & 2024          & C/C++                 & Precision, Recall, F1, Accuracy, AUC                              & \begin{tabular}[c]{@{}c@{}}5,528 vulnerable and \\ 1,682,713 uncertain functions\end{tabular}                                                                                                                  \\ \hline
\rowcolor[HTML]{EFEFEF} \junda{PairVul}~\cite{du2024vul}                  & 2024          & C/C++                 & Accuracy, Pairwise Accuracy, F1-score, MCC                        & 14,082 functions                                                                                                                                                    \\ \hline
\junda{VulSmart}~\cite{alam2024detection}                                 & 2024          & Solidity              & Precision, Recall, F1, Accuracy                                   & 4,218 Solidity smart contracts                                                                                                                                      \\ \hline
\rowcolor[HTML]{EFEFEF} \junda{KernJC}~\cite{ruan2024kernjc}              & 2024          & C/C++                 & TP, TN, FP, FN, Precision, Recall, F1, Accuracy                   & 36 Linux kernel vulnerabilities                                                                                                                                     \\ \hline
\rowcolor[HTML]{EFEFEF} \junda{LLM4Vuln}~\cite{sun2024llm4vuln}           & 2025          & C/C++, Java, Solidity & TP,TN, FP, FN, F1-score                                           & 294 code pieces, 3,528 scenarios                                                                                                                                    \\ \hline
\junda{VULZOO}~\cite{ruan2024vulzoo}                                      & 2025          & C/C++                 & Precision, Recall, F1, Accuracy                                   & 17 projects                                                                                                                                                         \\ \hline
\junda{CWE-Bench-Java}~\cite{li2405llm}                                   & 2025          & Java                  & \#Detected, Avg. False Discovery Rate, Avg. F1, Precision, Recall & 120 projects                                                                                                                                                        \\ \hline
\rowcolor[HTML]{EFEFEF} \junda{CASTLE}~\cite{dubniczky2025castle}         & 2025          & C                     & CASTLE Score, Combination Score, Precision, Recall, Accuracy      & 250 C programs                                                                                                                                                      \\ \hline
\junda{SecVulEval}~\cite{ahmed2025secvuleval}                              & 2025          & C/C++                 & Precision, Recall, F1, Accuracy                                   & 25,440 function                                                                                                                                                     \\ \hline
\rowcolor[HTML]{EFEFEF} \junda{VADER}~\cite{liu2025vader}                 & 2025          & 15 languages          & human-evaluated scoring rubric                                    & 174 real-world vulnerability cases                                                                                                                                  \\ \hline
\junda{JITVUL}~\cite{yildiz2025benchmarking}                              & 2025          & C/C++                 & Precision, Recall, F1, Accuracy                                   & 1,758 functions                                                                                                                                                     \\ \hline
\rowcolor[HTML]{EFEFEF} \junda{Li et al.}~\cite{zhang2025benchmarking}    & 2025          & Python, Java, JS      & Precision, Recall, F1, Accuracy                                   & \begin{tabular}[c]{@{}c@{}}Python: 8k vuln, 90k+ non-vuln functions\\ Java: 7.5k vuln, 70k+ non-vuln functions\\ JS: 29k vuln, 25k+ non-vuln functions\end{tabular} \\ \hline
\junda{BinPool}~\cite{arasteh2025binpool}                                 & 2025          & C/C++                 & Precision, Recall, F1, Accuracy                                   & 8,750 binaries                                                                                                                                                      \\ \hline
\rowcolor[HTML]{EFEFEF} \junda{ICVul}~\cite{lu2025icvul}                  & 2025          & C/C++                 & Precision, Recall, F1, Accuracy                                   & 15,396 functions                                                                                                                                                    \\ \hline
\end{tabular}%
}
\end{table}

\begin{table}[t]
\caption{Existing Benchmarks for Evaluating LLMs Related to Vulnerability Repair}
\label{tab:vul-repair}
\resizebox{\columnwidth}{!}{%
\begin{tabular}{ccccc}
\hline
Benchmarks & Year & language & Evaluation Metrics & \# Cases\\ \hline
\rowcolor[HTML]{EFEFEF}
\junda{ManyBugs}~\cite{le2015manybugs} & 2015 & C/C++ & Repair Rate & 26 vulnerabilities \\
BigVul~\cite{fan2020bigvul} & 2020 & C/C++ & Perfection Prediction, BLEU, Meteor & 189k functions \\
\rowcolor[HTML]{EFEFEF}
Chen et al.~\cite{chen2022neural} & 2020 & C/C++ & Prediction Accuracy, BLEU, CodeBLEU, EM & 8,482 functions \\
VulDeeLocator~\cite{li2021vuldeelocator} & 2021 & C/C++ & BLEU, Rouge-L & 40,382 vulnerable snippets\\
\rowcolor[HTML]{EFEFEF}
\junda{ExtractFix}~\cite{gao2021beyond} & 2021 & C/C++ & Repair Rate & 30 vulnerabilities \\
CVE-Fixes~\cite{bhandari2021cvefixes} & 2021 & C/C++, Java & CodeBLEU & 1,754 projects \\
\rowcolor[HTML]{EFEFEF}
Vul4J~\cite{bui2022vul4j} & 2022 & Java & \begin{tabular}[c]{@{}c@{}}Compilation Rate\\ Correctly-fixed Rate\\ Plausibly-fixed Rate\end{tabular} & 79 vulnerabilities \\
SecurityEval~\cite{siddiq2022securityeval} & 2022 & Python & Security Rate & 130 samples \\
\rowcolor[HTML]{EFEFEF}
VulRep~\cite{wei2023vulrep} & 2023 & C/C++ & (Repair) Precision, Recall & 116 pairs of commits\\
VJBench~\cite{wu2023vjbench} & 2023 & Java & \begin{tabular}[c]{@{}c@{}}Compilation Rate\\ Correctly-fixed Rate\\ Plausibly-fixed Rate\end{tabular} & 42 vulnerabilities \\
\rowcolor[HTML]{EFEFEF}
Sven~\cite{he2023sven} & 2023 & C/C++, Python & Security Rate & 1,606 programs \\
Pearce et al.~\cite{pearce2023examining} & 2023 & C, Python, Verilog & Repair Rate & 58,500 patches \\
\rowcolor[HTML]{EFEFEF}
\junda{SmartFix}~\cite{so2023smartfix} & 2023 & Solidity & Repair Rate, Success Rate, Accuracy & 361 smart contracts \\
\junda{ManyVuls4J}~\cite{lin2024there} & 2024 & Java & Repair Rate & 103 vulnerabilities \\
\rowcolor[HTML]{EFEFEF}
\junda{ARVO}~\cite{mei2024arvo} & 2024 & C/C++ & Repair Rate & 1,001 vulnerabilities \\
\junda{Le et al.}~\cite{le2024study} & 2024 & JavaScript & Repair Rate & 20 vulnerabilities \\
\rowcolor[HTML]{EFEFEF}
\junda{Khan et al.}~\cite{khan2024code} & 2024 & C/C++ & Repair Rate & 56 code snippets \\
\junda{VADER}~\cite{liu2025vader} & 2025 & 15 languages & Human Evaluation & 174 cases \\
\hline
\end{tabular}%
}
\end{table}

\subsubsection{Evaluation Metrics.}

To evaluate the effectiveness of vulnerability detection and repair techniques, prior studies have adopted a wide variety of metrics depending on the nature of the task. These metrics can be broadly categorized into five groups: classification-oriented, ranking-based, generation-based, code validity, and security-specific metrics.

 \phead{\ding{42} Classification-Oriented Metrics} are commonly used in binary or multi-class vulnerability detection tasks to assess prediction quality. They include Accuracy, Precision, Recall, F1-score, False Positive Rate (FPR), False Negative Rate (FNR), and Matthews Correlation Coefficient (MCC). Additionally, many benchmarks report raw classification outcomes such as TP (True Positive), TN (True Negative), FP (False Positive), and FN (False Negative). In some multi-class scenarios, Macro-F1 is also employed. For LLM-based evaluations, metrics like Response Rate and Prediction Accuracy have been used to measure general response correctness.

 \phead{\ding{42} Ranking-Based Metrics.}
In vulnerability localization and prioritization tasks, the ability to rank potential vulnerabilities is crucial. Ranking-based metrics include AUC-ROC, AUC-PR, Mean Average Precision (MAP), Mean Reciprocal Rank (MRR), Precision@K, Recall@K, Top-N Accuracy, and Precision@k/Recall@k, depending on the granularity and format of evaluation.

 \phead{\ding{42} Generation-Based Metrics.}
For tasks involving code repair or patch generation, evaluation often requires comparing generated patches with reference patches. Common metrics in this category include BLEU, CodeBLEU, Rouge-L, Meteor, and Exact Match (EM). These metrics evaluate lexical or semantic similarity between generated and ground truth patches.

 \phead{\ding{42} Code Validity Metrics.}
To ensure that generated patches are syntactically and functionally valid, some benchmarks adopt metrics that assess code compilation or plausibility. These include Compilation Rate, Correctly-fixed Rate, and Plausibly-fixed Rate, particularly in Java-based benchmarks.

 \phead{\ding{42} Security-Specific Metrics.}
Several benchmarks focus on evaluating whether the generated fix effectively eliminates the underlying vulnerability. Metrics such as Security Rate, Repair Rate, Fix Rate, and Perfection Prediction are employed for this purpose. Additionally, in property-based benchmarks, the Average Property Violations Per File/Line is used to quantify the residual security risks after repair.

\subsubsection{Benchmark Construction}
We summarize the construction of the vulnerability management benchmarks in terms of their data sources and labeling methods.

\noindent \textbf{Source.} 
In general, vulnerability benchmarks can be classified into three categories based on the source of the data: \textbf{synthetic}, \textbf{real-world}, and \textbf{AI-generated} datasets. 
\textbf{Synthetic} benchmarks are constructed using predefined vulnerability patterns. They are often used to evaluate traditional vulnerability management tools, such as static and dynamic analysis systems. Notable examples include the Juliet dataset~\cite{okun2013juliet} and OWASP~\cite{owasp}. However, these datasets often fail to reflect the complexities of real-world vulnerabilities due to their simplified and isolated nature.
\textbf{Real-world} benchmarks aim to overcome these limitations by mining data from open-source repositories, security patches, and issue trackers. They provide a more accurate representation of real-world vulnerabilities, making them suitable for assessing practical performance. Prominent examples include Big-Vul~\cite{fan2020bigvul}, Draper~\cite{russell2018draper}, Devign~\cite{zhou2019devign}, REVEAL~\cite{chakraborty2021reveal}, and DiverseVul~\cite{chen2023diversevul}, which consist of vulnerable real-world functions from open-source software. However, these datasets may lack the dependencies required for compilation.
\textbf{AI-generated} benchmarks are collections of programs generated by LLMs, annotated with detailed vulnerability classifications. These datasets serve as a new benchmark category, offering unique advantages and addressing some limitations of synthetic and real-world datasets. Unlike synthetic datasets, which rely on predefined vulnerability patterns, AI-generated datasets capture a broader spectrum of vulnerabilities introduced by the generative nature of LLMs. Further, the programs in these datasets are often verified for compilability. For instance, the FormAI dataset~\cite{tihanyi2023formai}, includes a large collection of compilable C programs generated GPT-3.5. Each program is systematically labeled for vulnerabilities using formal verification techniques like Efficient SMT-based Bounded Model Checking (ESBMC). FormAI-v2 expands on this by integrating programs from multiple LLMs and enhancing diversity through dynamic prompting methods.

\noindent \textbf{Labelling. }
We classify the labeling methods for vulnerability benchmarks into four categories: \textbf{manual labeling}, \textbf{static analyzer}, \textbf{security patches}, and \textbf{formal verification}.

\textbf{Manual labeling} involves human experts reviewing code to identify vulnerabilities. For example, the Devign~\cite{zhou2019devign} dataset was labeled by three security researchers through an intensive process. 
Devign provides high-quality labels but comes with significant costs. The labeling process required approximately 600 man-hours, demonstrating the resource-intensive nature of manual labeling. Despite its accuracy, this method is less scalable for larger datasets.
\textbf{Static analyzers} generate alerts for potential vulnerabilities in code, which can be used as labels. For example, the Draper~\cite{russell2018draper} dataset, uses labels generated from alerts by three static analyzers: Clang, Cppcheck, and Flawfinder. While static analyzer labels are scalable and automated, they suffer from low accuracy due to false positives and false negatives inherent in static analysis tools. This reduces the reliability of the generated labels.
\textbf{Security patches} method identifies vulnerabilities by analyzing code changes associated with security patches or vulnerability-fixing commits. 
The ReVeal dataset~\cite{chakraborty2021reveal} uses security patches from Chromium and Debian. Datasets like BigVul~\cite{fan2020bigvul}, CrossVul~\cite{nikitopoulos2021crossvul}, and CVEfixes~\cite{bhandari2021cvefixes} utilized vulnerability-fixing commits from CVE records in the National Vulnerability Database (NVD).
\textbf{Formal verification} employs rigorous verification techniques to label programs by checking their adherence to safety and security properties. For example, the FormAI dataset~\cite{tihanyi2023formai} uses the Efficient SMT-based Bounded Model Checker (ESBMC) for labeling. ESBMC combines model checking, abstract interpretation, constraint programming, and satisfiability modulo theories (SMT) to analyze programs. It systematically identifies vulnerabilities by generating counterexamples for violated properties and minimizes false positives and false negatives during the verification process.

\subsubsection{Discussions}

For vulnerability detection, the most widely used benchmarks are \textbf{Devign}~\cite{zhou2019devign}, \textbf{BigVul}~\cite{fan2020bigvul}, and \textbf{ReVeal}~\cite{chakraborty2021reveal}, with one literature finding that 53 out of 81 papers in top venues used at least one of them~\cite{risse2025top}. The widespread adoption of benchmarks can be attributed to several key factors: 
(i) align with the dominant experimental design of the field. A recent survey~\cite{risse2025top} shows that 88\% of studies formulate vulnerability detection as a binary classification problem at the function-level. This simplification reduces a complex security analysis to a direct question of whether an isolated function contains the vulnerability. Datasets such as Devign~\cite{zhou2019devign} and BigVul~\cite{fan2020bigvul} are structured to perfectly match this formulation.
(ii) The focus on C/C++ is another key reason for their widespread use. C/C++ are the dominant languages in system programming and are notoriously susceptible to high-impact vulnerabilities, e.g., memory-related flaws. This composition makes them highly relevant to the SE community and ensures that research targets a well-recognized and critical problem area.
(iii) The substantial size of these datasets also drove their adoption. Because LLMs require vast amounts of data for effective training, the sheer scale of these benchmarks made them one of the few viable resources available, especially during the initial stages of this research area.

Recent trends in vulnerability detection research highlight a significant shift in how benchmarks are utilized, with a clear move towards more comprehensive, realistic, and diverse datasets.  Devign~\cite{zhou2019devign}, BigVul~\cite{fan2020bigvul}, and ReVeal~\cite{chakraborty2021reveal} remain popular benchmarks, while DiverseVul~\cite{chen2023diversevul} has also gained significant use in many recently published articles. A key trend is the move away from relying on a single benchmark for model evaluation. Researchers are increasingly using multiple datasets to provide a more thorough assessment of their vulnerability detection models. Furthermore, the field is expanding into new, high-impact domains. Given the financial stakes of blockchain applications, benchmarks for smart contract vulnerabilities, such as SmartBugs~\cite{ferreira2020smartbugs} and FELLMVP~\cite{luo2024fellmvp}, are gaining prominence.

For vulnerability repair, CVEFixes~\cite{bhandari2021cvefixes} and Big-Vul~\cite{fan2020bigvul} are two of the most widely adopted benchmarks.  Primarily because
(i) Both benchmarks are sourced from a multitude of open-source projects, providing a large volume of realistic data needed to train data-intensive LLMs.
(ii) Another reason for their popularity is that they effectively filled a critical vacuum. The field of automated vulnerability repair has long faced a shortage of high-quality, large-scale, real-world datasets. Big-Vul and CVEFixes were among the first to provide a large corpus of real-world vulnerable code paired with its corresponding fix. 
(iii) as we mentioned previously, the vulnerability issues of C and C++ are highly relevant and valuable to the software engineering research community.

\nff{For a long time, vulnerability repair benchmarks were focused on C and C++ programming languages~\cite{le2015manybugs, fan2020bigvul, li2018vuldeepecker}. A major recent trend is the push to overcome this monolingual constraint. Researchers recognize that modern software is multilingual and have begun creating and utilizing benchmarks that reflect this reality. For example VADER~\cite{liu2025vader} covers 15 languages, including Python, Java, JavaScript, and Go, providing a comprehensive evaluation ground for cross-language vulnerability detection and repair. This shift toward multilingual benchmarks enables more generalizable and robust model evaluation, bridging the gap between traditional single-language datasets and the diverse ecosystems of real-world software.}

\subsubsection{Limitations and Opportunities.}
Despite the advancements in vulnerability management benchmarks, several limitations still exist:

\noindent\textbf{Quality of Datasets.} A major challenge in vulnerability research is the lack of high-quality benchmarks. Common issues include low accuracy in labeling and high rates of duplication~\cite{fan2020bigvul}. Automated methods for collecting vulnerability data often fail to ensure accurate labeling.
\nff{For instance, critical benchmarks like BigVul and CrossVul have been found to have a correctness rate below 50\%. Furthermore, data duplication is rampant, with datasets like BigVul and CVEFixes containing over 10\% duplicate entries~\cite{ding2024vulnerability}. }
Manual verification can enhance quality but is impractical for large datasets due to its high cost and labor intensity.
Additionally, the potential for data contamination—where evaluation datasets might overlap with LLM training corpora—complicates the development of reliable benchmarks, emphasizing the need for datasets without such overlaps.

\noindent \textbf{Limitations in Reflecting Real-World Vulnerability Scenarios.}
The existing datasets still do not effectively represent real-world scenarios, primarily due to several key limitations. First, the narrow focus of these datasets often overlooks broader and more complex code structures. \nff{For instance, popular benchmarks like BigVul~\cite{fan2020bigvul} and Devign~\cite{zhou2019devign} are function-level benchmarks.}
Second, the real-world deployment of these models faces challenges from imbalanced test sets, where certain types of vulnerabilities are overrepresented while others are scarce, skewing model performance and reliability. \nff{For instance, Datasets like BigVul~\cite{fan2020bigvul} and PrimeVul~\cite{ding2024vulnerability} are dominated by memory-related issues, such as buffer errors (CWE-119, CWE-125, CWE-787).}
Lastly, vulnerabilities can exist at various structural levels within the software, from single lines to entire modules, and the current datasets’ limited scope in this regard can result in missed or inaccurately identified vulnerabilities. Together, these issues highlight a significant gap between dataset conditions and the actual demands of vulnerability management in real-world settings.

%% file: Challenges_Opportunities.tex
\section{Challenges and Opportunities}\label{sec:challengesandopportunities}
\subsection{Challenges}

\subsubsection{Ambiguities in Benchmark Design}
\textbf{Unclear Task Definitions}: Some benchmarks have task descriptions that are not sufficiently clear, leading to inconsistencies in evaluation standards. For example, in code generation tasks, different evaluators may have varying definitions of ``correct generation'', which can result in unreliable evaluation outcomes.
\textbf{Single Evaluation Metric}: Many benchmarks focus on only one performance metric (e.g., accuracy), while overlooking other important factors (e.g., code maintainability, execution efficiency), making it difficult to fully reflect the model's practical utility.

\subsubsection{Scalability Issues}
\textbf{Scalability of Benchmarks}: As model complexity increases and task diversity expands, existing benchmarks may struggle to keep up with development needs. For example, when tasks involve multiple programming languages or complex codebases, current benchmarks may be too simple or narrow.
\textbf{Computational Costs of Large-Scale Data Processing}: Running and evaluating large benchmarks can be computationally expensive, especially for tasks that require processing large amounts of code samples or long training periods.

\subsubsection{\nff{Data Contamination Issue.}}
\nff{\textbf{Overlap Between Training and Evaluation Data}: A critical challenge in benchmark development is the risk of overlap between benchmark samples and the pretraining corpora of LLMs. Since many benchmarks are derived from widely available sources such as open-source repositories, forums, or academic datasets, they are likely to be included in large-scale training data, which compromises the validity of evaluation results.
\textbf{Erosion of Benchmark Reliability}: As LLMs are iteratively trained on increasingly larger and more diverse corpora, older benchmarks quickly lose their effectiveness. Once benchmark items have been exposed during training, the resulting contamination inflates performance scores and prevents a fair assessment of genuine generalization ability.
\textbf{Difficulty in Sustaining Long-Term Benchmark Utility}: The continuous evolution of training datasets means that maintaining contamination-free benchmarks requires ongoing effort. Static benchmarks eventually become saturated and unusable, demanding the design of new, dynamically updated evaluation frameworks to ensure fair and trustworthy assessment.}

\subsubsection{Lack of Generalization}

\textbf{Overfitting}: Models may overfit to specific tasks within a benchmark, leading to poor generalization in real-world scenarios. Benchmarks do not always fully reflect the complexity of the real world, causing models to struggle when faced with new problems.
\textbf{Difficulty in Cross-Domain Transfer}: Benchmarks often cover only a single domain or task, but software engineering tasks can involve knowledge and skills from multiple domains. Performance on specific benchmarks may not generalize well to other related tasks or domains.

\subsubsection{Lag in Updates and Evolution}
\textbf{Delayed Benchmark Updates}: The field of software engineering evolves rapidly, with new technologies, languages, and frameworks constantly emerging. However, benchmarks often lag behind these developments, rendering some benchmarks less effective at reflecting current technological needs.
\textbf{Maintenance Costs}: As datasets grow in size and complexity, maintaining and updating benchmarks becomes increasingly challenging, requiring significant resources to ensure their continued relevance and effectiveness.

\subsubsection{Data Privacy Concerns} Some benchmarks use data that may contain sensitive information, particularly when involving real codebases or internal company projects. This raises significant concerns about data privacy and ethics.

\subsection{Opportunities}

\subsubsection{Development of More Comprehensive and Diverse Benchmarks}

\textbf{Broader Task Coverage}: Future benchmarks can cover a wider variety of software engineering tasks, including complex multi-step tasks like requirements, debugging, and software design. This will help ensure that models are evaluated on a more representative set of real-world challenges.
\textbf{Cross-Language Benchmarks}: Expanding benchmarks to support multiple programming languages and paradigms will allow for better evaluation of models’ versatility and ability to generalize across different software ecosystems.
\textbf{Inclusion of Edge Cases}: Incorporating edge cases and more challenging scenarios in benchmarks can help in assessing the robustness of LLMs, pushing models to handle rare or difficult situations better.

\subsubsection{\nff{Mitigating Data Contamination Risks.}}
\nff{\textbf{Contamination Detection and Auditing}: A central challenge in benchmark development is preventing overlap between benchmark data and the pretraining corpus of LLMs. Recent studies have introduced systematic methods for detecting such overlaps~\cite{samuel2024towards, singh2024evaluation}. Integrating these techniques into benchmark pipelines can enhance the reliability of contamination audits and enable more effective removal of overlapping samples.
\textbf{Transparent Benchmark Construction}: Benchmark creators should clearly document the data sources, collection timeframes, and filtering procedures used to minimize contamination risks. Publicly releasing contamination audit reports can enhance transparency and trustworthiness of the evaluation.
\textbf{Robust Benchmark Design}: To further reduce contamination, benchmarks can be constructed from private or newly curated datasets, or generated synthetically through expert-designed tasks and adversarial methods. Dynamic and evolving benchmarks, which continuously introduce novel tasks and datasets, can also help ensure fair and forward-looking evaluations.
}

\subsubsection{Multi-Metric Evaluation}

\textbf{Holistic Performance Metrics}: Future benchmarks can go beyond accuracy and include metrics for code quality, maintainability, performance, and security. These additional metrics can help in evaluating not just whether the model produces correct outputs, but also how practical and efficient those outputs are in real-world use.

\subsubsection{Continuous and Dynamic Benchmarks}
Future benchmarks should evolve over time to keep pace with advancements in software engineering practices and emerging technologies. This could involve regularly adding new datasets, tasks, and evaluation criteria.

\subsubsection{Real-World Integration}
\textbf{Benchmarks with Real-World Data}: Incorporating real-world codebases and real-time software engineering tasks in benchmarks will lead to more practical evaluations. This will allow LLMs to be tested in environments that closely resemble the actual contexts in which they will be deployed.
\textbf{Industry Collaboration}: Engaging with industry stakeholders to develop benchmarks based on actual development processes and challenges can result in more relevant evaluations. This collaboration can help align academic research with industry needs.

\subsubsection{Scalability and Efficiency}
\textbf{Scalable Benchmarks}: As models and datasets grow, future benchmarks must be designed to scale efficiently. This involves creating benchmarks that can handle large-scale data processing and provide meaningful results without excessive computational overhead.
\textbf{Efficient Evaluation Techniques}: Developing new techniques to evaluate models more efficiently, perhaps by sampling or using approximations, can reduce the time and cost of benchmarking large LLMs while still providing accurate assessments.

\subsubsection{Incorporation of Human Feedback}
\textbf{Human-in-the-Loop Benchmarks}: Incorporating human feedback into benchmark evaluations can provide more nuanced assessments of LLM performance. Human evaluators can assess whether the code generated is not just syntactically correct but also aligns with good coding practices and user requirements.
\textbf{Crowdsourced Benchmarking}: Engaging the developer community in benchmark creation and evaluation can lead to more diverse and robust benchmarks, as well as continuous improvement through community contributions.

\subsubsection{Ethical and Responsible AI Considerations}
Future benchmarks should evaluate how LLMs handle ethical issues such as producing harmful code, detecting malicious intent, and respecting user privacy. By including ethics-focused benchmarks, researchers can ensure that models are not only functional but also responsible.

\subsubsection{Explainability and Interpretability}
\textbf{Benchmarks for Explainability}: Developing benchmarks that evaluate a model's ability to explain its outputs can promote transparency. This is especially important in software engineering, where understanding the rationale behind a generated code segment or recommendation is crucial. While two existing studies have explored explainability in fault localization~\cite{widyasari2024demystifying} and code review~\cite{widyasari2023explaining}, a broader set of benchmarks is still needed to capture diverse real-world use cases and foster trustworthy model behavior.
\textbf{Debugging and Error Explanation}: Creating benchmarks that assess a model’s ability to diagnose and explain errors in code would enhance their utility in real-world debugging and development tasks.

%% file: threats.tex
\section{Threats to Validity}\label{sec:threats}

In this section, we conclude the threats to validity from four parts: construction, conclusion, internal, and external threats.

\subsection{Construct Validity}
Construct validity concerns whether we are accurately capturing what we intend to study—in this case, benchmark datasets designed for evaluating LLMs in SE tasks.

We followed Kitchenham et al.'s guidelines~\cite{kitchenham2004procedures} for conducting systematic surveys. To ensure comprehensive coverage, we searched benchmarks based on prior studies~\cite{DBLP:journals/tosem/HouZLYWLLLGW24, fan2023large, zhang2023survey} and searched major SE and AI repositories, including IEEE Xplore, ACM Digital Library, arXiv, and Google Scholar. We also employed backward and forward snowballing to supplement initial results.

A threat arises from the possibility that our search strategy missed relevant benchmarks due to terminology or publication venue variations. Moreover, the application of inclusion and exclusion criteria introduces potential subjectivity. To reduce this risk, two authors independently assessed all candidate benchmarks, and any disagreements were resolved through structured discussion.

\subsection{Internal Validity}
Internal validity addresses whether the results of our analysis could be affected by errors or biases during the execution of the study.
Although benchmark selection was guided by well-defined criteria, the manual classification of benchmark properties (e.g., task type, language, modality) may involve subjective judgment. To minimize bias, all metadata extraction and coding were performed independently by at least two authors, followed by reconciliation through discussion.
Additionally, since benchmarks can evolve (e.g., updated versions, expanded tasks), there is a risk that some information may be outdated or inconsistent at the time of analysis. We attempted to mitigate this by using the latest available versions and consulting official benchmark documentation when available.

\subsection{Conclusion Validity}
Conclusion validity concerns the strength of the relationship between the evidence and the claims we make based on that evidence.
Our findings, such as the distribution of benchmarks across tasks and programming languages, are directly derived from the data we curated. However, small differences in interpretation or scope (e.g., how to classify multi-task benchmarks) could lead to slightly different conclusions in a replication. To strengthen conclusion validity, all analysis steps were documented, and quantitative statistics were cross-checked by multiple authors.

\subsection{External Validity}
External validity relates to how our findings can be generalized beyond this study.
This review specifically focuses on benchmarks used to evaluate LLMs in SE contexts. As such, our results may not generalize to benchmarks used in other domains (e.g., natural language understanding, computer vision, or multi-agent systems), or to LLM benchmarks that do not explicitly target SE tasks.
Moreover, due to the fast-evolving nature of LLM research, some newer benchmarks released after our data collection cutoff may not have been included. We encourage future replications to expand this work as the landscape evolves.

%% file: conclusion.tex
\section{Conclusion and Future Work}\label{sec:conclusion}
In this paper, we conducted a comprehensive survey of benchmarks used to evaluate LLMs in SE tasks. By systematically collecting and categorizing \papers~benchmarks across a variety of task domains and programming languages, we provided a structured overview of their key characteristics, including modalities, task coverage, and evaluation metrics. Our analysis highlights both well-established areas and notable gaps in the current benchmark landscape, offering insights into the maturity, focus, and diversity of LLM evaluation within the SE domain.

We hope that this survey serves as a valuable foundation for understanding the current state of LLM evaluation in SE and encourages further efforts to systematically study and enhance evaluation practices in this rapidly evolving field.
As part of our future work, we aim to address the aforementioned challenges. We also plan to develop a unified platform that consolidates diverse benchmarks to enable more comprehensive, flexible, and accessible evaluation of LLMs across multiple dimensions.

%% file: output.bbl

%% file: appendix.tex
\appendix

\section{Summary of studied benchmarks} \label{appendix:A}

\textbf{Requirements and Design}
\begin{multicols}{3}
    \begin{itemize}
        \item \href{https://tinyurl.com/y7tj9lkq}{NFR-Review~\cite{wang2018aspects}}
        \item \href{}{Rahman and Zhu~\cite{DBLP:journals/corr/abs-2404-01558}}
        \item \href{}{Habib et al.~\cite{habib2025reqbrain}}
        \item \href{https://figshare.com/s/46661b70336a46d66ac8}{Voria et al.~\cite{voria2025recover}}
        \item \href{https://terapromise.csc.ncsu.edu/!/#repo/view/head/requirements/nfr}{PROMISE~\cite{Sayyad-Shirabad+Menzies:2005}}
        \item \href{https://zenodo.org/records/4530183}{SecReq~\cite{SecReq}}
        \item \href{http://fmt.isti.cnr.it/nlreqdataset/}{PURE~\cite{ferrari2017pure}}
        \item \href{}{Dalpiaz et al.~\cite{dalpiaz2019requirements}}
        \item \href{https://zenodo.org/records/4471411}{ReqEval~\cite{abualhaija20203rd}}
        \item \href{}{DAMIR~\cite{ezzini2022automated}}
        \item \href{}{NFR-SO~\cite{luo2022prcbert}}
        \item \href{}{Gärtner and Göhlich~\cite{DBLP:journals/ase/GartnerG24}}
        \item \href{https://zenodo.org/records/10817760}{Preda et al.~\cite{preda2024supporting}}
        \item \href{https://zenodo.org/records/10652222}{Koltoff et al.~\cite{DBLP:conf/re/KolthoffKBMP24}}
        \item \href{https://github.com/albertogoffi/toradocu}{Jdoctor-dataset~\cite{blasi2018translating}}
        \item \href{}{DocTer-dataset~\cite{xie2022docter}}
        \item \href{https://tinyurl.com/RSA-Results}{Poudel et al.~\cite{poudel2023leveraging}}
        \item \href{}{Mandal et al.~\cite{mandal2023large}}
        \item \href{}{SV-Benchmarks~\cite{svbenchmarks2024}}
        \item \href{}{SpecGenBench~\cite{ma2024specgen}}
        \item \href{}{Reinpold et al.~\cite{DBLP:journals/corr/abs-2411-11582}}
        \item \href{}{Krishna et al.~\cite{DBLP:conf/re/KrishnaGVJ24}}
        \item \href{https://github.com/lishangyu- hkust/OSVBench}{OSVBench~\cite{DBLP:journals/corr/abs-2504-20964}}
        \item \href{https://github.com/MeloFancy/DeepCoref}{Wang et al.[2020]~\cite{wang2020deep}}
        \item \href{}{Helmeczi et al.~\cite{helmeczi2023few}}
    \end{itemize}
\end{multicols}

\textbf{Coding Assistant}
\begin{multicols}{3}
\begin{itemize}
    \item \href{}{Lin et al.~\cite{chen2023effectiveness}}
    \item \href{https://leetcode.com}{Leetcode~\cite{leetcode}}
    \item \href{}{Example-Check~\cite{zhang2018code}}
    \item \href{https://github.com/sriniiyer/concode}{CONCODE~\cite{iyer2018mapping}}
    \item \href{http://conala-corpus.github.io/}{CoNaLa~\cite{yin2018learning}}
    \item \href{https://github.com/TellinaTool/nl2bash/tree/master/data}{NL2Bash~\cite{lin2018nl2bash}}
    \item \href{https://yale-lily.github.io/spider}{Spider~\cite{yu2018spider}}
    \item \href{https://github.com/github/CodeSearchNet}{CodeSearchNet~\cite{DBLP:journals/corr/abs-1909-09436}}
    \item \href{https://github.com/hendrycks/apps}{APPS~\cite{hendrycksapps2021}}
    \item \href{https://huggingface.co/datasets/google-research-datasets/mbpp}{MBPP~\cite{DBLP:journals/corr/abs-2108-07732}}
    \item \href{https://github.com/microsoft/CodeXGLUE}{CodeXGLUE~\cite{lu2021codexglue}}
    \item \href{https://github.com/openai/human-eval}{HumanEval~\cite{chen2021codex}}
    \item \href{https://github.com/openai/miniF2F/tree/v1}{miniF2F~\cite{zheng2021minif2f}}
    \item \href{https://github.com/LIANGQINGYUAN/Lyra}{Lyra~\cite{liang2021lyra}}
    \item \href{https://github.com/LiuZeJie97/Code-Generation-FromFlowcharts-with-Texts-A-Benchmark-Dataset-and-AnApproach}{FC2Code~\cite{liu2022code}}
    \item \href{https://github.com/google-deepmind/code_contests}{CodeContests~\cite{li2022alphacode}}
    \item \href{https://github.com/aixcoder-plugin/nl2code-dataset}{AixBench~\cite{hao2022aixbench}}
    \item \href{https://github.com/amazon-science/recode}{ReCode~\cite{wang2022recode}}
    \item \href{https://github.com/s2e-lab/SecurityEval}{SecurityEval~\cite{siddiq2022securityeval}}
     \item \href{https://github.com/ejones313/codex-cog-biases}{MathEquations~\cite{jones2022capturing}}
     \item \href{https://github.com/amazon-research/mxeval}{MBXP~\cite{athiwaratkun2022multi}}
     \item \href{https://github.com/microsoft/PyCodeGPT}{NumpyEval~\cite{zan2022cert}}
     \item \href{https://github.com/microsoft/PyCodeGPT}{PandasEval~\cite{zan2022cert}}
     \item \href{https://github.com/microsoft/PyCodeGPT/tree/main/apicoder}{TorchData-Eval~\cite{zan2022language}}
     \item \href{https://github.com/microsoft/PyCodeGPT/tree/main/apicoder}{MonkeyEval~\cite{zan2022language}}
     \item \href{https://github.com/microsoft/PyCodeGPT/tree/main/apicoder}{BeatNumEval~\cite{zan2022language}}
     \item \href{https://github.com/salesforce/CodeGen/tree/main}{MTBP~\cite{nijkamp2022codegen}}
     \item \href{https://github.com/amazon-research/mxeval}{Multi-HumanEval~\cite{athiwaratkun2022multi}}
     \item \href{https://github.com/microsoft/DataScienceProblems}{DSP~\cite{chandel2022training}}
    \item \href{https://github.com/Jun-jie-Huang/ExeDS}{ExeDS~\cite{huang2022execution}}
    \item \href{https://github.com/reddy-lab-code-research/XLCoST}{XLCoST~\cite{DBLP:journals/corr/abs-2206-08474}}
    \item \href{}{Qing et al.~\cite{huang2023adaptive}}
    \item \href{https://github.com/FudanSELab/ClassEval}{ClassEval~\cite{du2023classeval}}
    \item \href{https://github.com/FlagOpen/TACO}{TACO~\cite{li2023taco}}
    \item \href{https://github.com/ntunlp/xCodeEval}{xCodeEval~\cite{khan2023xcodeeval}}
    \item \href{https://apex.sjtu.edu.cn/codeapex/}{CodeApex~\cite{fu2023codeapex}}
    \item \href{https://github.com/ChuyueSun/Clover}{CloverBench~\cite{sun2023clover}}
    \item \href{https://github.com/antonio-mastropaolo/robustness-copilot}{Mastropaolo et al.~\cite{mastropaolo2023robustness}}
    \item \href{https://github.com/CoderEval/CoderEval}{CoderEval~\cite{yu2024codereval}}
    \item \href{https://github.com/evalplus/evalplus}{EvalPlus~\cite{liu2024your}}
    \item \href{}{Shapkin et al.~\cite{shapkin2023entity}}
    \item \href{https://github.com/NougatCA/CrossCodeBench}{CrossCode-Bench~\cite{niu2023crosscodebench}}
    \item \href{http://github.com/nuprl/MultiPL-E}{MultiPL-E~\cite{cassano2023multipl}}
    \item \href{https://huggingface.co/datasets/wellesley-easel/StudentEval}{StudentEval~\cite{babe2023studenteval}}
    \item \href{}{TorchDataComplexEval~\cite{zan2023private}}
    \item \href{https://ds1000-code-gen.github.io/}{DS-1000~\cite{lai2023ds}}
    \item \href{https://github.com/gersteinlab/ML-bench}{ML-Bench~\cite{tang2023ml}}
    \item \href{https://github.com/lowcoder-org/lowcoder}{LowCoder~\cite{rao2024ai}}
    \item \href{https://github.com/goodchar123/KPC}{Ren et al.~\cite{ren2023misuse}}
    \item \href{https://github.com/sahil280114/codealpaca}{CodeAlpaca (Py)~\cite{chaudhary2023code}}
    \item \href{}{CoLadder~\cite{yen2023coladder}}
    \item \href{https://github.com/shailja-thakur/VGen}{VeriGen~\cite{thakur2024verigen}}
    \item \href{https://paperswithcode.com/dataset/soeval}{SOEval~\cite{siddiq2023lightweight}}
    \item \href{}{Decept-Prompt~\cite{wu2023deceptprompt}}
    \item \href{https://github.com/THUDM/CodeGeeX}{HumanEval-X~\cite{codegeex}}
    \item \href{}{ARCADE~\cite{yin2023natural}}
    \item \href{https://github.com/zorazrw/multilingual-conala}{MCoNaLa~\cite{wang2023mconala}}
    \item \href{https://crosscodeeval.github.io/}{CrossCode-Eval~\cite{ding2024crosscodeeval}}
    \item \href{}{Pisces~\cite{yang2023syntax}}
    \item \href{https://github.com/Dingjz/CodeScore}{MBPP/HumanEval/APPS-ET~\cite{dong2023codescore}}
    \item \href{https://livecodebench.github.io}{LiveCode-Bench~\cite{jain2024livecodebench}}
    \item \href{https://github.com/Elfsong/Mercury}{Mercury~\cite{du2024mercury}}
    \item \href{https://huggingface.co/spaces/EffiBench/effibench-leaderboard}{EffiBench~\cite{huang2024effibench}}
    \item \href{https://github.com/Mondego/dafny-synthesis}{MBPP-san-DFY~\cite{misu2024towards}}
    \item \href{https://github.com/WisdomShell/ujb}{CoderUJB~\cite{DBLP:journals/corr/abs-2403-19287}}
    \item \href{https://anonymous.4open.science/r/PythonSaga}{PythonSaga~\cite{yadav2024boldly}}
    \item \href{https://github.com/seketeam/DevEval}{DevEval~\cite{li2024deveval}}
    \item \href{https://github.com/Veronicium/CodeBenchGen}{Exec-CSN~\cite{xie2024codebenchgen}}
    \item \href{}{Wang et al.~\cite{wang2024teaching}}
    \item \href{https://github.com/seketeam/EvoCodeBench}{EvoCode-Bench~\cite{li2024evocodebench}}
    \item \href{https://github.com/sfikas/rusteval}{RustEval~\cite{pan2024enhancing}}
    \item \href{https://github.com/open-compass/DevBench}{Devbench~\cite{li2024devbench}}
    \item \href{https://bigcode-bench.github.io/}{BigCode-Bench~\cite{zhuo2024bigcodebench}}
    \item \href{https://github.com/alphadl/OOP-eval}{OOPEval~\cite{wang2024oop}}
    \item \href{https://github.com/zorazrw/odex}{ODEX~\cite{wang2022execution}}
    \item \href{https://github.com/THUDM/NaturalCodeBench}{NaturalCodeBench~\cite{zhang2024naturalcodebench}}
    \item \href{https://github.com/parallelcodefoundry/ParEval}{PAREval~\cite{nichols2024can}}
    \item \href{https://drive.google.com/drive/folders/1i0UWqy1K4WwaCLnb7yhyQfV8UqjdXSkl}{CAASD~\cite{zhang2024experimenting}}
    \item \href{https://github.com/WeixiangYAN/CodeScope}{CodeScope~\cite{DBLP:conf/acl/YanLWL0W0ZZSD24}}
    \item \href{https://github.com/zkcpku/CodeAgent}{CodeAgent-Bench~\cite{zhang2024codeagent}}
    \item \href{https://github.com/java-bench/JavaBench}{JavaBench~\cite{cao2024javabench}}
    \item \href{https://github.com/Ryan-Holben/holbench}{Chart2Code-160k~\cite{DBLP:conf/icse/Rodriguez-Cardenas24}}
    \item \href{https://github.com/ythere-y/PoorCodeSumEval}{PoorCodeSumEval~\cite{DBLP:conf/kbse/HuCZM0024}}
    \item \href{https://github.com/ComplexCodeEval/ComplexCodeEval}{ComplexCodeEval~\cite{DBLP:conf/kbse/FengLGCWG024}}
    \item \href{https://github.com/ProsusAI/stack-eval}{StackEval~\cite{DBLP:conf/nips/ShahGA24}}
    \item \href{https://github.com/wanghanbinpanda/CodeVision?tab=readme-ov-file}{Code-Vision~\cite{DBLP:journals/corr/abs-2502-11829}}
    \item \href{https://github.com/zhu-zhu-ding/CodeIF-Bench}{CodeIF-Bench~\cite{DBLP:journals/corr/abs-2503-22688}}
    \item \href{https://github.com/lin-rany/codeIF}{CodeIF~\cite{DBLP:journals/corr/abs-2502-19166}}
    \item \href{https://lib-evolution-eval.github.io}{LibEvolutionEval~\cite{DBLP:conf/naacl/KuharAWJQRRMD25}}
    \item \href{https://github.com/JohnnyPeng18/Coffe?tab=readme-ov-file}{COFFE~\cite{DBLP:journals/corr/abs-2502-02827}}
    \item \href{https://anonymous.4open.science/r/DL-Bench-D65E/}{Deep-Bench~\cite{DBLP:journals/corr/abs-2502-18726}}
    \item \href{https://github.com/HWH-2000/DynaCode}{DynaCode~\cite{DBLP:journals/corr/abs-2503-10452}}
    \item \href{https://github.com/microsoft/FEA-Bench}{FEA-Bench~\cite{DBLP:journals/corr/abs-2503-06680}}
    \item \href{}{MaintainCoder~\cite{DBLP:journals/corr/abs-2503-24260}}
    \item \href{https://github.com/mraihan-gmu/mHumanEval-Benchmark}{mHumanEval~\cite{DBLP:conf/naacl/RaihanAZ25}}
    \item \href{https://github.com/FSoft-AI4Code/RepoExec}{REPOEXEC~\cite{DBLP:conf/naacl/HaiNB25}}
    \item \href{https://huggingface.co/datasets/TencentARC/Plot2Code}{Plot2Code~\cite{DBLP:conf/naacl/WuLGGLWSL25}}
    \item \href{https://github.com/RyanLoil/ProjectEval/}{ProjectEval~\cite{DBLP:journals/corr/abs-2503-07010}}
    \item \href{https://anonymous.4open.science/r/SolEval-1C06/README.MD}{SolEval~\cite{DBLP:journals/corr/abs-2502-18793}}
    \item \href{https://huggingface.co/spaces/ConvCodeWorld/ConvCodeWorld}{ConvCodeWorld~\cite{DBLP:conf/iclr/HanHSH25}}
    \item \href{https://huggingface.co/datasets/bytedance-research/Web-Bench}{Web-Bench~\cite{xu2025web}}
    \item \href{https://huggingface.co/datasets/lt-asset/REPOCOD}{REPOCOD~\cite{repocod}}
    \item \href{https://github.com/Avmb/code-docstring-corpus}{PCSD~\cite{DBLP:conf/ijcnlp/BaroneS17}}
    \item \href{https://github.com/xinghu/TL-CodeSum}{JCSD~\cite{DBLP:conf/ijcai/HuLXLLJ18}}
    \item \href{https://github.com/huxingfree/DeepCom}{Deepcom~\cite{DBLP:conf/iwpc/HuLXLJ18}}
    \item \href{}{Funcom~\cite{DBLP:conf/icse/LeClairJM19}}
    \item \href{https://github.com/apcl-research/jam}{Funcom-java-long~\cite{DBLP:conf/sigsoft/SuBJGM23}}
    \item \href{https://github.com/RosenZhang/CroCoSum}{CroCoSum~\cite{DBLP:conf/coling/ZhangE24}}
    \item \href{https://github.com/AISE-TUDelft/Capybara-BinT5}{CAPYBARA~\cite{DBLP:journals/corr/abs-2301-01701}}
    \item \href{https://github.com/xinjin95/BinSum}{BinSum~\cite{DBLP:journals/corr/abs-2312-09601}}
    \item \href{https://github.com/Linshuhuai/P-CodeSum}{P-CodeSum~\cite{DBLP:journals/jss/YunLGS24}}
    \item \href{https://github.com/DeepSoftwareAnalytics/SparseCoder}{FILE-CS~\cite{DBLP:conf/wcre/WangHGZZ24}}
    \item \href{https://github.com/github/CodeSearchNet}{CodeNet~\cite{DBLP:conf/nips/Puri0JZDZD0CDTB21}}
    \item \href{https://github.com/reddy-labcode-research/MuST-CoST}{CoST~\cite{DBLP:conf/aaai/Zhu0R22}}
    \item \href{}{Nova~\cite{DBLP:journals/corr/abs-2311-13721}}
    \item \href{}{SUT~\cite{DBLP:conf/emnlp/QiHWYLGCS23}}
    \item \href{https://github.com/WeixiangYAN/CodeTransOcean}{CodeTransOcean~\cite{DBLP:conf/emnlp/YanTLCW23}}
    \item \href{https://github.com/PolyEval/G-TransEval}{G-TransEval~\cite{DBLP:conf/kbse/JiaoYLQGS23}}
    \item \href{https://github.com/wasiahmad/AVATAR}{AVATAR~\cite{DBLP:conf/acl/AhmadTCC23}}
    \item \href{https://anonymous.4open.science/r/CoTran}{AVATAR-TC~\cite{jana2023cotran}}
    \item \href{https://github.com/SYSUSELab/RustRepoTrans/}{RustRepoTrans~\cite{ou2024repository}}
    \item \href{https://crux-eval.github.io}{CRUXEval~\cite{REva}}
    \item \href{https://r-eval.github.io}{REval~\cite{REva}}
    \item \href{https://codekaleidoscope.github.io/dycodeeval.html}{DyCodeEval~\cite{chen2025dynamic}}
\end{itemize}
\end{multicols}

\textbf{Software Testing}
\begin{multicols}{3}
    \begin{itemize}
    \item \href{https://www.evosuite.org/experimental-data/sf100/}{Evosuite SF110~\cite{DBLP:journals/tosem/FraserA14}}
    \item \href{https://github.com/rjust/defects4j}{Defects4J~\cite{just2014defects4j}}
    \item \href{https://www.evosuite.org/experimental-data/evolutionary-algorithm-study/}{DynaMOSA~\cite{DBLP:journals/tse/PanichellaKT18}}
    \item \href{https://github.com/soarsmu/BugsInPy}{BugsInPy~\cite{DBLP:conf/sigsoft/WidyasariSLQPTT20}}
    \item \href{https://github.com/openai/human-eval}{HumanEval~\cite{chen2021codex}}
    \item \href{https://huggingface.co/datasets/google-research-datasets/mbpp}{MBPP~\cite{DBLP:journals/corr/abs-2108-07732}}
    \item \href{https://github.com/hendrycks/apps}{APPS~\cite{hendrycksapps2021}}
    \item \href{https://github.com/google-deepmind/code_contests}{CodeContests~\cite{li2022alphacode}}
    \item \href{https://github.com/THUDM/CodeGeeX}{HumanEval-X~\cite{codegeex}}
    \item \href{https://github.com/WisdomShell/ujb}{CoderUJB~\cite{DBLP:journals/corr/abs-2403-19287}}
    \item \href{https://github.com/logic-star-ai/SWT-Bench}{SWT-Bench~\cite{swt-bench}}
    \item \href{https://github.com/iSEngLab/TestBench}{TestBench~\cite{DBLP:journals/corr/abs-2409-17561}}
    \item \href{https://github.com/LLM4SoftwareTesting/TestEval}{TestEval~\cite{wang-etal-2025-testeval}}
    \item \href{https://github.com/YiboWANG214/ProjectTest}{ProjectTest~\cite{DBLP:journals/corr/abs-2502-06556}}
    \item \href{https://sites.google.com/view/atlas-nmt/home}{ATLAS~\cite{DBLP:conf/icse/WatsonTMBP20}}
    \item \href{https://github.com/the-themis-benchmarks/home}{Themis~\cite{DBLP:conf/sigsoft/0001WS21}}
    \item \href{https://github.com/franklinbill/QTypist}{QTypist~\cite{DBLP:conf/icse/LiuCWCHHW23}}
    \item \href{https://github.com/panda-re/lava}{LAVA-M~\cite{DBLP:conf/sp/Dolan-GavittHKL16}}
    \item \href{https://github.com/unifuzz/unibench}{Unibench~\cite{DBLP:conf/uss/LiJCLLCLWBCL021}}
    \item \href{https://google.github.io/fuzzbench}{FuzzBench~\cite{DBLP:conf/sigsoft/MetzmanSSSA21}}
    \item \href{https://github.com/ise-uiuc/FuzzGPT}{FuzzGPT~\cite{DBLP:journals/corr/abs-2304-02014}}
    \item \href{https://github.com/TestingResearchIllinois/idoft}{IDoFT~\cite{DBLP:conf/icst/LamOSM019}}
    \item \href{https://github.com/AlshammariA/FlakeFlagger}{FlakeFlagger~\cite{DBLP:conf/icse/AlshammariMH021}}
    \item \href{https://figshare.com/articles/dataset/_b_TaRBench_A_Comprehensive_Benchmark_for_Automated_Test_Case_Repair_b_/25008893}{TARBENCH~\cite{DBLP:journals/tse/YaraghiHKB25}}
    \item \href{https://sites.google.com/view/utfix}{Syn-Bench~\cite{DBLP:journals/pacmpl/RahmanKCGWMDR25}}
\end{itemize}
\end{multicols}

\textbf{AIOps}
\begin{multicols}{3}
\begin{itemize}
    \item \href{https://github.com/antonio-mastropaolo/LANCE}{LANCE~\cite{mastropaolo2022using}}
    \item \href{https://github.com/LoggingResearch/LoggingEmpirical}{LogBench~\cite{li2023exploring}}
    \item \href{https://github.com/YichenLi00/SCLogger}{SCLoger~\cite{li2024go}}
    \item \href{https://github.com/shuaijiumei/logging-benchmark}{AL-Bench~\cite{tan2025bench}}
    \item \href{https://github.com/logpai/loghub}{Loghub~\cite{zhu2023loghub}}
    \item \href{https://zenodo.org/records/8275861}{Loghub-2.0~\cite{jiang2024large}}
\end{itemize}
\end{multicols}

\textbf{Maintenance}
\begin{multicols}{3}
\begin{itemize}
    \item \href{https://zenodo.org/records/5387856#.YTDrPZ4zZyo}{CodeReview~\cite{tufano2022using}}
    \item \href{https://zenodo.org/records/6900648}{CodeReviewer~\cite{li2022automating}}
    \item \href{https://gitlab.com/ai-for-se-public-data/auger-fse-2022}{AUGER~\cite{li2022auger}}
    \item \href{https://figshare.com/s/135201b8f87ab705448b}{Review-Explaining~\cite{widyasari2023explaining}}
    \item \href{https://sites.google.com/view/chatgptcodereview}{CodeReview-New~\cite{guo2024exploring}}
   \item \href{https://zenodo.org/records/7533156}{Code-Review-Assist~\cite{sghaier2023multi}}
   \item \href{https://zenodo.org/records/13150598}{ManualReviewComment~\cite{DBLP:journals/corr/abs-2502-02757}}
   \item \href{https://github.com/clonebench/BigCloneBench}{BigCloneBench~\cite{svajlenko2015evaluating}}
   \item \href{https://drive.google.com/file/d/0B2i-vWnOu7MxVlJwQXN6eVNONUU/view?resourcekey=0-Po3kiAifLfCCYnanCBDMHw}{POJ-104~\cite{mou2016convolutional}}
   \item \href{https://github.com/SFI-Lero/SSCD/tree/main}{Company-C/C++~\cite{chochlov2022using}}
   \item \href{https://github.com/srlabUsask/GPTCloneBench}{GPTCloneBench~\cite{alam2023gptclonebench}}
   \item \href{https://github.com/LLM4CodeClone/LLM4CodeClone}{Curated CodeNet~\cite{dou2023towards}}
   \item \href{https://drive.google.com/drive/folders/1aw2yiUTXwB3gJrDcFWeDpYvGgJNYjt51}{JavaRef~\cite{liu2023refbert}}
\end{itemize}
\end{multicols}

 \textbf{Quality Management}
\begin{multicols}{3}
    \begin{itemize}
    \item \href{https://github.com/bugs-dot-jar/bugs-dot-jar}{Bugs.jar~\cite{saha2018bugs}}
    \item \href{https://github.com/bears-bugs}{Bears~\cite{madeiral2019bears}}
    \item \href{https://github.com/ZZR0/ISSTA21-JIT-DP}{Zeng et al.~\cite{zeng2021deep}}
    \item \href{https://github.com/jacknichao/JIT-Fine}{JIT-defects4j~\cite{ni2022best}}
    \item \href{}{Opu et al.~\cite{opu2025llm}}
    \item \href{http://dx.doi.org/10.6084/m9.figshare.951967}{Ye et al.~\cite{ye2014learning}}
    \item \href{https://github.com/rjust/defects4j}{Defects4J~\cite{just2014defects4j}}
    \item \href{https://github.com/exatoa/Bench4BL}{Bench4BL~\cite{lee2018bench4bl}}
    \item \href{https://github.com/epicosy/devign}{Devign~\cite{zhou2019devign}}
    \item \href{https://github.com/soarsmu/BugsInPy}{BugsInPy~\cite{DBLP:conf/sigsoft/WidyasariSLQPTT20}}
    \item \href{https://github.com/pkuzqh/Recoder}{Zhu et al.~\cite{zhu2021syntax}}
    \item \href{https://zenodo.org/records/6900648}{CodeReviewer~\cite{li2022automating}}
    \item \href{https://anonymous.4open.science/r/fbl-bert-700C/README.md}{Ciborowska et al.~\cite{ciborowska2022fast}}
    \item \href{https://lamda.nju.edu.cn/mayf/sgAttention-Code.zip}{Ma et al.~\cite{ma2023capturing}}
    \item \href{https://github.com/hkust-zhiyao/RTLLM}{RTLLM~\cite{lu2024rtllm}}
    \item \href{https://zenodo.org/records/15122980}{BeetleBox~\cite{chakraborty2024blaze}}
    \item \href{https://www.swebench.com}{SWE-Bench~\cite{jimenez2024swe}}
    \item \href{}{Chandramohan et al.~\cite{chandramohan2024supporting}}
    \item \href{}{Stracquadanio et al.~\cite{stracquadanio2024veribug}}
    \item \href{https://github.com/Ruchira-1/Theia?tab=readme-ov-file}{Manke et al.~\cite{manke2024leveraging}}
    \item \href{}{D580~\cite{yan2024better}}
    \item \href{https://zenodo.org/records/12692994}{Saha et al.~\cite{saha2024toward}}
    \item \href{https://figshare.com/s/bce02cb607a6c30043ad?file=47354473}{Widyasari et al.~\cite{widyasari2024demystifying}}
    \item \href{https://github.com/FudanSELab/LinuxFLBench}{LINUXFLBENCH~\cite{zhou2025benchmarkingenhancingllmagents}}
    \item \href{https://github.com/zhenlongDai/AdaPatcher}{ACPR~\cite{dai2025less}}
    \item \href{https://github.com/jkoppel/QuixBugs}{QuixBugs~\cite{QuixBugs}}
    \item \href{https://github.com/zhiyufan/apr4codex}{LMDefects~\cite{fan2023automated}}
    \item \href{https://github.com/microsoft/InferredBugs}{InferredBugs~\cite{InferredBugs}}
    \item \href{}{ARHE~\cite{kang2023explainable}}
    \item \href{}{Leetcode-debug~\cite{sakib2023extending}}
    \item \href{https://anonymous.4open.science/r/TOSEM-API-Misuse}{API-Misuse-Repair~\cite{API-Misuse-Repair}}
    \item \href{https://github.com/thunlp/DebugBench}{DebugBench~\cite{tian2024debugbench}}
    \item \href{https://www.swebench.com/multimodal}{SWE-bench Multimodal~\cite{yang2024swe}}
    \item \href{https://github.com/openai/SWELancer-Benchmark}{SWE-Lancer~\cite{miserendino2025swe}}
    \item \href{https://github.com/multi-swe-bench/multi-swe-bench/tree/main}{Multi-SWE-bench~\cite{zan2025multi}}
    \item \href{https://github.com/mjc92/buffer_overrun_memory_networks}{Choi et al.~\cite{jeong2017end}}
    \item \href{}{Lin et al.~\cite{lin2017poster}}
    \item \href{https://dbgbench.github.io}{DGBBench~\cite{bohme2017developers}}
    \item \href{https://samate.nist.gov/SARD/test-suites/112}{Juliet~\cite{black2018juliet}}
    \item \href{https://github.com/CGCL-codes/VulDeePecker}{VulDeePecker~\cite{li2018vuldeepecker}}
    \item \href{https://osf.io/d45bw/}{Draper~\cite{russell2018draper}}
    \item \href{https://github.com/SAP/project-kb}{Ponta et al.~\cite{ponta2019manually}}
    \item \href{https://github.com/ZeoVan/MSR_20_Code_vulnerability_CSV_Dataset}{BigVul~\cite{fan2020bigvul}}
    \item \href{https://github.com/VulDetProject/ReVeal}{ReVeal~\cite{chakraborty2021reveal}}
    \item \href{https://smartbugs.github.io}{SmartBugs~\cite{ferreira2020smartbugs}}
    \item \href{https://zenodo.org/records/3954944}{Great~\cite{hellendoorn2019global}}
    \item \href{https://hexhive.epfl.ch/magma/}{Magma~\cite{hazimeh2020magma}}
    \item \href{https://github.com/DependableSystemsLab/SolidiFI-benchmark}{SolidiFI~\cite{ghaleb2020effective}}
    \item \href{https://github.com/SySeVR/SySeVR}{SySeVR~\cite{li2021sysevr}}
    \item \href{https://github.com/ibm/D2A}{D2A~\cite{zheng2021d2a}}
    \item \href{https://github.com/SunLab-GMU/PatchDB}{PatchDB~\cite{wang2021patchdb}}
    \item \href{https://github.com/secureIT-project/CVEfixes}{CVEFixes~\cite{bhandari2021cvefixes}}
    \item \href{https://doi.org/10.5281/zenodo.4734050}{CrossVul~\cite{nikitopoulos2021crossvul}}
    \item \href{https://figshare.com/s/0f3ed11f9348e2f3a9f8}{VCmatch~\cite{wang2022vcmatch}}
    \item \href{https://github.com/LauraWartschinski/VulnerabilityDetection?tab=readme-ov-file}{VUDENC~\cite{wartschinski2022vudenc}}
    \item \href{https://samate.nist.gov/SARD}{SARD~\cite{black2017sard}}
    \item \href{https://github.com/wagner-group/diversevul}{DiverseVul~\cite{chen2023diversevul}}
    \item \href{https://github.com/ZhangZhuoSJTU/Web3Bugs}{Web3Bugs~\cite{zhang2023web3bugs}}
    \item \href{https://wooded-meter-1d8.notion.site/0e85e02c5ed34df3855ea9f3ca40f53b?v=22e5e2c506ef4caeb40b4f78e23517ee}{DeFi Hacks~\cite{defi_hacks_analysis}}
    \item \href{https://github.com/Hustcw/VulBench}{VulBench~\cite{gao2023vulbench}}
    \item \href{https://owasp.org/www-project-benchmark}{OWASP~\cite{owasp}}
    \item \href{https://figshare.com/articles/online_resource/TreeVul_-_Replication_Package/19727050}{TreeVul~\cite{pan2023fine}}
    \item \href{https://github.com/FormAI-Dataset}{FormAI~\cite{tihanyi2023formai}}
    \item \href{https://github.com/git-disl/GPTLens}{Hu et al.~\cite{hu2023large}}
    \item \href{}{FalconVulnDB~\cite{ferrag2023securefalcon}}
    \item \href{https://github.com/FormAI-Dataset}{FormAI-v2~\cite{tihanyi2024neutral}}
    \item \href{https://github.com/JafarAkhondali/Morefixes}{MoreFixes~\cite{akhoundali2024morefixes}}
    \item \href{}{VulEval~\cite{wen2024vuleval}}
    \item \href{https://github.com/CGCLcodes/VulTrigger}{InterPVD~\cite{li2024effectiveness}}
    \item \href{https://github.com/Eshe0922/ReposVul}{ReposVul~\cite{wang2024reposvul}}
    \item \href{https://github.com/Icyrockton/MegaVul}{MegaVul~\cite{ni2024megavul}}
    \item \href{https://github.com/ai4cloudops/SecLLMHolmes}{SecLLMHolmes~\cite{ullah2024llms}}
    \item \href{https://github.com/Sweetaroo/VulDetectBench}{VulDetectBench~\cite{liu2024vuldetectbench}}
    \item \href{}{SC-LOC~\cite{zhang2024empirical}}
    \item \href{https://sites.google.com/view/iaudittool/home}{Ma et al.~\cite{ma2024combining}}
    \item \href{https://drive.google.com/drive/folders/1uSXaY7vOvcwQIwXs5JwD9C2hxK9bFMsZ}{FELLMVP~\cite{luo2024fellmvp}}
    \item \href{}{Yıldırım et al.~\cite{yildirim2024evaluating}}
    \item \href{}{Vulcorpus~\cite{kouliaridis2024assessing}}
    \item \href{}{Fang et al.~\cite{fang2024llm}}
    \item \href{}{SLFHunter~\cite{ye2024detecting}}
    \item \href{https://zenodo.org/records/10975439}{Guo et al.~\cite{guo2024outside}}
    \item \href{https://github.com/niklasrisse/VPP}{VulnPatchPairs~\cite{risse2024uncovering}}
    \item \href{https://zenodo.org/records/12707476}{Real-Vul~\cite{chakraborty2024revisiting}}
    \item \href{}{PairVul~\cite{du2024vul}}
    \item \href{}{VulSmart~\cite{alam2024detection}}
    \item \href{}{KernJC~\cite{ruan2024kernjc}}
    \item \href{https://anonymous.4open.science/r/LLM4Vuln/README.md}{LLM4Vuln~\cite{sun2024llm4vuln}}
    \item \href{https://github.com/NUS-Curiosity/VulZoo}{VULZOO~\cite{ruan2024vulzoo}}
    \item \href{https://github.com/iris-sast/iris}{CWE-Bench-Java~\cite{li2405llm}}
    \item \href{https://github.com/CASTLE-Benchmark}{CASTLE~\cite{dubniczky2025castle}}
    \item \href{https://huggingface.co/datasets/arag0rn/SecVulEval}{SecVulEval~\cite{ahmed2025secvuleval}}
    \item \href{https://huggingface.co/datasets/AfterQuery/vader}{VADER~\cite{liu2025vader}}
    \item \href{https://anonymous.4open.science/r/JitVul-C6C7/}{JITVUL~\cite{yildiz2025benchmarking}}
    \item \href{https://github.com/soarsmu/SVD-Bench}{Li et al.~\cite{zhang2025benchmarking}}
    \item \href{https://docs.google.com/spreadsheets/d/1qztIwB8xJ10H-2HLX15vI29Ze7yFDOrv7kDQ4JUi1g8/edit?gid=1679944928#gid=1679944928}{BinPool~\cite{arasteh2025binpool}}
    \item \href{https://github.com/Chaomeng-Lu/ICVul}{ICVul~\cite{lu2025icvul}}
    \item \href{https://repairbenchmarks.cs.umass.edu}{ManyBugs~\cite{le2015manybugs}}
    \item \href{https://github.com/ASSERT-KTH/VRepair}{Chen et al.~\cite{chen2022neural}}
    \item \href{https://github.com/VulDeeLocator/VulDeeLocator}{VulDeeLocator~\cite{li2021vuldeelocator}}
    \item \href{https://extractfix.github.io}{ExtractFix~\cite{gao2021beyond}}
    \item \href{https://zenodo.org/records/13118970}{CVE-Fixes~\cite{bhandari2021cvefixes}}
    \item \href{https://github.com/tuhh-softsec/vul4j}{Vul4J~\cite{bui2022vul4j}}
    \item \href{https://github.com/s2e-lab/SecurityEval}{SecurityEval~\cite{siddiq2022securityeval}}
    \item \href{https://github.com/lin-tan/llm-vul}{VulRep~\cite{wei2023vulrep}}
    \item \href{https://github.com/lin-tan/llm-vul}{VJBench~\cite{wu2023vjbench}}
    \item \href{https://github.com/eth-sri/sven}{Sven~\cite{he2023sven}}
    \item \href{https://zenodo.org/records/7199939}{Pearce et al.~\cite{pearce2023examining}}
    \item \href{https://zenodo.org/records/8256377}{SmartFix~\cite{so2023smartfix}}
    \item \href{https://zenodo.org/records/11084782}{ManyVuls4J~\cite{lin2024there}}
    \item \href{}{ARVO~\cite{mei2024arvo}}
    \item \href{https://zenodo.org/records/10783763}{Le et al.~\cite{le2024study}}
    \item \href{}{Khan et al.~\cite{khan2024code}}
\end{itemize}
\end{multicols}

\section{Descriptions of Benchmark Categories} \label{appendix:B}

\begin{table}[h]
\centering
\caption{Descriptions of Benchmark Categories.}
\begin{tabular}{p{4cm}p{9cm}}
\hline
\textbf{Category} & \textbf{Description} \\
\hline
Requirements and Design & Benchmarks for extracting, analyzing, validating, and managing software requirements and design artifacts. \\ \hline
Coding Assistant & Benchmarks assessing LLMs’ ability to generate, recommend, translate, and reason about source code. \\ \hline 
Software Testing & Benchmarks for automatically generating and repairing test artifacts to ensure correctness and reliability. \\ \hline
AIOps & Benchmarks focused on operational tasks, such as generating and parsing logs for monitoring and debugging. \\ \hline
Maintenance & Benchmarks for maintaining software systems, including code review, clone detection, and code refactoring. \\ \hline
Quality Management & Benchmarks for predicting, detecting, localizing, and repairing defects or vulnerabilities to ensure software quality. \\
\hline
\end{tabular}
\end{table}

\section{Granularity Levels and Definitions for Code Generation and Recommendation} \label{appendx:C}

\begin{table}[h]
\centering
\caption{Granularity Levels and Definitions for Code Generation and Recommendation.}
\begin{tabular}{p{3.2cm} p{11cm}}
\hline
\textbf{Granularity} & \textbf{Definition} \\
\hline
Token-level & The smallest granularity, where the task is to predict or classify the next token in a sequence of code or text. \\
Line-level & Tasks focused on generating, repairing, or evaluating a single line of code. \\
Snippet-level & Involves short code fragments (smaller than a full function), typically capturing a small logical unit. \\
Statement-level & Operates at the granularity of a complete statement (larger than a token/line but smaller than a full function), often requiring semantic correctness beyond syntax. \\
Instruction-level & Benchmarks where each item is framed as an instruction or prompt that guides code generation or modification. The model must follow natural language instructions. \\
Function-level & Requires generating or understanding a complete function that implements a specified behavior. \\
API-level & Tasks designed around the correct usage and integration of APIs, assessing whether the model can call, compose, or adapt APIs properly. \\
Solution-level & Involves producing or analyzing a full solution to a programming problem, which may span multiple functions. \\
Repo-level & Benchmarks that operate at the repository level, requiring reasoning across multiple files and modules. \\
Library-level & Evaluation at the library or framework level, requiring the model to understand APIs and interactions across large-scale libraries. \\
Contest-level & The most challenging granularity, involving complex programming contest problems that demand optimization and multi-step reasoning. \\
Turducken-Style & A composite granularity where tasks are nested across multiple levels (e.g., combining token, function, and repo reasoning), designed to stress-test models across contexts. \\
\hline
\end{tabular}
\end{table}